\documentclass[12pt]{article}
\usepackage{amsmath,amssymb,bm,epsf,epsfig,graphicx,hyperref,physics, cleveref, subfig, slashed, stmaryrd, tensor, float}
\usepackage[backend=bibtex,natbib,style=numeric-comp,sorting=none, doi=false, isbn=false,url=false]{biblatex}
\input epsf.sty
\usepackage[]{todonotes}
\usepackage{tikz}
\usetikzlibrary{arrows.meta}

\usepackage[weather, alpine]{ifsym}

\definecolor{bisque}{rgb}{1.0, 0.89, 0.77}
\definecolor{forestgreen(web)}{rgb}{0.13, 0.55, 0.13}

\def\be{\begin{equation}}
\def\ee{\end{equation}}
\def\bea{\begin{eqnarray}}
\def\eea{\end{eqnarray}}
\def\ie{\begin{equation}\begin{aligned}}
\def\fe{\end{aligned}\end{equation}}
\numberwithin{equation}{section}

\renewcommand{\title}[1]{\vbox{\center\LARGE{#1}}\vspace{5mm}}
\renewcommand{\author}[1]{\vbox{\center#1}\vspace{5mm}}
\newcommand{\address}[1]{\vbox{\center\em#1}}
\newcommand{\email}[1]{\vbox{\center\tt#1}\vspace{5mm}}

\newcommand{\mmu}{\boldsymbol{\mu}}
\newcommand{\ww}{\boldsymbol{w}}
\newcommand{\XX}{\mathbf{X}}

\newcommand{\A}{{\alpha}}

\newcommand{\annulus}{\bigcirc\mkern-13.5mu\circ}

\begin{document}

\unitlength = .8mm

\begin{titlepage}

\begin{center}

\hfill \\
\hfill \\
\vskip 1cm

\title{D-instanton Effects on a D3-brane}

\author {Jaroslav Scheinpflug, Yuchen Wang, Xi Yin}

\address{Jefferson Physical Laboratory, Harvard University,
Cambridge, MA 02138 USA
}

\email{jscheinpflug@g.harvard.edu, yuchen\_wang@fas.harvard.edu, xiyin@fas.harvard.edu}

\end{center}

\abstract{It has been proposed by Sen that the D-instanton effects in string theory can be systematically determined through the framework of open-closed string field theory. We apply the latter formalism to analyze the D-instanton corrections to the quantum effective action of a D3-brane in type IIB superstring theory, and determine the leading single and multi-instanton contributions to the $D^4 F^4$ effective coupling which is unprotected by supersymmetry. Notably, while we find that the one-instanton contribution agrees with a conjecture of Green and Gutperle, the multi-instanton contribution disagrees with the conjecture.}

\vfill

\end{titlepage}

\eject

\begingroup
\hypersetup{linkcolor=black}

\tableofcontents

\endgroup

\section{Introduction}
\label{intro}

The worldsheet formalism of string theory captures, in addition to the perturbative genus expansion, a specific type of non-perturbative effects known as D-instantons \cite{Polchinski:1994fq}. In the 10-dimensional Minkowskian vacuum of type IIB superstring theory, a basic class of D-instantons are D(-1)-branes, which may be characterized as pointlike defects in the Euclidean spacetime that support open strings, and mediate non-perturbative processes by absorbing and emitting closed strings. In particular, the D-instanton contributions to closed string amplitudes are essential for the consistency of S-duality \cite{Green:1997tv}.

In this paper, we investigate the D-instanton effects in the open string dynamics on a D3-brane. The D(-1)-D3 system was studied extensively from the perspective of the massless effective gauge theory in \cite{Witten:1995gx, Douglas:1995bn, Douglas:1996uz, Billo:2002hm, Billo:2004zq, Billo:2005fg, Bianchi:2015cta, Maccaferri:2018vwo, Maccaferri:2019ogq, Mattiello:2019gxc, Vosmera:2019mzw, Berkovits:2021eny}, where D-instanton effects in the $\A'\to 0$ limit were identified with those of gauge instantons on the D3-brane. The D(-1)-D3-brane system was further analyzed in the on-shell worldsheet formalism by Green and Gutperle \cite{Green:2000ke}. However, these earlier works are subject to ambiguities of the on-shell formalism, particularly concerning the measure on the D-instanton moduli space $\mathcal{M}_\text{inst}$ and open string IR divergences \cite{Polchinski:1994fq, Balthazar:2019rnh}.
The D-instanton perturbation theory in the framework of open-closed string field theory was introduced by Sen to resolve these ambiguities \cite{Sen:2019qqg, Sen:2020eck, Sen:2020cef, Sen:2021qdk, Sen:2021tpp}. In particular, the effect of a single D-instanton is captured by the following contribution to the string field effective action $\Gamma[\Psi]$ \cite{Agmon:2022vdj, Sen:2024zqr},
\ie\label{ocsftpathint}
\left.e^{-\Gamma[\Psi]}\right|_{\text{1-inst}}=\mathcal{N}\int [\mathcal{D}\Psi_\text{inst}]\ e^{-S[\Psi, \Psi_\text{inst}]},
\fe
where $\Psi$ stands for the open and closed string fields in the absence of the D-instanton, $\mathcal{N}$ is a suitable normalization constant, and $\Psi_\text{inst}$ stands for the open string fields with at least one end of the string attached to the D-instanton. $S[\Psi, \Psi_\text{inst}]$ is the gauge fixed action of open string field theory (OSFT), and $[\mathcal{D}\Psi_\text{inst}]$ is the functional measure on the gauge-fixing Lagrangian submanifold of the space of string fields (in the sense of Batalin-Vilkovisky formalism).

The aforementioned IR divergences in the on-shell formalism can be understood in the string field theory framework as due to the singular propagator of certain zero modes of $\Psi_\text{inst}$ in the Siegel gauge. There are two types of zero modes of $\Psi_{\rm inst}$ that admit the interpretation as the collective coordinates of the D-instanton and the Faddeev-Popov ghost associated with the gauge symmetry on the D-instanton respectively. The latter can be treated by slightly relaxing the Siegel gauge condition and accounting for the gauge redundancy by dividing by the volume of the D-instanton gauge group \cite{Sen:2020cef}. The integration over the remaining zero modes of $\Psi_\text{inst}$ can then be treated non-perturbatively and turned into an integral over $\mathcal{M}_\text{inst}$ via the open string background independence. Finally, the normalization factor $\mathcal{N}$ is determined by a careful rewriting of the annulus diagram in the naive on-shell formalism in terms of a Gaussian functional integral over the open string fields \cite{Sen:2021qdk, Sen:2021tpp}.

It is well known that the massless open string degrees of freedom on a D3-brane are characterized by a supersymmetric Born-Infeld action in the low energy limit \cite{Cederwall:1996pv, Cederwall:1996ri, Aganagic:1996pe, Aganagic:1996nn, Bergshoeff:1996tu}. The leading interaction term in the derivative expansion is a BPS-protected four-derivative coupling, which we refer to as the $F^4$ term, that does not receive string loop nor D-instanton corrections \cite{Sethi:1999qv, Wang:2015jna, Lin:2015ixa}. In this paper, we will focus on the eight-derivative $D^4 F^4$ effective coupling on the D3-brane, which is not captured by the Born-Infeld theory nor protected by supersymmetry, and receives perturbative as well as D-instanton corrections. Based on perturbative results up to 1-loop order and the $SL(2,\mathbb{Z})$ duality, Green and Gutperle \cite{Green:2000ke} conjectured that the coefficient of the $D^4F^4$ effective coupling is proportional to the following modular invariant function of the axion-dilaton expectation value $\tau$,
\ie
h(\tau, \bar{\tau})=\log |\tau_2\eta(\tau)^4|=-\frac{\pi}{3} \tau_2+\ln \tau_2-2 \sum_{k=1}^{\infty} \sum_{d \mid k} \frac{1}{d}\left(e^{2 \pi i k \tau}+e^{-2 \pi i k \bar{\tau}}\right),
\label{gg}
\fe
which may be obtained by summing over $SL(2,\mathbb{Z})$ images of the leading perturbative, \textit{i.e.} tree-level, contribution. In contrast to the $R^4$ effective coupling of the closed string sector \cite{Green:1998by}, however, the $D^4 F^4$ effective coupling is not subject to any known non-renormalization property, and (\ref{gg}) is by no means the unique $SL(2,\mathbb{Z})$-invariant completion of the leading perturbative result.

Based on the D(-1)-D3 OSFT, we perform a first-principle computation of the leading $k$-instanton contribution to the massless open string 4-point amplitude on the D3-brane, and find the result
\ie\label{ourresult}
\hat{\mathcal{A}}^{D^4F^4}_{k\text{-inst}}=4\pi^4(\alpha')^4\tau_2^{-2}e^{2\pi i k\tau} \frac{p(k)}{k} K(\{e_i, k_i\})(s^2+t^2+u^2),
\fe
where $p(k)$ is the partition number of $k$, and $K(\{e_i, k_i\})=t_8^{\mu_1\nu_1...\mu_4\nu_4}\prod_{i=1}^4 e_{i\mu_i}k_{\nu_i}$ with $t_8$ being the Lorentz-invariant tensor defined as in \cite{Green:2000ke}. In the $k=1$ case, our result (\ref{ourresult}) remarkably agrees with the prediction of (\ref{gg}) after accounting for the normalization. However, for $k>1$ (\ref{ourresult}) disagrees with the prediction of (\ref{gg}) as can be seen from the $k$-dependence of the coefficient of $e^{2\pi i k \tau}$ in these expressions.

The rest of this paper is organized as follows. Section \ref{sec2} reviews D-instanton perturbation theory and determines the normalization of the zero mode integration measure of the D(-1)-D3 system. The one-instanton and multi-instanton corrections to the $D^4 F^4$ effective coupling are computed in Section \ref{sec3} and Section \ref{sec4} respectively. In Section \ref{sec5}, we compare our result to the conjecture of \cite{Green:2000ke} and discuss potential implications. Our conventions for the open string perturbation theory are summarized in Appendix \ref{appA}.

\section{Normalization of D(-1)-D3 OSFT measure}

\label{sec2}

\subsection{Normalization for D(-1)-D(-1) open strings}

We begin by reviewing Sen's prescription for the path integral measure of open string fields on the D-instanton \cite{Sen:2021qdk, Sen:2021tpp}. This is achieved through a careful interpretation of the annulus diagram with boundaries on the D-instanton. In the on-shell approach to string perturbation theory, the annulus diagram contributes to any D-instanton mediated amplitude by the factor\footnote{Note that the D-instanton mediated amplitude is computed by summing over worldsheets with disconnected components that share the same D-instanton boundary condition \cite{Polchinski:1994fq}.}
\ie
\exp \mathcal{A}_{\annulus}&=\exp\left[\int_0^\infty \frac{dt}{2t}\operatorname{Tr}_{\mathcal{H}^o_\text{NS}\ominus\mathcal{H}^o_\text{R}}\left(\frac{1+(-1)^F}{2}(-1)^{N_{bc}+N_{\beta\gamma}}b_0c_0e^{-2\pi t L_0}\right)\right],
\label{annulus0}
\fe
	where ${\cal H}_{\rm NS/R}^o$ is the space of NS/R sector states of the worldsheet CFT on the strip subject to the D-instanton boundary condition on both sides, $F$ is the fermion number that counts the number of worldsheet spinor field excitations, $N_{bc}$ and $N_{\beta\gamma}$ are the ghost numbers in the $bc$ and $\beta\gamma$ system respectively. The (super-)trace appearing in the exponent of the RHS of (\ref{annulus0}) in fact vanishes due to spacetime supersymmetry. However, the contribution from individual zero weight ($L_0=0$) states to the $t$-integral is logarithmically divergent, and must be carefully regularized and interpreted.

	A suitably regularized version of (\ref{annulus0}) can be obtained starting from the analogous annulus amplitude of open strings stretched between a pair of D-instantons, which we refer to as D(-1) and D(-1)$'$, that are separated by a small distance $\delta x$ in the Euclidean spacetime. This is such that the GSO-projected ground state acquires conformal weight
	\ie\label{hweight}
	h=\frac{(\delta x)^2}{4\pi^2 \alpha'}.
	\fe
	While the contributions from the nonzero modes cancel by spacetime supersymmetry, the contributions from the zero modes are appropriately re-interpreted as\footnote{The factor $C_{D^2}$ is the normalization factor for disk correlators. Our convention for string perturbation theory is summarized in appendix \ref{appA}.}
	\ie
	\exp \mathcal{A}_{\annulus}&=\exp\left[\int_0^\infty \frac{dt}{2t}\left( 8e^{-2\pi t h}-8e^{-2\pi t h}\right)\right]=\sqrt{\frac{h^8}{h^8}}\\
	&=2^{-8}(C_{D^2})^{-4}\int \frac{d^{10}\tilde{x}}{(2\pi)^5}d^{16}\tilde{\theta}d\xi_1d\xi^2\exp\left[-C_{D^2}\left(\frac{1}{2}h\tilde{x}^2+h\xi_1\xi^2+\frac{1}{2}g_{\mathcal{A}\mathcal{B}}\tilde{\theta}^{\mathcal{A}} \tilde{\theta}^{\mathcal{B}}\right)\right].
	\label{annulus}
	\fe
	The last line takes the form of the functional integral over the open string field zero modes on the D-instanton. In particular, $\tilde x^M, \tilde\theta^{\cal A}$ are bosonic and fermionic variables associated with the collective coordinates of the D-instanton, and $\xi_1, \xi^2$ are fermionic variables associated with the Faddeev-Popov ghosts on the D-instanton; they appear in the open string field $|\Phi^{(-1,-1)}\rangle$ as
\ie\label{zmexpan}
|\Phi^{(-1,-1)} \rangle = \tilde{x}_M c_1 \psi^M_{-\frac{1}{2}}|e^{-\phi}\rangle+\xi^1\beta_{-\frac{1}{2}}c_0c_1|e^{-\phi}\rangle+\xi^2\beta_{-\frac{1}{2}}c_1|e^{-\phi}\rangle+\xi_1\gamma_{-\frac{1}{2}}c_1|e^{-\phi}\rangle+\tilde{\theta}^{\mathcal{A}}c_1|e^{-\phi/2}\mathcal{S}_{\mathcal{A}}\rangle.
\fe
Here $\mathcal{S}_{\mathcal{A}}$ is the 10-dimensional spin field. The $16\times 16$ matrix $g_{\mathcal{A}\mathcal{B}}$ appearing in the action in the last line of (\ref{annulus}) satisfies $g^2=4h^2$, where $h$ is given by (\ref{hweight}). Note that the mode $\xi^1$ would be excluded by the Siegel gauge condition $b_0|\phi\rangle = 0$, but will be kept by a slight relaxation of the Siegel gauge to be considered below.

The kinetic term of the open string field action takes the form
\ie
S_\text{kin} = -\frac{1}{2}\langle \langle \tilde{\Phi}^{(-1,-1)}|Q_\text{B}\mathcal{G}|\tilde{\Phi}^{(-1,-1)}\rangle + \langle \langle \tilde{\Phi}^{(-1,-1)}|Q_\text{B}|\Phi^{(-1,-1)}\rangle ,
\fe
where $\tilde\Phi^{(-1,-1)}$ is an auxiliary open string field of picture number $-1$ in the NS sector and $-{3\over 2}$ in the R sector, and ${\cal G}$ is defined to be identity in the NS sector and the picture-raising operator ${\cal X}_0$ (\ref{zmpcodef}) in the R sector. Upon integrating out $\tilde\phi$, the kinetic term involving the modes appearing in (\ref{zmexpan}) is
\ie
S_\text{kin} = C_{D^2}\left(\frac{1}{2}h\tilde{x}^2-\frac{1}{4} (\xi^1)^2 +h\xi_1\xi^2+\frac{1}{2}g_{\mathcal{A}\mathcal{B}}\tilde{\theta}^{\mathcal{A}} \tilde{\theta}^{\mathcal{B}}\right).
\fe
Formally, imposing the Siegel gauge condition $\xi^1=0$ leads to an open string field functional integral proportional to (\ref{annulus}). Matching with (\ref{annulus}) then fixes the normalization of the OSFT path integral to be of the form
\ie
2^{-8}(C_{D^2})^{-4}\int \prod_{\text{bosonic modes}}\frac{d\phi_b}{\sqrt{2\pi}}\prod_{\text{fermionic modes}}d\phi_f\ \exp\left(-S\right).
\label{sg}
\fe
To arrive at the correct OSFT of the D-instanton, we should take the $h\to 0$ limit, in which the kinetic term $S_{\rm kin}$ vanishes. The dependence of the full string field action $S$ on $\tilde{x}, \tilde{\theta}$ can in principle be absorbed into a deformation of the D-instanton boundary condition, upon which the integral over $\tilde{x}, \tilde{\theta}$ is recast into an integration over the D-instanton moduli space \cite{Sen:2019qqg, Sen:2020cef}.\footnote{This is possible due to the background independence of OSFT \cite{Sen:1990hh,Sen:1993kb, Sen:2017szq, Agmon:2022vdj}.}

The zero modes $\xi_1, \xi^2$, on the other hand, amount to Faddeev-Popov ghosts associated with the $U(1)$ gauge transformation
\ie\label{thegaugf}
\delta |\Psi\rangle=  \delta \vartheta \, Q_\text{B}|c e^{-2\phi}\partial \xi\rangle+\cdots
\fe
generated by the ghost number 0 string field $|c e^{-2\phi}\partial \xi\rangle$, and the Siegel gauge fixes this $U(1)$ gauge symmetry by imposing the gauge condition $\xi^1=0$. In the $h\rightarrow 0$ limit, the Faddeev-Popov determinant vanishes and the Siegel gauge is singular. This is treated by relaxing the Siegel gauge condition on the zero modes via the replacement
\ie
(C_{D^2})^{-1} \left.\int d\xi^2 d\xi_1\right|_{\xi^1=0} \rightarrow \frac{1}{\int d\vartheta}\left.\int d\xi^1\right|_{\xi^2=\xi_1=0},
\label{repl}
\fe
where the integration range of $\vartheta$ can be determined by relating $\delta \vartheta$ of (\ref{thegaugf}) to the variation $\delta\vartheta_0$ of the $2\pi$-periodic angular parameter $\vartheta_0$ of the $U(1)$ gauge group. The latter is achieved by introducing a ``spectator'' D-instanton, which we refer to as D$^s$, and inspecting the $U(1)$ gauge transformation on the D-D$^s$ open string states.

The result at the leading order in open string field perturbation theory turns out to be $\int d\vartheta = -8\pi i$ \cite{Sen:2020cef}, leading to the replacement rule
\ie
\int d\xi^2 d\xi_1 \rightarrow \frac{i}{8\pi} C_{D^2} \int d\xi^1.
\fe

We can also relate the open string field modes $\tilde{x}, \tilde{\theta}$ to the D-instanton collective coordinates $x, \theta$ by\footnote{The relation (\ref{supercoord}) is only expected to hold at the leading order in D-instanton perturbation theory.}
\ie
\tilde{x}^M=\frac{1}{\pi}\sqrt{\frac{2}{\alpha'}}\, x^M,\quad \tilde{\theta}^{\mathcal{A}}=\frac{\theta^{\mathcal{A}}}{2\pi i}.
\label{supercoord}
\fe
Here, the collective coordinate $x$ is defined by the boundary condition $X^M|_{\partial^{(-1)} \Sigma}=x^M$ of the worldsheet CFT, where $\partial^{(-1)}\Sigma$ denotes the D(-1) boundary. The fermionic collective coordinate $\theta$ is defined by deforming the D-instanton boundary condition with the spacetime SUSY current $j_{\mathcal{A}}=e^{-\phi/2}\mathcal{S}_{\mathcal{A}}$:
\ie
\left\langle \cdots \right\rangle_{\Sigma, (x,\theta)}\equiv \left\langle \mathcal{G}\exp\left(\int_{\partial^{(-1)}\Sigma} \frac{dz}{2\pi i}\theta^{\mathcal{A}} j_{\mathcal{A}}\right) \cdots \right\rangle_{\Sigma, (x,0)},
\label{cc}
\fe
where $\mathcal{G}$ is the picture-adjusting operator, defined by inserting appropriate number of PCOs on the Riemann surface $\Sigma$. We should point out that there is a subtlety associated with this prescription of picture-adjustment: If one inserts off-shell vertex operators on the Riemann surface, the correlator will have nontrivial dependence on the PCO positions. Therefore, strictly speaking, the boundary condition defined above at nonzero $\theta$ is not well-defined as a BCFT. The rigorous prescription is to compute the amplitude with arbitrary number of $\theta$ vertex operators as external states, and then integrate over $\theta$ \cite{Sen:2020cef}. However, in the discussions below, all vertex operators inserted on the Riemann surface will be on-shell, so in this case the rigorous prescription is equivalent to our picture-adjustment prescription.

Taking all of the above into account, one finds that the correct measure on the D-instanton moduli space takes the form:
\ie
2^{-8}e^{2\pi i \tau}&(C_{D^2})^{-4}\int \frac{d^{10}\tilde{x}}{(2\pi)^5}d^{16}\tilde{\theta}d\xi_1d\xi^2\\
&= ie^{2\pi i \tau}2^{-16}\pi^{-6}(C_{D^2})^{-3}\int d^{10}\tilde{x}d^{16}\tilde{\theta}d\xi^1\exp\left[\frac{1}{4}C_{D^2}(\xi^1)^2\right]\\
&=ie^{2\pi i \tau} 2^{6} \pi^{-10}\alpha'^{-5}\tau_2^{-7/2}\int d^{10}xd^{16}\theta \equiv ie^{2\pi i \tau}\mathcal{N}_\text{inst}\int d^{10}xd^{16}\theta,
\fe
where we have included the D-instanton action $e^{2\pi i \tau}$ to the normalization and performed the replacement \eqref{repl} on the second line. In the final line, we have integrated out the mode $\xi^1$ and used the fact that the normalization factor $C_{D^2}$ of disk diagram is given by $C_{D^2}=\pi^3\tau_2$.

\subsection{Normalization for D(-1)-D3 open strings}

We can apply the same logic to the annulus diagram with mixed D(-1) and D3 boundaries to fix the normalization for stretched (-1)-3-strings. In Siegel gauge, a general stretched string field in the zero mode sector takes the form:
\begin{equation} \begin{gathered}
|\Phi_\text{ZM}^{(-1,3)}\rangle=(C_{D^2})^{-1/2}\left({w}_{\dot{\alpha}}|ce^{-\phi}{\Delta} S^{\dot{\alpha}}\rangle+{\mu}^A|ce^{-\phi/2}{\Delta}\Theta_A\rangle\right)\\
|\Phi_\text{ZM}^{(3,-1)}\rangle=(C_{D^2})^{-1/2}\left(\bar{w}_{\dot{\alpha}}|ce^{-\phi} \bar{\Delta} S^{\dot{\alpha}}\rangle+\bar{\mu}^A|ce^{-\phi/2}\bar{\Delta}\Theta_A\rangle\right)
\label{zm}
\end{gathered} \end{equation}
Here, $\Delta$ is the product of 4 Dirichlet-Neumann twist fields, $\bar{\Delta}$ is the product of 4 Neumann-Dirichlet twist fields, $S$ is the anti-chiral $SO(4)$ spin field, and $\Theta$ is the $SO(6)$ spin field. The modes ${w}, \bar{w}$ are bosonic, while the modes ${\mu},\bar{\mu}$ are fermionic, since they are in the NS and R sector, respectively. Importantly, there are no additional off-shell zero modes analogous to $\xi^2$ and $\xi_1$ in the stretched string sector, so we can continue to use Siegel gauge for stretched strings.

By assigning suitable Chan-Paton factors, the full zero mode string field of $\Psi_\text{inst}$ can be written in the following matrix form:
\begin{equation} |\Psi_\text{inst}\rangle=\left(\begin{matrix} |\Phi^{(-1,-1)}\rangle & |\Phi^{(-1,3)}\rangle \\ |\Phi^{(3,-1)}\rangle & 0 \end{matrix}\right). \end{equation}

Again, we move the D-instanton slightly away from the D3-brane by $\delta x^a$, so that the zero modes of $\Phi^{(3,-1)}, \Phi^{(-1,3)}$ acquire conformal weight $h=\frac{(\delta x)^2}{4\pi^2 \alpha'}$. The kinetic terms of these stretched zero modes can then be calculated straightforwardly:
\ie
S_\text{kin}=h \bar{w}^{\dot{\alpha}} w_{\dot{\alpha}}+g_{AB}\bar{\mu}^A\mu^B,
\label{Skinstr}
\fe
where the $4\times 4$ matrix $g_{AB}$ satisfies $g^2=-4h$. Following the same logic, the annulus diagram with mixed D(-1) and D3 boundary conditions can be computed by the path integral over $w,\bar{w}$ and $\mu,\bar{\mu}$ with only the kinetic terms in the exponent. Since the mixed annulus diagram also vanishes, we must have:
\ie
\exp \mathcal{A}_{\annulus\, ,\text{mixed}}=\mathcal{N}_\text{str}\int d^2w d^2\bar{w} d^4\mu d^4\bar{\mu} \exp\left(-h \bar{w}^{\dot{\alpha}} w_{\dot{\alpha}}-g_{AB}\bar{\mu}^A\mu^B\right)=1,
\fe
which fixes the normalization factor $\mathcal{N}_\text{str}$ to be:\footnote{We define the integral measure $d^4w\equiv d^2w d^2\bar{w}$ such that $\int d^4w \exp(-\bar{w}^{\dot{\alpha}}M_{\dot{\alpha}}{}^{\dot{\beta}}w_{\dot{\beta}})=(2\pi)^2(\det M)^{-1}$. Similarly $d^8\mu$ is defined by $\int d^8\mu \exp(-\bar{\mu}^A M_{AB} \mu^B)=\det M$.}
\ie
\mathcal{N}_\text{str}=2^{-6}\pi^{-2}.
\fe

The full zero mode integral therefore has the following normalization:
\begin{equation}
\int [\mathcal{D}\Psi_\text{inst}]_\text{ZM}=ie^{2\pi i \tau}\mathcal{N}_\text{inst}\mathcal{N}_{\text{str}}\int d^{10}x d^{16}\theta d^2 w d^2 \bar{w} d^4\mu d^4\bar{\mu}.
\end{equation}

This is not the end of the story. 6 of the 10 moduli $x^a$ interact with the stretched strings, since exciting these moduli moves the D-instanton away from the D3-brane, giving a nonzero mass to the stretched strings. This effect is governed by the higher-point interactions of zero modes in the OSFT action, and these interactions should be included as a part of the instanton measure. The 3-point interactions include terms like $\bar{\mu}\mu x^a$ and $\theta \bar{\mu}w$, which can be computed from disk diagrams:
\begin{equation} S\supset -\frac{i}{\pi \sqrt{\alpha'}} x^a \bar{\mu}^A (\Sigma^a)_{AB} \mu^B+\frac{1}{2\pi}\left(\bar{\mu}^A w^{\dot{\alpha}} \theta_{\dot{\alpha}A}+\text{c.c.}\right). \end{equation}
Note that we have split the $SO(10)$ spinor $\theta^{\mathcal{A}}$ into $(\theta_{\dot{\alpha}A}, \theta_{\alpha }^A)$. Correspondingly, the spin field $e^{-\phi/2}\mathcal{S}_{\mathcal{A}}$ decomposes into $(e^{-\phi/2}\Theta^{A} S^{\dot{\alpha}}, e^{-\phi/2}\Theta_{A} S^{\alpha})$.

Note that only the zero modes $\theta_{\dot{\alpha}A}$, but not $\theta_{\alpha}^A$, appear in the interactions. This is consistent with the supersymmetries preserved by the D-instanton and the D3-brane: the 32 supercharges $Q_{\mathcal{A}}, \tilde{Q}_{\mathcal{A}}$ of type IIB decompose into $Q^{\dot{\alpha}A}$ and $Q^{\alpha}_A$ under $SO(10)\rightarrow SO(4)\times SO(6)$, and can be further regrouped into:
\ie
Q^{\pm \alpha}_A=Q^{\alpha}_A\pm i \tilde{Q}^{\alpha}_A,\quad Q^{\pm \dot{\alpha}A}=Q^{\dot{\alpha}A}\pm i \tilde{Q}^{\dot{\alpha}A}.
\fe
The 16 supercharges $Q^{-\alpha}_A, Q^{-\dot{\alpha}A}$ are preserved by the D-instanton, while the other 16 supercharges $Q^{+\alpha}_A, Q^{+\dot{\alpha}A}$ are broken. Acting the broken supercharges on the D-instanton boundary moves the 16 D-instanton collective coordinates $(\theta_{\alpha}^A, \theta_{\dot{\alpha} A})$. The D3-brane preserves the supercharges $Q^{+\alpha}_A$ and $Q^{-\dot{\alpha}A}$, so the effect of turning on $\theta_{\alpha}^A$ can be cancelled by a supersymmetry transformation on the D3. This explains why only $\theta_{\dot{\alpha}A}$'s are involved in the interactions.

The 4-point interactions $\bar{w}wx^ax^a$ and $w^4$ also contribute to D-instanton amplitudes at leading order. The honest calculations of those interactions involve integrating out the mode $\xi^1$. However, we can determine the $\bar{w}wx^ax^a$ 4-point interaction by going to the Coulomb branch and comparing with the kinetic term \eqref{Skinstr}: if we turn on a VEV of $x^a$, the field $w$ acquires the following mass term:
\begin{equation}
	S_\text{mass}= h \bar{w}^{\dot{\alpha}} w_{\dot{\alpha}}= \frac{x^a x_a}{4\pi^2 \alpha'} \bar{w}^{\dot{\alpha}} w_{\dot{\alpha}},
\end{equation}
which fixes the $\bar{w}wx^ax^a$ interaction. The $w^4$ interaction can only be computed directly, and the details are provided in Appendix \ref{w4}. In the end, one finds that the zero mode sector of the OSFT path integral takes the following form at the leading order in $g_s$:
\begin{equation} \begin{aligned}
&ie^{2\pi i \tau}\mathcal{N}_\text{inst}\mathcal{N}_{\text{str}}\int d^{10}x d^{16}\theta d^2 w d^2 \bar{w} d^4\mu d^4\bar{\mu}\\
 &\quad\times \exp\left[-\frac{\alpha'^2 (g_\text{YM}^{(-1)})^2}{2}(\bar{w}^{\dot{\alpha}}{w}_{\dot{\alpha}})^2+\frac{i}{\pi \sqrt{\alpha'}}x^a \bar{\mu}^A (\Sigma^a)_{AB} \mu^B-\frac{1}{2\pi}\left(\bar{\mu}^A w^{\dot{\alpha}} \theta_{\dot{\alpha}A}+\text{c.c.}\right)-\frac{1}{4\pi^2 \alpha'} x^a x_a \bar{w}^{\dot{\alpha}} w_{\dot{\alpha}}\right],
\label{pi}
\end{aligned} \end{equation}
where $g_\text{YM}^{(-1)}$ is the Yang-Mills coupling on the D-instanton. It can be related to $\alpha', \tau_2$ by:
\ie
g_\text{YM}^{(-1)}=(2\pi)^{-3/2}\alpha'^{-1}\tau_2^{-1/2}.
\label{gym}
\fe

We can further redefine:
\ie
\chi^a\equiv \frac{x^a}{2\pi\sqrt{\alpha'}},\quad \lambda_{\dot{\alpha}}^A\equiv \frac{\theta^A_{\dot{\alpha}}}{2\pi i}
\fe
and \eqref{pi} can be rewritten as:
\ie
&i2^{-2}\pi^{-14}\alpha'^{-2}\tau_2^{-7/2}\int d^{4}x d^6 \chi d^{8}\theta d^8\lambda d^4 w d^8\mu \\
 &\quad\times \exp\left[2\pi i \tau-\frac{\alpha'^2 (g_\text{YM}^{(-1)})^2}{2}(\bar{w}^{\dot{\alpha}}{w}_{\dot{\alpha}})^2+2i\chi^a \bar{\mu}^A (\Sigma^a)_{AB} \mu^B-i\left(\bar{\mu}^A w^{\dot{\alpha}} \lambda_{\dot{\alpha}A}+\mu^A \bar{w}^{\dot{\alpha}}\lambda_{\dot{\alpha}A}\right)-\chi^a \chi_a \bar{w}^{\dot{\alpha}}w_{\dot{\alpha}}\right].
\label{pipi}
\fe
This is our final result for the normalization of D-instanton amplitudes on a single D3-brane. The statistics and representation contents of open string fields $x, \theta, \chi, \lambda, w, \mu$ are summarized in table \ref{table}.
\begin{table}[H]
\centering
\begin{tabular}{ c c c c } \hline \hline
Open string field & Statistics & $\mathfrak{so}(4)=\mathfrak{su}(2)_L\times \mathfrak{su}(2)_R$ rep & $\mathfrak{so}(6)$ rep \\ \hline \hline
$x^\mu$ & Boson & $(\mathbf{2}, \mathbf{2})$ & $\mathbf{1}$ \\ \hline
$\chi^a$ & Boson & $(\mathbf{1}, \mathbf{1})$ & $\mathbf{6}$ \\ \hline
$\theta_{\alpha}^A$ & Fermion & $(\mathbf{2}, \mathbf{1})$ & $\mathbf{4}$ \\ \hline
$\lambda_{\dot{\alpha}A}$ & Fermion & $(\mathbf{1}, \mathbf{2})$ & $\overline{\mathbf{4}}$ \\ \hline
$w_{\dot{\alpha}}, \bar{w}_{\dot{\alpha}}$ & Boson & $(\mathbf{1}, \mathbf{2})$ & $\mathbf{1}$ \\ \hline
$\mu^A, \bar{\mu}^A$ & Fermion & $(\mathbf{1}, \mathbf{1})$ & $\mathbf{4}$ \\ \hline \hline
\end{tabular}
\caption{Properties of open string fields in the zero mode sector.}
\label{table}
\end{table}

\begin{figure}[ht!]
\centering
\begin{tikzpicture}
\filldraw[color=black, fill=black!20, thick] (-1,0) circle (1.25);
\draw (-2.7,0) node[] {$\mathcal{V}_1$};
\draw (-2.3,1) node[] {\color{gray} D3};
\draw (-2.3,-1) node[] {\color{gray} D3};
\draw (0.6,-1) node[] {\color{gray} D(-1)};
\draw (-1,1.7) node[] {$\Phi_\text{ZM}^{(-1,3)}$};
\draw (-1,-1.7) node[] {$\Phi_\text{ZM}^{(3,-1)}$};
\draw[fill] (-2.25,0) circle [radius=0.1];
\draw[fill] (-1,1.25) circle [radius=0.1];
\draw[fill] (-1,-1.25) circle [radius=0.1];

\filldraw[color=black, fill=black!20, thick] (3.5,0) circle (1.25);
\draw (1.8,0) node[very thick] {$\mathcal{V}_2$};
\draw (2.2,1) node[] {\color{gray} D3};
\draw (2.2,-1) node[] {\color{gray} D3};
\draw (5.1,-1) node[] {\color{gray} D(-1)};
\draw (3.5,1.7) node[very thick] {$\Phi_\text{ZM}^{(-1,3)}$};
\draw (3.5,-1.7) node[very thick] {$\Phi_\text{ZM}^{(3,-1)}$};
\draw[fill] (2.25,0) circle [radius=0.1];
\draw[fill] (3.5,1.25) circle [radius=0.1];
\draw[fill] (3.5,-1.25) circle [radius=0.1];

\draw[fill] (6,0) circle [radius=0.025];
\draw[fill] (6.25,0) circle [radius=0.025];
\draw[fill] (6.5,0) circle [radius=0.025];

\filldraw[color=black, fill=black!20, thick] (9.5,0) circle (1.25);
\draw (7.8,0) node[very thick] {$\mathcal{V}_n$};
\draw (8.2,1) node[] {\color{gray} D3};
\draw (8.2,-1) node[] {\color{gray} D3};
\draw (11.1,-1) node[] {\color{gray} D(-1)};
\draw (9.5,1.7) node[very thick] {$\Phi_\text{ZM}^{(-1,3)}$};
\draw (9.5,-1.7) node[very thick] {$\Phi_\text{ZM}^{(3,-1)}$};
\draw[fill] (8.25,0) circle [radius=0.1];
\draw[fill] (9.5,1.25) circle [radius=0.1];
\draw[fill] (9.5,-1.25) circle [radius=0.1];

\end{tikzpicture}
\caption{The worldsheet diagram for the leading D-instanton correction to the amplitude $\mathcal{A}[\mathcal{V}_1,\ldots,\mathcal{V}_n]$ of $n$ 3-3 string modes, with all D(-1) boundaries sharing the same collective coordinates $(x,\theta,\lambda)$.}
\label{npt-diagram}
\end{figure}
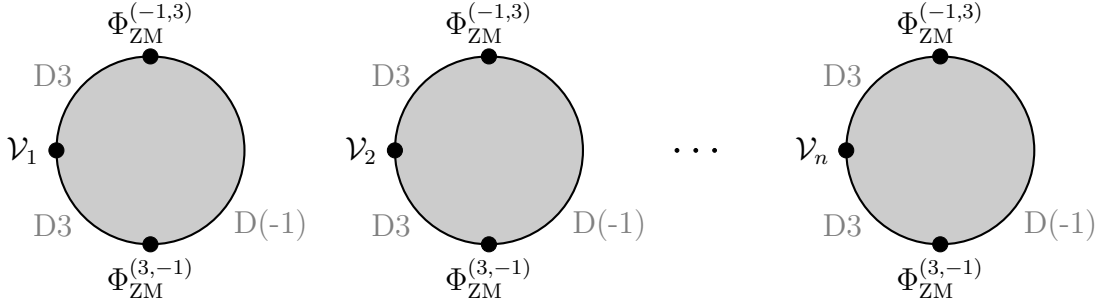

At the leading order in the $g_s$-expansion, the 1-instanton correction to the amplitude $\mathcal{A}[\mathcal{V}_1,...,\mathcal{V}_n]$ is given by the worldsheet diagram consisting of $n$ disconnected disks, each with a D3 boundary and a D(-1) boundary. On each disk, a 3-3 string vertex operator $\mathcal{V}_i$ is inserted on the D3 boundary, and suitable stretched string zero modes are inserted between the D(-1) and D3 boundaries, as shown in figure \ref{npt-diagram}.  Therefore, the leading 1-instanton contribution to this amplitude can be written as:
\ie
&\mathcal{A}_\text{1-inst}[\mathcal{V}_1,...,\mathcal{V}_n]=i2^{-2}\pi^{-14}\alpha'^{-2}\tau_2^{-7/2}\int d^{4}x d^6 \chi d^{8}\theta d^8\lambda d^4 w d^8\mu\ \prod_{i=1}^n \left\langle \mathcal{V}_i \Phi_\text{ZM}^{(3,-1)} \Phi_\text{ZM}^{(-1,3)}\right\rangle_{D^2, (x,\theta,\lambda)} \\
&\quad\times \exp\left[2\pi i \tau-\frac{\alpha'^2 (g_\text{YM}^{(-1)})^2}{2}(\bar{w}^{\dot{\alpha}}{w}_{\dot{\alpha}})^2+2i\chi^a \bar{\mu}^A (\Sigma^a)_{AB} \mu^B-i\left(\bar{\mu}^A w^{\dot{\alpha}} \lambda_{\dot{\alpha}A}+\mu^A \bar{w}^{\dot{\alpha}}\lambda_{\dot{\alpha}A}\right)-\chi^a \chi_a \bar{w}^{\dot{\alpha}}w_{\dot{\alpha}}\right],
\label{amp}
\fe
where $\Phi_\text{ZM}^{(3,-1)}$ and $\Phi_\text{ZM}^{(-1,3)}$ are the operators that correspond to the state in equation \eqref{zm}, and the subscript $(x,\theta,\lambda)$ of the disk correlator indicates that the D-instanton boundary is of collective coordinates $(x,\theta,\lambda)$, as defined in \eqref{cc}. The $\Phi_\text{ZM}$ insertions follow from the usual background field method of computing the effective action \eqref{ocsftpathint}.

\section{Single-instanton corrections to the $D^4 F^4$ coupling}

\label{sec3}

We compute the leading single-instanton contribution to the 4-gauge boson amplitude using the formalism outlined in Section \ref{sec2}. The gauge boson vertex operators in the $(-1)$-picture are normalized as:
\ie
\mathcal{V}_A^{(-1)}(e,k)=g_o^{(3)} e_\mu ce^{-\phi}\psi^\mu e^{ikX},
\fe
where $g_o^{(3)}$ is the open string coupling on the D3-brane. The polarizations $e_1,...,e_4$ are chosen such that $\mathcal{V}_{A}(e_1,k_1), \mathcal{V}_{A}(e_2,k_2)$ carries helicity $+1$, while $\mathcal{V}_{A}(e_3,k_3), \mathcal{V}_{A}(e_4,k_4)$ carries helicity $-1$. We will denote these vertex operators by $\mathcal{V}_+$ and $\mathcal{V}_-$ respectively. Equation \eqref{amp} then implies that the leading instanton amplitude is given by:
\ie
&\mathcal{A}_\text{1-inst}=i2^{-2}\pi^{-14}\alpha'^{-2}\tau_2^{-7/2}\int d^{4}x d^6 \chi d^{8}\theta d^8\lambda d^4 w d^8\mu\\
&\quad\times \exp\left[2\pi i \tau-\frac{\alpha'^2 (g_\text{YM}^{(-1)})^2}{2}(\bar{w}^{\dot{\alpha}}{w}_{\dot{\alpha}})^2+2i\chi^a \bar{\mu}^A (\Sigma^a)_{AB} \mu^B-i\left(\bar{\mu}^A w^{\dot{\alpha}} \lambda_{\dot{\alpha}A}+\mu^A \bar{w}^{\dot{\alpha}}\lambda_{\dot{\alpha}A}\right)-\chi^a \chi_a \bar{w}^{\dot{\alpha}}w_{\dot{\alpha}}\right]\\
&\quad \times \prod_{i=1}^2 \left\langle \mathcal{V}_+(e_i, k_i) \Phi_\text{ZM}^{(3,-1)} \Phi_\text{ZM}^{(-1,3)}\right\rangle_{D^2, (x,\theta,\lambda)}\prod_{j=3}^4 \left\langle \mathcal{V}_-(e_j, k_j) \Phi_\text{ZM}^{(3,-1)} \Phi_\text{ZM}^{(-1,3)}\right\rangle_{D^2, (x,\theta,\lambda)}.
\fe

\begin{figure}[ht!]
\centering
\begin{tikzpicture}
\filldraw[color=black, fill=black!20, thick] (0,0) circle (1.25);
\draw (-1.7,0) node[] {$\mathcal{V}_+$};
\draw (-1.3,1) node[] {\color{gray} D3};
\draw (-1.3,-1) node[] {\color{gray} D3};
\draw (1.6,-1) node[] {\color{gray} D(-1)};
\draw (1.6,-1.5) node[] {\color{gray} $(x,\theta,0)$};
\draw (0,1.7) node[] {$\mathcal{V}_{w}$};
\draw (0,-1.7) node[] {$\mathcal{V}_{\bar{w}}$};
\draw[fill] (-1.25,0) circle [radius=0.1];
\draw[fill] (0,1.25) circle [radius=0.1];
\draw[fill] (0,-1.25) circle [radius=0.1];

\filldraw[color=black, fill=black!20, thick] (4,0) circle (1.25);
\draw (2.3,0) node[] {$\mathcal{V}_+$};
\draw (2.7,1) node[] {\color{gray} D3};
\draw (2.7,-1) node[] {\color{gray} D3};
\draw (5.6,-1) node[] {\color{gray} D(-1)};
\draw (5.6,-1.5) node[] {\color{gray} $(x,\theta,0)$};
\draw (4,1.7) node[] {$\mathcal{V}_{w}$};
\draw (4,-1.7) node[] {$\mathcal{V}_{\bar{w}}$};
\draw[fill] (2.75,0) circle [radius=0.1];
\draw[fill] (4,1.25) circle [radius=0.1];
\draw[fill] (4,-1.25) circle [radius=0.1];

\filldraw[color=black, fill=black!20, thick] (8,0) circle (1.25);
\draw (6.3,0) node[very thick] {$\mathcal{V}_-$};
\draw (6.7,1) node[] {\color{gray} D3};
\draw (6.7,-1) node[] {\color{gray} D3};
\draw (9.6,-1) node[] {\color{gray} D(-1)};
\draw (9.6,-1.5) node[] {\color{gray} $(x,\theta,0)$};
\draw (8,1.7) node[very thick] {$\mathcal{V}_w$};
\draw (8,-1.7) node[very thick] {$\mathcal{V}_{\bar{w}}$};
\draw[fill] (6.75,0) circle [radius=0.1];
\draw[fill] (8,1.25) circle [radius=0.1];
\draw[fill] (8,-1.25) circle [radius=0.1];

\filldraw[color=black, fill=black!20, thick] (12,0) circle (1.25);
\draw (10.3,0) node[] {$\mathcal{V}_-$};
\draw (10.7,1) node[] {\color{gray} D3};
\draw (10.7,-1) node[] {\color{gray} D3};
\draw (13.6,-1) node[] {\color{gray} D(-1)};
\draw (13.6,-1.5) node[] {\color{gray} $(x,\theta,0)$};
\draw (12,1.7) node[] {$\mathcal{V}_{w}$};
\draw (12,-1.7) node[] {$\mathcal{V}_{\bar{w}}$};
\draw[fill] (10.75,0) circle [radius=0.1];
\draw[fill] (12,1.25) circle [radius=0.1];
\draw[fill] (12,-1.25) circle [radius=0.1];

\end{tikzpicture}
\caption{Leading D-instanton contribution to the 4-gauge boson amplitude $\mathcal{A}[\mathcal{V}_+,\mathcal{V}_+,\mathcal{V}_-,\mathcal{V}_-]$.}
\label{4pt-diagram}
\end{figure}
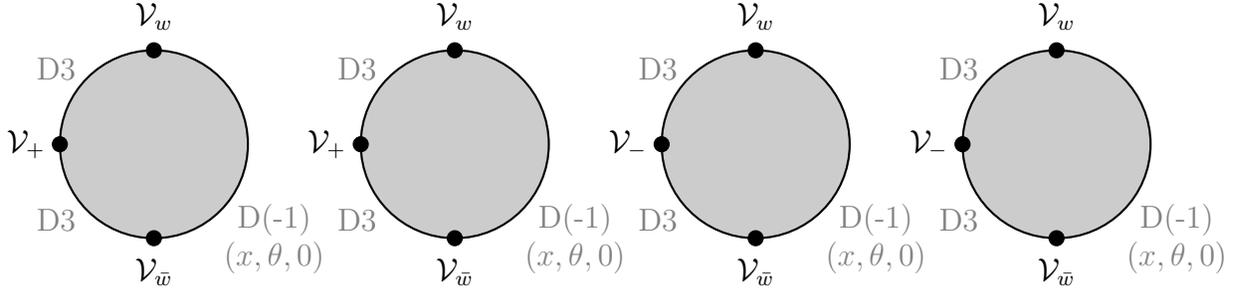

To obtain a nonzero result, the integrand must have exactly 8 $\theta$'s, 8 $\lambda$'s and 8 $\mu$'s to soak up the Grassmann integrals. The leading contribution in $g_s$ comes from the sector where only $w, \bar{w}$ terms are kept from all the $\Phi_\text{ZM}^{(3,-1)}, \Phi_\text{ZM}^{(-1,3)}$'s, the D-instanton sits at $\lambda=0$, and the $\lambda, \mu$ integrals are soaked up solely by the zero mode action \cite{Green:2000ke}, giving a total contribution of:
\begin{equation} \int d^8\lambda d^8\mu \exp\left[2i\chi^a \bar{\mu}^A (\Sigma^a)_{AB} {\mu}^B-i\left(\bar{{\mu}}^A w^{\dot{\alpha}} \lambda_{\dot{\alpha}A}+\mu^A \bar{w}^{\dot{\alpha}}\lambda_{\dot{\alpha}A}\right)\right]=(\bar{w}w)^4 \end{equation}
and hence (as shown in figure \ref{4pt-diagram}):
\ie
&\mathcal{A}_\text{1-inst}=i2^{-6}\pi^{-14}\alpha'^{-2}\tau_2^{-7/2}\int d^{4}x d^6 \chi d^{8}\theta d^4 w \, (\bar{w}w)^4\\
&\quad\times \exp\left[2\pi i \tau-\frac{\alpha'^2 (g_\text{YM}^{(-1)})^2}{2}(\bar{w}^{\dot{\alpha}}{w}_{\dot{\alpha}})^2-\chi^a \chi_a \bar{w}^{\dot{\alpha}}w_{\dot{\alpha}}\right]\\
&\quad \times \prod_{i=1}^2 \left\langle \mathcal{V}_+(e_i, k_i) (\bar{w}_{\dot{\alpha}}\mathcal{V}_{\bar{w}}^{\dot{\alpha}})({w}_{\dot{\beta}}\mathcal{V}_{w}^{\dot{\beta}})\right\rangle_{D^2, (x,\theta,0)}\prod_{j=3}^4 \left\langle \mathcal{V}_-(e_j, k_j) (\bar{w}_{\dot{\alpha}}\mathcal{V}_{\bar{w}}^{\dot{\alpha}})({w}_{\dot{\beta}}\mathcal{V}_{w}^{\dot{\beta}})\right\rangle_{D^2, (x,\theta,0)},
\label{amp1}
\fe
where the $w,\bar{w}$ vertex operators $\mathcal{V}_w, \mathcal{V}_{\bar{w}}$ are defined as:
\ie
\mathcal{V}_w^{\dot{\alpha}}&=(C_{D^2})^{-\frac{1}{2}}ce^{-\phi} \Delta S^{\dot{\alpha}}\\
\mathcal{V}_{\bar{w}}^{\dot{\alpha}}&=(C_{D^2})^{-\frac{1}{2}}ce^{-\phi} \bar{\Delta} S^{\dot{\alpha}}.
\fe

\begin{figure}[ht!]
\center
\begin{tikzpicture}

\filldraw[color=black, fill=black!20, thick] (0,0) circle (2);
\filldraw[color=black, fill=black, thick] (-2,0) circle (0.1);
\draw[color = blue, line width = 1pt] (0,-2) arc[start angle=-85, end angle = 85, radius=2];
\draw[color = blue, line width = 1pt, arrows=-{Stealth}] (1.82,-0.1) -- (1.82,0.1);
\draw[color = blue] (3,1.7) node {$\mathcal{G}\int \frac{dz}{2\pi i}\theta_{\alpha}^A  j^{\alpha}_A$};
\draw (-2.5,0) node {$\mathcal{V}_\pm$};
\filldraw[color=black, fill=black, thick] (0, 2) circle (0.1);
\draw (0,2.5) node {$\mathcal{V}_{{w}}$};
\filldraw[color=black, fill=black, thick] (0,-2) circle (0.1);
\draw (0,-2.5) node {$\mathcal{V}_{\bar{w}}$};
\draw (2.8,0.25) node[color=gray] {D(-1)};
\draw (2.8,-0.25) node[color=gray] {$(x,0,0)$};

\draw[line width = 1pt] (4,0.1) -- (4.5,0.1);
\draw[line width = 1pt] (4,-0.1) -- (4.5,-0.1);

\filldraw[color=black, fill=black!20, thick] (8,0) circle (2);
\filldraw[color=black, fill=black, thick] (6,0) circle (0.1);
\draw[color = blue, line width = 1pt] (6.08,-0.5) arc[start angle=-90, end angle = 90, radius=0.5];
\draw[color = blue, line width = 1pt, arrows=-{Stealth}] (6.57, 0) -- (6.57, 0.1);
\draw (5.5,0) node {$\mathcal{V}_\pm$};
\filldraw[color=black, fill=black, thick] (8, 2) circle (0.1);
\draw (8,2.5) node {$\mathcal{V}_{{w}}$};
\filldraw[color=black, fill=black, thick] (8,-2) circle (0.1);
\draw (8,-2.5) node {$\mathcal{V}_{\bar{w}}$};
\draw (10.8,0.25) node[color=gray] {D(-1)};
\draw (10.8,-0.25) node[color=gray] {$(x,0,0)$};
\draw[color = blue] (8,0.5) node {$\mathcal{G}\int \frac{dz}{2\pi i}\theta_{\alpha}^A  j^{\alpha}_A$};

\end{tikzpicture}
\caption{Contour pull used in equation \eqref{pulleq}. This relates the deformation of the D-instanton boundary condition to the SUSY transformation of the $\mathcal{V}_\pm$ vertex operator.}
\label{pull}
\end{figure}
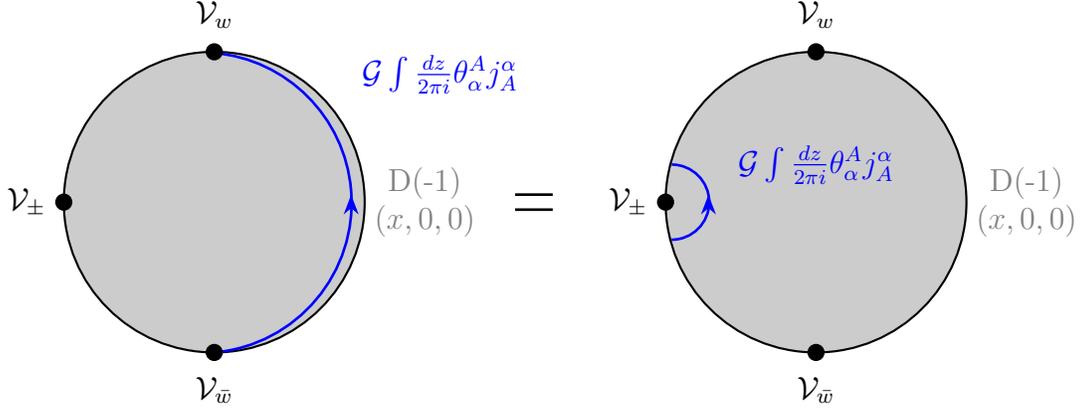

The correlators with D-instanton boundary at collective coordinates $(x,\theta,0)$ can be related to those at $(x,0,0)$ by deforming the boundary condition as in \eqref{cc}. Such a deformation can be further related to the SUSY transformation of the $\mathcal{V}_\pm$ vertex operator by pulling the $\theta$ contour, as illustrated in figure \ref{pull}:
\ie
&\left\langle \mathcal{V}_\pm (e_j, k_j) (\bar{w}_{\dot{\alpha}}\mathcal{V}_{\bar{w}}^{\dot{\alpha}})({w}_{\dot{\beta}}\mathcal{V}_{w}^{\dot{\beta}})\right\rangle_{D^2, (x,\theta,0)}=\left\langle \mathcal{G}\exp\left(\theta_{\alpha}^A\int_{\partial^{(-1)} \Sigma}\frac{dz}{2\pi i} j^{\alpha}_A\right)\mathcal{V}_\pm (e_j, k_j) (\bar{w}_{\dot{\alpha}}\mathcal{V}_{\bar{w}}^{\dot{\alpha}})({w}_{\dot{\beta}}\mathcal{V}_{w}^{\dot{\beta}})\right\rangle_{D^2, (x,0,0)}\\
&\quad =\sum_{n=0}^{\infty} \frac{1}{2^{2n}(2n)!}\left \langle \left[\mathcal{X}_0^{n+1} (\theta_{\alpha}^A Q^{\alpha}_A)^{2n}\cdot \mathcal{V}_\pm(e_j,k_j) \right](\bar{w}_{\dot{\alpha}}\mathcal{V}_{\bar{w}}^{\dot{\alpha}})({w}_{\dot{\beta}}\mathcal{V}_{w}^{\dot{\beta}})\right \rangle_{D^2, (x,0,0)},
\label{pulleq}
\fe
where $j^{\alpha}_A=e^{-\phi/2} S^\alpha \Theta_A$ is the holomorphic SUSY current in the $(-\frac{1}{2})$-picture.

The operator $\mathcal{V}_+$ is annihilated by all the supercharges $Q^+$, and hence:
\ie
\left\langle \mathcal{V}_+ (e, k) (\bar{w}_{\dot{\alpha}}\mathcal{V}_{\bar{w}}^{\dot{\alpha}})({w}_{\dot{\beta}}\mathcal{V}_{w}^{\dot{\beta}})\right\rangle_{D^2, (x,\theta,0)}=2^{-\frac{3}{2}}\pi^{-\frac{3}{4}}g_s^{\frac{1}{2}}e^{ikx}e_\mu k_\nu (\bar{w}\bar{\sigma}^{\mu\nu} w).
\label{plus}
\fe
The other operator $\mathcal{V}_-$ can be converted to $\mathcal{V}_+$ by acting with 4 $Q^+$ supercharges consecutively, and the relevant disk correlator turns out to be:
\ie
\left\langle \mathcal{V}_- (e, k) (\bar{w}_{\dot{\alpha}}\mathcal{V}_{\bar{w}}^{\dot{\alpha}})({w}_{\dot{\beta}}\mathcal{V}_{w}^{\dot{\beta}})\right\rangle_{D^2, (x,\theta,0)}=-\frac{1}{4!}2^{-\frac{19}{2}}\pi^{-\frac{3}{4}}g_s^{\frac{1}{2}}\alpha' e^{ikx}e_\rho k_\sigma k_\nu k^\lambda  (\theta \sigma^{\rho\sigma} \Sigma^a \theta)  (\theta \sigma_{\lambda \mu}\Sigma^a \theta) (\bar{w}\bar{\sigma}^{\mu \nu}w).
\label{minus}
\fe

Plugging \eqref{plus} and \eqref{minus} into the amplitude \eqref{amp1}, it can then be rewritten as:
\ie
\mathcal{A}_\text{1-inst}&=i\frac{1}{(4!)^2}2^{-24}\pi^{-17}g_s^2 \tau_2^{-7/2} e^{2\pi i\tau}\\
&\times\int d^{4}x d^6 \chi d^{8}\theta d^4 {w} \exp\left[-\frac{\alpha'^2 (g_\text{YM}^{(-1)})^2}{2}({\bar{w}}{w})^2-\chi^a \chi_a \bar{{w}}^{\dot{\alpha}} {w}_{\dot{\alpha}}\right] e^{i(k_1+...+k_4)x} \\
&\times (\bar{w}\bar{\sigma}^{\mu_1\nu_1}w)(\bar{w}\bar{\sigma}^{\mu_2\nu_2}w)(\bar{w}\bar{\sigma}^{\mu_3\nu_3}w)(\bar{w}\bar{\sigma}^{\mu_4\nu_4}w)(\bar{w}w)^4\\
&\times e_{1\mu_1}k_{1\nu_1}e_{2\mu_2}k_{2\nu_2} e_{3\rho_3} k_{3\sigma_3}e_{4\rho_4} k_{4\sigma_4}k_3^{\lambda_3}k_{3\nu_3}k_4^{\lambda_4}k_{4\nu_4}(\theta \sigma^{\rho_3\sigma_3}\Sigma^{a}\theta)(\theta \sigma^{\rho_4\sigma_4}\Sigma^{b}\theta) (\theta \sigma_{\lambda_3 \mu_3}\Sigma^{a}\theta) (\theta \sigma_{\lambda_4 \mu_4}\Sigma^{b}\theta).
\label{amp2}
\fe

The remaining part of our calculation can be worked out in the same manner as in \cite{Green:2000ke}. We now perform the integrals in \eqref{amp2}. The $\theta$ integral can be evaluated using the following formula:\footnote{Mathematica code used to verify the formulas in this paper is available in the GitHub repository: \\ \url{https://github.com/Yuchen-Wang-Harvard/D-instanton-on-D3}.}
\begin{equation} \int d^8\theta (\theta \Sigma^{a}\tau^{c_1}\theta)(\theta \Sigma^{b}\tau^{c_2}\theta)(\theta \Sigma^{a}\tau^{c_3}\theta)(\theta \Sigma^{b}\tau^{c_4}\theta) = 3 \cdot 2^9(3\delta^{c_1c_2}\delta^{c_3c_4}+3\delta^{c_1c_4}\delta^{c_2c_3}-2\delta^{c_1c_3}\delta^{c_2c_4})
\label{thetaint}
\end{equation}
where the Pauli matrices $\tau^c$ and the $SO(4)$ generators $\sigma^{\mu\nu}$ are related by the 't Hooft symbols:
\begin{equation} (\sigma_{\mu\nu})_{\alpha}{}^{\beta}=i\eta^c_{\mu\nu} (\tau^c)_{\alpha}{}^{\beta},\quad (\bar{\sigma}_{\mu\nu})^{\dot{\alpha}}{}_{\dot{\beta}}=i\bar{\eta}^c_{\mu\nu} (\tau^c)^{\dot{\alpha}}{}_{\dot{\beta}}, \end{equation}
and we have used the identity (\ref{sigmasigmaeps}).

The $\chi$-integral is a simple Gaussian:
\begin{equation} \int d^6\chi \exp\left(-\chi^a \chi_a \bar{w}w\right)=\pi^3(\bar{w}w)^{-3}.
\label{chiint} \end{equation}

To perform the $w$-integral, we can first define new variables $W_0\equiv \bar{w}w,\ W^c\equiv \bar{w}\tau^c w$, which satisfy the identity $W_0^2=W^cW^c$. One can then replace the $d^4w$ integral with an integral over the new variables:
\ie
\int &d^4w \exp\left[-\frac{\alpha'^2 (g_\text{YM}^{(-1)})^2}{2}({\bar{w}}{w})^2\right](w\bar{\sigma}^{\mu_1\nu_1}w)(w\bar{\sigma}^{\mu_2\nu_2}w)(w\bar{\sigma}^{\mu_3\nu_3}w)(w\bar{\sigma}^{\mu_4\nu_4}w)(\bar{w}w)\\
&={\pi}\int dW^1dW^2dW^3 \exp\left[-\frac{\alpha'^2 (g_\text{YM}^{(-1)})^2}{2}W_0^2\right] W^{c_1}W^{c_2}W^{c_3}W^{c_4}\bar{\eta}^{c_1\mu_1\nu_1}\bar{\eta}^{c_2\mu_2\nu_2}\bar{\eta}^{c_3\mu_3\nu_3}\bar{\eta}^{c_4\mu_4\nu_4}\\
&=\frac{4\pi^2}{15}\int dW_0 W_0^6\exp\left[-\frac{\alpha'^2 (g_\text{YM}^{(-1)})^2}{2}W_0^2\right]
\\&\quad \quad \times \left(\delta^{c_1c_2}\delta^{c_3c_4}+\delta^{c_1c_3}\delta^{c_2c_4}+\delta^{c_1c_4}\delta^{c_2c_3}\right) \bar{\eta}^{c_1\mu_1\nu_1}\bar{\eta}^{c_2\mu_2\nu_2}\bar{\eta}^{c_3\mu_3\nu_3}\bar{\eta}^{c_4\mu_4\nu_4}\\
&=2^{3/2}\pi^{5/2}\alpha'^{-7}(g_{\text{YM}}^{(-1)})^{-7}\left(\delta^{c_1c_2}\delta^{c_3c_4}+\delta^{c_1c_3}\delta^{c_2c_4}+\delta^{c_1c_4}\delta^{c_2c_3}\right)\bar{\eta}^{c_1\mu_1\nu_1} \bar{\eta}^{c_2\mu_2\nu_2}\bar{\eta}^{c_3\mu_3\nu_3}\bar{\eta}^{c_4\mu_4\nu_4}.
\label{wint}
\fe

Plugging \eqref{thetaint}, \eqref{chiint} and \eqref{wint} into \eqref{amp2}, and using the relations \eqref{gym} and \eqref{gB}, we finally arrive at:
\ie
\mathcal{A}_\text{1-inst}&=i\frac{3\cdot2^5}{(4!)^2}\pi^4 (\alpha')^4 \tau_2^{-2}e^{2\pi i\tau}(2\pi)^4 \delta^4(k_1+...+k_4)\\
&\quad \times \left(\delta^{d_1d_2}\delta^{d_3d_4}+\delta^{d_1d_3}\delta^{d_2d_4}+\delta^{d_1d_4}\delta^{d_2d_3}\right)(3\delta^{c_1c_2}\delta^{c_3c_4}+3\delta^{c_1c_4}\delta^{c_2c_3}-2\delta^{c_1c_3}\delta^{c_2c_4})\\
&\quad \times \bar{\eta}^{d_1\mu_1\nu_1} \bar{\eta}^{d_2\mu_2\nu_2}\bar{\eta}^{d_3\mu_3\nu_3}\bar{\eta}^{d_4\mu_4\nu_4}\eta^{c_1\rho_3\sigma_3}\eta^{c_2\rho_4\sigma_4}\eta^{c_3}_{\lambda_3\mu_3}\eta^{c_4}_{\lambda_4\mu_4}\\
&\quad \times e_{1\mu_1}k_{1\nu_1}e_{2\mu_2}k_{2\nu_2} e_{3\rho_3} k_{3\sigma_3}e_{4\rho_4} k_{4\sigma_4}k_3^{\lambda_3}k_{3\nu_3}k_4^{\lambda_4}k_{4\nu_4},\\
\fe
which simplifies to
\ie
\mathcal{A}_\text{1-inst}=4i\pi^4(\alpha')^4\tau_2^{-2}e^{2\pi i \tau}(2\pi)^4\delta^4(k_1+...+k_4)t_8^{\mu_1\nu_1...\mu_4\nu_4}e_{1\mu_1}k_{1\nu_1}...e_{4\mu_4}k_{4\nu_4}(s^2+t^2+u^2).
\fe
Namely, the 1-instanton contribution to the $D^4F^4$ term is given by:
\begin{equation} \begin{aligned}
\hat{\mathcal{A}}^{D^4F^4}_{\text{1-inst}}&=4\pi^4(\alpha')^4\tau_2^{-2}e^{2\pi i \tau}K(\{e_i, k_i\})(s^2+t^2+u^2)\\
&=4\pi^4M_\text{pl}^{-4}e^{2\pi i \tau}K(\{e_i, k_i\})(\underline{s}^2+\underline{t}^2+\underline{u}^2),
\label{result}
\end{aligned} \end{equation}
where $\hat{\mathcal{A}}$ denotes the amplitude with the kinematic factor $i(2\pi)^4\delta^4(\sum_i k_i)$ stripped off. The 10-dimensional Planck mass $M_\text{pl}$ is related to $\alpha'$ and $\tau_2$ by $M_\text{pl}^2=\tau_2^{\frac{1}{2}}\alpha'^{-1}$, and the dimensionless Mandelstam variables $\underline{s}, \underline{t}, \underline{u}$ are defined by $\underline{s}=M_\text{pl}^{-2}s$, and so on. The tensor structure $K(\{e_i, k_i\})$ is the standard one defined by $K(\{e_i, k_i\})=t_8^{\mu_1\nu_1...\mu_4\nu_4}\prod_{i=1}^4 e_{i\mu_i}k_{\nu_i}$.

\section{Multi-instanton corrections to the $D^4 F^4$ coupling}
\label{sec4}

The formalism used above can be generalized to the case of multiple D-instantons \cite{Sen:2021jbr}. For $k$-instantons, the open strings now carry a $U(k)$ Chan-Paton factor. After applying Sen's relaxation from Siegel gauge, a general D(-1)-D(-1) open string field in the zero mode sector can be written as
\ie
|\Phi_\text{ZM}^{(-1,-1)}\rangle &= (\tilde{X}_M)_{ij} c_1 \psi^M_{-\frac{1}{2}}|e^{-\phi}\rangle\otimes |ij\rangle+\xi^1_{ij}\beta_{-\frac{1}{2}}c_0c_1|e^{-\phi}\rangle\otimes |ij\rangle\\
&+\xi^2_{ij}\beta_{-\frac{1}{2}}c_1|e^{-\phi}\rangle\otimes |ij\rangle+(\xi_1)_{ij}\gamma_{-\frac{1}{2}}c_1|e^{-\phi}\rangle\otimes |ij\rangle+\tilde{\theta}^{\mathcal{A}}_{ij} c_1|e^{-\phi/2}\mathcal{S}_{\mathcal{A}}\rangle \otimes |ij\rangle,
\fe
where $|ij\rangle$ denotes the Chan-Paton factor. The annulus diagram with both D(-1) boundary conditions is now given by:
\ie
\exp \mathcal{A}_{\bigcirc\mkern-13.5mu\circ}=2^{-8k^2}(C_{D^2})^{-4k^2}&\int \frac{d^{10k^2}\tilde{X}}{(2\pi)^{5k^2}}d^{16k^2}\tilde{\theta}d^{k^2}\xi_1d^{k^2}\xi^2\\
&\quad \quad \times\exp\left[-C_{D^2}\mathrm{Tr}\left(\frac{1}{2}h\tilde{X}^2+h\xi_1\xi^2+\frac{1}{2}g_{\mathcal{A}\mathcal{B}}\tilde{\theta}^{\mathcal{A}} \tilde{\theta}^{\mathcal{B}}\right)\right].
\label{hannu}
\fe
Here, we used the normalization $d^{k^2}M=\prod_{\mathtt{a}=1}^{k^2} dM^\mathtt{a}$ for the integration over a $k\times k$ matrix $M$, where $M=M^\mathtt{a}t^\mathtt{a}$. The generators $t^\mathtt{a}$ are the Hermitian generators of $U(k)$, normalized such that $\mathrm{Tr}(t^\mathtt{a} t^\mathtt{b})=\delta^{\mathtt{ab}}$.\footnote{Notably, in our normalization, the $U(1)$ generator is given by $t_\text{U(1)}=k^{-1/2}\mathrm{diag}(1,...,1)$.}

Similar to the $U(1)$ case, the modes $(\xi^2)_{ij}$ and $(\xi_1)_{ij}$ are the Faddeev-Popov ghosts associated with the gauge-fixing condition $(\xi^1)_{ij}=0$, which is used to fix a $U(k)$ gauge symmetry in Siegel gauge of the form:
\ie
\delta |\Psi\rangle = \delta \vartheta_{ij}|ce^{-2\phi}\partial \xi\rangle \otimes |ij\rangle.
\fe
After fixing this $U(k)$ gauge symmetry by dividing out the $\int d^{k^2}\vartheta$ integral and integrating out the $\xi^1$ modes, the annulus diagram can be written as
\ie
\exp \mathcal{A}_{\bigcirc\mkern-13.5mu\circ}=\frac{2^{-12k^2}\pi^{-\frac{9}{2}k^2}(C_{D^2})^{-\frac{7}{2}k^2}}{\int d^{k^2}\vartheta}\int d^{10k^2}\tilde{X} d^{16k^2}\tilde{\theta}.
\label{instannu}
\fe
The $\vartheta$-integral can be related to the volume of $U(k)$ by introducing a spectator D-instanton again, and it turns out that:
\ie
\int d^{k^2}\vartheta = (-4i)^{k^2}\operatorname{Vol}U(k).
\fe
Here, the volume $\operatorname{Vol}U(k)$ is defined via the Haar measure, normalized such that the volume element is locally $\prod_{\mathtt{a}=1}^{k^2} d\vartheta_{0}^\mathtt{a}$ near the identity, where a $U(k)$ element is parametrized as $\exp(i\vartheta_{0}^\mathtt{a}t^\mathtt{a})$. Using the relation $U(k)=(U(1)\times SU(k))/\mathbb{Z}_k$, one can find
\ie
\int d^{k^2}\vartheta=(-4i)^{k^2}\cdot 2\pi\sqrt{k}\operatorname{Vol}(SU(k)/\mathbb{Z}_k),
\fe

The stretched (-1)-3 and 3-(-1) string zero modes are given by
\begin{equation} \begin{gathered}
|\Phi_\text{ZM}^{(-1,3)}\rangle=(C_{D^2})^{-1/2}\left({w}_{\dot{\alpha}, i}|ce^{-\phi} \Delta S^{\dot{\alpha}}\rangle+{\mu}^{iA}|ce^{-\phi/2}\Delta\Theta_A\rangle\right)\otimes |i\rangle\\
|\Phi_\text{ZM}^{(3,-1)}\rangle=(C_{D^2})^{-1/2}\left({\bar{w}}_{\dot{\alpha}}^i|ce^{-\phi}\bar{\Delta} S^{\dot{\alpha}}\rangle+{\bar{\mu}}^{iA}|ce^{-\phi/2}\bar{\Delta}\Theta_A\rangle\right)\otimes |\bar{i}\rangle
\end{gathered} \end{equation}
and the annulus diagram with mixed D(-1) and D3 boundaries reads:
\ie
\exp \mathcal{A}_{\annulus\, ,\text{mixed}}=2^{-6k}\pi^{-2k}\int d^{4k}w d^{8k}\mu.
\label{mixannu}
\fe

The instanton measure also includes higher-point interactions among the zero modes. For open string fields on the D-instantons, these interactions are given by the IKKT matrix model \cite{Ishibashi:1996xs}:
\ie
S_\text{int}[\tilde{X}, \tilde{\theta}]=-C_{D^2}\mathrm{Tr}\left(\frac{1}{32}[\tilde{X}^M, \tilde{X}^N][\tilde{X}_M, \tilde{X}_N]+\frac{1}{2\sqrt{2}}\Gamma^M_{\mathcal{A}\mathcal{B}}\tilde{\theta}^{\mathcal{A}}[\tilde{X}_M, \tilde{\theta}^{\mathcal{B}}]\right),
\label{IKKT}
\fe
whose coefficients can be obtained by going to the Coulomb branch and comparing with the kinetic term \eqref{hannu}. The form of the interaction terms involving stretched strings can be obtained by dimensional reduction of the D5/D9 system, as in \cite{Dorey:2002ik}. The coefficient for each term can be read off by comparing with the single-instanton measure \eqref{pipi} and the IKKT matrix model \eqref{IKKT}. It turns out that after introducing 3 $k\times k$ matrix-valued auxiliary fields $D^c$, the full interaction of the zero modes takes the following form:
\ie
S_{k\text{-inst}}=2C_{D^2} S_G+S_K+S_D,
\label{kaction}
\fe
where the individual terms are given by:
\ie
\begin{aligned}
& S_G=\operatorname{Tr}\bigg[-\left[\chi_a, \chi_b\right]^2- \frac{1}{2}(\bar{\Sigma}^a)^{A B}\lambda_A[\chi_a, \lambda_B]-D^c D^c\bigg], \\
& S_K=\operatorname{Tr}\bigg[-[\chi_a, a_n]^2+\chi_a \bar{w}^{\dot{\alpha}} w_{\dot{\alpha}} \chi_a-({\Sigma}^{a})_{A B} \theta^{\alpha A}[\chi_a, \theta_{\alpha}^B]-2i ({\Sigma}^{a})_{A B} \bar{\mu}^A \mu^B \chi_a\bigg] \\
& S_D=\operatorname{Tr}\bigg[\frac{1}{\sqrt{2}}{D}^c \left(W^c+i\bar{\eta}^c_{\mu\nu}[a_\mu, a_\nu]\right)+\left(i\bar{\mu}^A w_{\dot{\alpha}}+i\bar{w}_{\dot{\alpha}} \mu^A-\left[\theta^{\alpha A}, a_{\alpha \dot{\alpha}}\right]\right) \lambda^{\dot{\alpha}}_A\bigg] .
\end{aligned}
\fe
Here, the combination $(W^c)_i{}^j\equiv \bar{w}^j\tau^c w_i $ should be regarded as a $k\times k$ matrix rather than a scalar. The same applies to all bilinears of the fundamental string fields. The string fields $a_\mu, \chi_a, \lambda_{\dot{\alpha}A}, \theta_{\alpha}^A$ can be related to $\tilde{X}_M, \tilde{\theta}_{\mathcal{A}}$ by the following rescalings:
\ie
\chi^a=\frac{1}{2\sqrt{2}}\tilde{X}^a,\quad a_\mu = \frac{1}{\sqrt{2}}(C_{D^2})^{1/2}\tilde{X}_\mu,\quad \lambda_{\dot{\alpha} A} = \tilde{\theta}_{\dot{\alpha}A},\quad \theta_{\alpha}^A=(C_{D^2})^{1/2}\tilde{\theta}_{\alpha}^A.
\fe
After integrating out the auxiliary field $D^c$, the terms involving $D^c$ are replaced by:
\ie
S_{k\text{-inst}}\supset \frac{1}{16C_{D^2}}\mathrm{Tr}(W^c W^c)-\frac{1}{8C_{D^2}}\mathrm{Tr}[a_\mu, a_\nu]^2 + \frac{i\bar{\eta}^c_{\mu\nu}}{8 C_{D^2}}\mathrm{Tr}(\bar{w} \tau^c w  [a^\mu, a^\nu]).
\fe

Putting \eqref{instannu}, \eqref{mixannu}, and \eqref{kaction} together, we obtain the following $k$-instanton measure:
\ie
\int d\mu_{k\text{-inst}}=&ik^{-\frac{7}{2}}e^{2\pi i k \tau}\frac{2^{-3k^2-6k+7}\pi^{-9k^2-2k-3}(\tau_2)^{-\frac{3}{2}k^2-2}(\alpha')^{-2}}{\operatorname{Vol}(SU(k)/\mathbb{Z}_k)}\\
&\quad \quad \times \int d^4x_\text{CM} d^8\theta_\text{CM} d^{4(k^2-1)}a d^{6k^2}\chi d^{8k^2}\lambda d^{8(k^2-1)}\theta d^{4k}w d^{8k}\mu \exp(-S_{k\text{-inst}}),
\fe
where we have isolated the integration over the center-of-mass supercoordinates $(x_\text{CM},\theta_\text{CM})$ of the D-instanton, which correspond to the $U(1)$ part of the matrix-valued fields $(a,\theta)$. These supercoordinates are related to $\tilde{x}, \tilde{\theta}$ by \eqref{supercoord}. We have also Wick rotated the Minkowskian $a^0$ mode via $a^0\rightarrow -ia^{4}$, which introduces a factor of $(-i)^{k^2-1}$.

Analogously to the arguments leading to \eqref{amp} and \eqref{amp1}, the leading $k$-instanton contribution to the $D^4 F^4$ term is given by:
\ie
\mathcal{A}_{k\text{-inst}}=\int d\mu_{k\text{-inst}} &\prod_{i=1}^2 \left\langle \mathcal{V}_+(e_i, k_i) \Phi_\text{ZM}^{(3,-1)} \Phi_\text{ZM}^{(-1,3)}\right\rangle_{D^2, (x_\text{CM},\theta_\text{CM},0)}
\\&\quad \times \prod_{j=3}^4 \left\langle \mathcal{V}_-(e_j, k_j) \Phi_\text{ZM}^{(3,-1)} \Phi_\text{ZM}^{(-1,3)}\right\rangle_{D^2, (x_\text{CM},\theta_\text{CM},0)},
\label{multiamp}
\fe
where each D-instanton boundary sits on the center-of-mass supercoordinates $(x_\text{CM},\theta_\text{CM},0)$.

The correlators appearing in \eqref{multiamp} are given by:
\ie
&\left\langle \mathcal{V}_+ (e, k) (\bar{w}_{\dot{\alpha}}\mathcal{V}_{\bar{w}}^{\dot{\alpha}})({w}_{\dot{\beta}}\mathcal{V}_{w}^{\dot{\beta}})\right\rangle_{D^2, (x_\text{CM},\theta_\text{CM},0)}=2^{-\frac{3}{2}}\pi^{-\frac{3}{4}}g_s^{\frac{1}{2}}e^{ikx_\text{CM}}e_\mu k_\nu \mathrm{Tr}(\bar{w}\bar{\sigma}^{\mu\nu} w)\\
&\left\langle \mathcal{V}_- (e, k) (\bar{w}_{\dot{\alpha}}\mathcal{V}_{\bar{w}}^{\dot{\alpha}})({w}_{\dot{\beta}}\mathcal{V}_{w}^{\dot{\beta}})\right\rangle_{D^2, (x_\text{CM},\theta_\text{CM},0)}\\
&\quad\quad\quad\quad=-\frac{1}{4!}2^{-\frac{19}{2}}\pi^{-\frac{3}{4}}g_s^{\frac{1}{2}}\alpha' e^{ikx_\text{CM}}e_\rho k_\sigma k_\nu k^\lambda \mathrm{Tr}(\bar{w}\bar{\sigma}^{\mu \nu}w)(\theta_\text{CM} \sigma^{\rho\sigma} \Sigma^a \theta_\text{CM})  (\theta_\text{CM} \sigma_{\lambda \mu}\Sigma^a \theta_\text{CM}).
\fe
These correlators follow from \eqref{plus} and \eqref{minus}, with $\bar{w}\bar{\sigma}^{\mu\nu}w$ replaced with $\mathrm{Tr}(\bar{w}\bar{\sigma}^{\mu\nu}w)=\bar{w}^i\bar{\sigma}^{\mu\nu}w_i$. Therefore,
\ie
\mathcal{A}_{k\text{-inst}}&=ik^{-\frac{7}{2}}e^{2\pi i k \tau}\frac{2^{-3k^2-6k+7}\pi^{-9k^2-2k-3}(\tau_2)^{-\frac{3}{2}k^2-2}(\alpha')^{-2}}{\operatorname{Vol}(SU(k)/\mathbb{Z}_k)}\\
&\quad \times \int d^4x_\text{CM} d^8\theta_\text{CM} d^{4(k^2-1)}a d^{6k^2}\chi d^{8k^2}\lambda d^{8(k^2-1)}\theta d^{4k}w d^{8k}\mu \exp(-S_{k\text{-inst}})\\
&\quad \times \frac{1}{(4!)^2} 2^{-22}\pi^{-3}g_s^2 (\alpha')^2 e^{i(k_1+k_2+k_3+k_4)x_\text{CM}} e_{1\mu_1}k_{1\nu_1}e_{2\mu_2}k_{2\nu_2} e_{3\rho_3} k_{3\sigma_3}e_{4\rho_4} k_{4\sigma_4}k_3^{\lambda_3}k_{3\nu_3}k_4^{\lambda_4}k_{4\nu_4}\\
&\quad \times \mathrm{Tr}(\bar{w}\bar{\sigma}^{\mu_1\nu_1}w)\mathrm{Tr}(\bar{w}\bar{\sigma}^{\mu_2\nu_2}w)\mathrm{Tr}(\bar{w}\bar{\sigma}^{\mu_3\nu_3}w)\mathrm{Tr}(\bar{w}\bar{\sigma}^{\mu_4\nu_4}w)\\
&\quad \times (\theta_\text{CM} \sigma^{\rho_3\sigma_3}\Sigma^{a}\theta_\text{CM})(\theta_\text{CM} \sigma^{\rho_4\sigma_4}\Sigma^{b}\theta_\text{CM}) (\theta_\text{CM} \sigma_{\lambda_3 \mu_3}\Sigma^{a}\theta_\text{CM}) (\theta_\text{CM} \sigma_{\lambda_4 \mu_4}\Sigma^{b}\theta_\text{CM}).
\label{bigampk}
\fe

The $x_\text{CM}, \theta_\text{CM}$ integrals are identical to the single-instanton case. To handle the factors $\mathrm{Tr}(\bar{w}\bar{\sigma}^{\mu\nu}w)$, we consider the following matrix integral with a source term $J^c\mathrm{Tr}(\bar{w}\tau^c w)$ added to the action \cite{Dorey:2001ym}:
\ie
\mathcal{Z}_k[J]\equiv \int d^{4(k^2-1)}a d^{6k^2}\chi d^{8k^2}\lambda d^{8(k^2-1)}\theta d^{3k^2} D d^{4k}w d^{8k}\mu \exp\left(-S_{k\text{-inst}}-\frac{1}{\sqrt{2}}iJ^c\mathrm{Tr}(\bar{w}\tau^c w)\right).
\fe
The source term can be absorbed by shifting the $U(1)$ part of the auxiliary field $D^c$ and by introducing a Fayet–Iliopoulos D-term $-4C_{D^2}J^c\mathrm{Tr}(D^c)\equiv \zeta^c \mathrm{Tr}(D^c)$ into the action $S_{k\text{-inst}}$. One can show that the matrix integral $\mathcal{Z}_k[0]$ is independent of the FI parameter $\zeta^c$,\footnote{This can be seen from the fact that such an FI term can be absorbed into a constant shift in $\mathcal{E}_\mathbb{R}$ and $\mathcal{E}_\mathbb{C}^{(1)}$ of \eqref{EEEEE}, and one can deform this term away by the usual logic of supersymmetric localization.} and hence the dependence on $J$ is simply a Gaussian factor after integrating out $D$:
\ie
\mathcal{Z}_k[J]=\exp\left(2kC_{D^2}J^c J^c\right)\mathcal{Z}_k[0].
\fe
Therefore, the effect of inserting $\mathrm{Tr}(\bar{w}\tau^{c_1} w)\mathrm{Tr}(\bar{w}\tau^{c_2} w)\mathrm{Tr}(\bar{w}\tau^{c_3} w)\mathrm{Tr}(\bar{w}\tau^{c_4} w)$ into the matrix integral is simply to multiply it by a constant factor:
\ie
&\int d\mu_{k\text{-inst}}\mathrm{Tr}(\bar{w}\tau^{c_1} w)\mathrm{Tr}(\bar{w}\tau^{c_2} w)\mathrm{Tr}(\bar{w}\tau^{c_3} w)\mathrm{Tr}(\bar{w}\tau^{c_4} w)e^{-S_{k\text{-inst}}}\\
&\quad\quad = \int d\mu_{k\text{-inst}} 2^6 k^2 (C_{D^2})^2(\delta^{c_1c_2}\delta^{c_3c_4}+\delta^{c_1c_3}\delta^{c_2c_4}+\delta^{c_1c_4}\delta^{c_2c_3})e^{-S_{k\text{-inst}}}.
\fe
Plugging this into \eqref{bigampk} and performing the integral over the center-of-mass supercoordinates using \eqref{thetaint}, we find:
\ie
\hat{\mathcal{A}}_{k\text{-inst}}&=\frac{1}{(4!)^2} k^{-\frac{3}{2}}e^{2\pi i k \tau}\frac{2^{-3k^2-6k+7}\pi^{-9k^2-2k+5}(\tau_2)^{-\frac{3}{2}k^2-2}(\alpha')^4}{\operatorname{Vol}(SU(k)/\mathbb{Z}_k)}\\
&\quad \times \int d^{4(k^2-1)}a d^{6k^2}\chi d^{8k^2}\lambda d^{8(k^2-1)}\theta d^{4k}w d^{8k}\mu \exp(-S_{k\text{-inst}})\\
&\quad \times \left(\delta^{d_1d_2}\delta^{d_3d_4}+\delta^{d_1d_3}\delta^{d_2d_4}+\delta^{d_1d_4}\delta^{d_2d_3}\right)(18\delta^{c_1c_2}\delta^{c_3c_4}+18\delta^{c_1c_4}\delta^{c_2c_3}-12\delta^{c_1c_3}\delta^{c_2c_4})\\
&\quad \times \bar{\eta}^{d_1\mu_1\nu_1} \bar{\eta}^{d_2\mu_2\nu_2}\bar{\eta}^{d_3\mu_3\nu_3}\bar{\eta}^{d_4\mu_4\nu_4} \eta^{c_1\rho_3\sigma_3}\eta^{c_2\rho_4\sigma_4}\eta^{c_3}_{\lambda_3\mu_3}\eta^{c_4}_{\lambda_4\mu_4}\\
&\quad \times e_{1\mu_1}k_{1\nu_1}e_{2\mu_2}k_{2\nu_2} e_{3\rho_3} k_{3\sigma_3}e_{4\rho_4} k_{4\sigma_4}k_3^{\lambda_3}k_{3\nu_3}k_4^{\lambda_4}k_{4\nu_4}\\
&=k^{-\frac{3}{2}}e^{2\pi i k \tau}\frac{2^{-3k^2-6k+5}\pi^{-9k^2-2k+5}(\tau_2)^{-\frac{3}{2}k^2-2}(\alpha')^4}{\operatorname{Vol}(SU(k)/\mathbb{Z}_k)}K(\{e_i, k_i\})(s^2+t^2+u^2)\\
&\quad \times \int d^{4(k^2-1)}a d^{6k^2}\chi d^{8k^2}\lambda d^{8(k^2-1)}\theta d^{4k}w d^{8k}\mu \exp(-S_{k\text{-inst}})\\
&\equiv k^{-\frac{3}{2}}e^{2\pi i k \tau}\frac{2^{-3k^2-6k+5}\pi^{-9k^2-2k+5}(\tau_2)^{-\frac{3}{2}k^2-2}(\alpha')^4}{\operatorname{Vol}(SU(k)/\mathbb{Z}_k)}K(\{e_i, k_i\})(s^2+t^2+u^2)\times \mathsf{Z}_k.
\label{bigbigequation}
\fe

As one can see, the $k$-instanton contribution to the $D^4 F^4$ effective interaction can be reduced to the following matrix integral:
\ie
\mathsf{Z}_k\equiv \int d^{4(k^2-1)}a d^{6k^2}\chi d^{8k^2}\lambda d^{8(k^2-1)}\theta d^{4k}w d^{8k}\mu \exp(-S_{k\text{-inst}}).
\fe
This matrix integral has been studied in the context of the instanton partition function in both $\mathcal{N}=2^*$ and $\mathcal{N}=4$ super Yang-Mills theory \cite{Dorey:2000zq, Dorey:2001ym, Hollowood:2002ds, Hollowood:2002zv}, and it has been computed by Bruzzo, Fucito, Morales, and Tanzini \cite{Bruzzo:2002xf} based on the localization method developed by Moore, Nekrasov, and Shatashvili \cite{Moore:1998et, Nekrasov:2002qd}. The details of the localization computation are reviewed in Appendix \ref{appC}. In our convention, their result is given by:
\ie
\mathsf{Z}_k=2^{3k^2+6k-3}\pi^{\frac{9}{2}k^2+2k-1}(C_{D^2})^{\frac{3k^2}{2}}\sqrt{k}p(k)\mathrm{Vol}(SU(k)/\mathbb{Z}_k),
\label{maintextZk}
\fe
where $p(k)$ is the partition number of $k$. Plugging this result into \eqref{bigbigequation}, we obtain:
\ie
\hat{\mathcal{A}}_{\text{k-inst}}&=4\pi^4(\alpha')^4 \tau_2^{-2}e^{2\pi i k \tau} \frac{p(k)}{k}K(\{e_i, k_i\})(s^2+t^2+u^2)\\
&=4\pi^4 M_\text{pl}^{-4}e^{2\pi i k \tau} \frac{p(k)}{k}K(\{e_i, k_i\})(\underline{s}^2+\underline{t}^2+\underline{u}^2).
\label{multiresult}
\fe
This is our final result for the leading $k$-instanton contribution to the $D^4F^4$ effective interaction on a D3-brane.

\section{The reasonable ineffectiveness of $SL(2,\mathbb{Z})$ images}
\label{sec5}

Green and Gutperle \cite{Green:2000ke} conjectured that the D-instanton contribution enters the 4-open-string amplitude on the D3-brane through the following modular-invariant factor:
\begin{equation} h(\tau, \bar{\tau})=\log |\tau_2\eta(\tau)^4|=-\frac{\pi}{3} \tau_2+\ln \tau_2-2 \sum_{k=1}^{\infty} \sum_{d \mid k} \frac{1}{d}\left(e^{2 \pi i k \tau}+e^{-2 \pi i k \bar{\tau}}\right).
\label{GGconjecture}
\end{equation}
This factor is obtained by summing over the $SL(2,\mathbb{Z})$ images of the tree-level perturbative contributions, namely:
\ie
h(\tau,\bar{\tau})=-\frac{1}{2\pi}\sum_{(m,n)\neq 0}\frac{\tau_2}{|m+n\tau|^2}.
\label{sumimages}
\fe
Note that this Poincare-type sum is divergent on its own, and it should be understood as the result after subtracting a $\tau$-independent logarithmic divergence from the right-hand side.

Assuming the conjecture holds, we can read off the leading $k$-instanton contribution from the tree-level perturbative result. It is well known that the tree-level 4-gauge boson amplitude is given by
\ie
 \hat{\mathcal{A}}&=-2(g_o^{(3)})^2 \alpha' K(\{e_i, k_i\})\left(\frac{\Gamma(-\alpha' s)\Gamma(-\alpha' t)}{\Gamma(1-\alpha' s-\alpha' t)}+\frac{\Gamma(-\alpha' s)\Gamma(-\alpha' u)}{\Gamma(1-\alpha' s-\alpha' u)}+\frac{\Gamma(-\alpha' t)\Gamma(-\alpha' u)}{\Gamma(1-\alpha' t-\alpha' u)}\right)\\
&=2^5 \pi M_\text{pl}^{-4} K(\{e_i, k_i\})\left(\frac{\pi^2}{2}+\frac{\pi^4}{48}\tau_2 (\underline{s}^2+\underline{t}^2+\underline{u}^2)+\cdots \right).
\fe
Therefore, if the $\tau$-dependence enters the $D^4 F^4$ term through the function \eqref{GGconjecture}, the full $D^4 F^4$ term should be:
\ie
\hat{\mathcal{A}}^{D^4 F^4}_{\text{conjectured}}=4 \pi^3 M_\text{pl}^{-4} K(\{e_i, k_i\})h(\tau, \bar{\tau})(\underline{s}^2+\underline{t}^2+\underline{u}^2)
\fe
and the $k$-instanton contribution to the $D^4 F^4$ term can be extracted:
\ie
\hat{\mathcal{A}}^{D^4F^4}_{\text{k-inst, conjectured}}=4\pi^4M_\text{pl}^{-4} e^{2\pi i k \tau} \bigg(\sum_{d|k}\frac{1}{d}\bigg) K(\{e_i, k_i\})(\underline{s}^2+\underline{t}^2+\underline{u}^2).
\label{prop}
\fe
Notably, the conjecture agrees with our $k=1$ result \eqref{result} obtained from D-instanton perturbation theory, but it disagrees with our result \eqref{multiresult} for $k>1$. Our calculation has thus ruled out the conjecture.

This shows that for unprotected couplings, summing over $SL(2,\mathbb{Z})$-images of the leading perturbative contribution does not yield the correct non-perturbative result. To make this point clearer, it is helpful to review why the \textit{protected} effective coupling $R^4$ in Type IIB supergravity \textit{can} be obtained by summing over $SL(2,\mathbb{Z})$ images in a manner similar to \eqref{sumimages}. The $R^4$ term in the 4-graviton amplitude in Type IIB is given by
\ie
\hat{\mathcal{A}}^{R^4}= \frac{1}{4}\kappa^2_{10} M_\text{pl}^{-6} K_\text{NS}(\{e_i,k_i\}) f_0(\tau,\bar{\tau}),
\fe
where $\kappa_{10}$ is the 10-dimensional Newton's constant, and the tensor structure is given by $K_\text{NS}(\{e_i,k_i\})=t_8^{\mu_1\nu_1...\mu_4\nu_4}t_8^{\rho_1\sigma_1...\rho_4\sigma_4}\prod_{i=1}^4 e_{i\mu_i\rho_i}k_{i\nu_i}k_{i\sigma_i}$. As we have noted, the function $f_0(\tau,\bar{\tau})$ can be obtained by summing over $SL(2,\mathbb{Z})$ images of the tree-level result:
\ie
f_0(\tau,\bar{\tau})=E_{3/2}(\tau,\bar{\tau})=\sum_{(m,n)\neq (0,0)}\frac{\tau_2^{3/2}}{|m+n\tau|^3}.
\label{eisenstein}
\fe

This is correct because the $R^4$ term is highly constrained by supersymmetry \cite{Green:1998by, Wang:2015jna}. Expanding the axion-dilaton field $\tau=\tau_0+\varphi$ around its VEV $\tau_0$, the $R^4$ interaction $f_0(\tau,\bar{\tau})R^4$ induces an interaction term $\partial_\tau \partial_{\bar{\tau}}f_0(\tau_0, \bar{\tau}_0)\varphi\bar{\varphi} R^4$, which will contribute to a 6-point amplitude of 4 gravitons and 2 soft axion-dilatons. Such a 6-point amplitude cannot be completed into a supervertex (\textit{i.e.}, a superamplitude without poles in the external momenta) that satisfies the SUSY Ward identities, and its behavior near the poles can be related to the $R^4$ coupling by unitarity. Therefore, there must be a linear relation between $f_0(\tau, \bar{\tau})$ and $\partial_\tau \partial_{\bar{\tau}} f_0(\tau,\bar{\tau})$, which is fixed simply by consistency with tree-level string amplitude to be
\ie
\tau_2^2\partial_\tau \partial_{\bar{\tau}} f_0(\tau,\bar{\tau}) = \frac{3}{16} f_0(\tau,\bar{\tau}).
\label{de}
\fe
This equation, together with the requirement that $f_0(\tau,\bar{\tau})$ is invariant under the $SL(2,\mathbb{Z})$ S-duality and the overall normalization of the string tree amplitude, fixes $f_0(\tau,\bar{\tau})$ to be the non-holomorphic Eisenstein series $E_{3/2}$ as given in \eqref{eisenstein}.

As one can see, the differential equation \eqref{de} is crucial in this line of reasoning. Without a differential equation like \eqref{de}, one may modify $f_0(\tau, \bar{\tau})$ by adding arbitrary Maass cusp forms. Since the cusp forms have no perturbative part (\textit{i.e.} zero Fourier mode in $\tau_1$), adding them to $f_0(\tau, \bar{\tau})$ does not modify the leading perturbative term, and hence the modular-invariant completion of the leading perturbative term is far from unique.

This absence of a differential equation is exactly what happens for our $D^4F^4$ effective coupling on a D3-brane. If we run the same argument for the interaction $h(\tau,\bar{\tau})D^4F^4$ in the language of \cite{Lin:2015ixa}, we would need to constrain the 6-point brane-bulk amplitude $\varphi\bar{\varphi} D^4F^4$ of 4 gluons and 2 soft axion-dilatons. However, there is an independent D-term supervertex that gives the $\varphi\bar{\varphi} D^4F^4$ coupling, which means that there are no differential equation constraints analogous to \eqref{de} for $D^4F^4$. The argument above then shows that S-duality alone cannot fix the $D^4F^4$ term to an Eisenstein series. This is the reason, at least for the unprotected $D^4F^4$ term, that summing over $SL(2,\mathbb{Z})$-images generically fails to give the correct non-perturbative completion for unprotected observables.

\section*{Acknowledgements}

We are grateful to Piotr Tourkine and Yutai Zhang for discussions. This work is supported by DOE grant DE-SC0007870. XY thanks the Aspen Center for Physics, which is supported by National Science Foundation grant PHY-2210452, ICTP-SAIFR, S\~ao Paulo, Brazil, the Yukawa Institute for Theoretical Physics at Kyoto University, and the organizers of the workshop “Progress of Theoretical Bootstrap” for their hospitality during the course of this work.

\appendix

\section{Conventions}

\label{appA}

\subsection{Spinors and Representation Theory}

\subsubsection*{Glossary of indices}

\begin{enumerate}
\item[$\bullet$] $SO(4)$ vector: $\mu,\nu=1,...,4$
\item[$\bullet$] $SO(4)$ chiral Weyl spinor: $\alpha,\beta=1,2$.
\item[$\bullet$] $SO(4)$ anti-chiral Weyl spinor: $\dot{\alpha},\dot{\beta}=1,2$
\item[$\bullet$] $SO(6)$ vector: $a,b = 5,...,10$
\item[$\bullet$] $SO(6)$ Weyl spinor: $A, B=1,...,4$
\item[$\bullet$] $SO(10)$ vector: $M,N=1,...,10$
\item[$\bullet$] $SO(10)$ chiral Weyl spinor: $\mathcal{A}, \mathcal{B}=1,...,16$
\item[$\bullet$] $SO(10)$ anti-chiral Weyl spinor: $\dot{\mathcal{A}}, \dot{\mathcal{B}}=1,...,16$
\item[$\bullet$] $SO(8)$ vector: $\mathbf{a}, \mathbf{b}=1,...,8$
\item[$\bullet$] $SO(8)$ chiral Weyl spinor: $\mathbf{A}, \mathbf{B}=1,...,8$
\item[$\bullet$] $SO(8)$ anti-chiral Weyl spinor: $\dot{\mathbf{A}}, \dot{\mathbf{B}}=1,...,8$
\item[$\bullet$] Spin$(7)$ vector: $\mathbf{i}, \mathbf{j}=1,...,7$
\item[$\bullet$] $U(k)$ vector: $i,j=1,...,k$
\item[$\bullet$] $U(k)$ adjoint: $\mathtt{a}, \mathtt{b}=1,...,k^2$
\item[$\bullet$] Cartan subalgebra of $\mathfrak{su}(k)$: $\mathtt{i}, \mathtt{j}=1,...,k-1$
\item[$\bullet$] Complex coordinates of $\mathbb{R}^8$: $s=1,...,4$
\end{enumerate}

\subsubsection*{$\mathfrak{so}(4)$ spinors}

We adopt the convention that 4D spinor indices are always contracted as ${}^\alpha{}_\alpha$ and ${}_{\dot{\alpha}}{}^{\dot{\alpha}}$. The spinors with raised and lowered index are defined by contracting with Levi-Civita symbols, which satisfy $\varepsilon^{12}=\varepsilon_{12}=-\varepsilon^{\dot{1} \dot{2}}=-\varepsilon_{\dot{1} \dot{2}}=+1$.

Note that we are working in Euclidean signature. The 4-dimensional chirality matrix $\gamma_5$ is defined as:
\ie
\gamma_5=-i\gamma^0\gamma^1\gamma^2\gamma^3.
\fe

A 4D Dirac spinor is decomposed as $\psi=(\psi_{L\alpha}, \psi^{\dot{\alpha}}_R)$. We take the following matrix representation for 4-dimensional Gamma matrices:
\begin{equation} \gamma^\mu=\left(\begin{array}{cc}
0 & \sigma^\mu \\
\bar{\sigma}^\mu & 0
\end{array}\right),\quad (\sigma^\mu)_{\alpha\dot{\beta}}=(1,i\sigma),\quad (\bar{\sigma}^\mu)^{\dot{\alpha}\beta}=(1,-i\sigma), \end{equation}
where the Pauli matrices satisfy $\sigma_\mu \bar{\sigma}_\nu+\sigma_\nu \bar{\sigma}_\mu = 2\delta_{\mu\nu}$. The $SO(4)$ generators are defined as:
\begin{equation} \sigma^{\mu\nu}=\frac{1}{2}(\sigma^\mu \bar{\sigma}^\nu-\sigma^\nu \bar{\sigma}^\mu),\quad \bar{\sigma}^{\mu\nu}=\frac{1}{2}(\bar{\sigma}^\mu {\sigma}^\nu-\bar{\sigma}^\nu {\sigma}^\mu).
\end{equation}
The 4-dimensional charge conjugation matrix is given by:
\begin{equation} C_{(4)}=\left(\begin{array}{cc}
C^{\alpha \beta} & 0 \\
0 & C_{\dot{\alpha} \dot{\beta}}
\end{array}\right)=\left(\begin{array}{cc}
-\varepsilon^{\alpha \beta} & 0 \\
0 & -\varepsilon_{\dot{\alpha} \dot{\beta}}
\end{array}\right).
\end{equation}

\subsubsection*{$\mathfrak{so}(6)_R\sim \mathfrak{su}(4)_R$ spinors}

We denote left-handed $SO(6)$ Weyl spinors by $\psi_A$ and right-handed Weyl spinors by $\psi^A$. A $SO(6)$ Dirac spinor is decomposed as $\Lambda=(\Lambda_{LA}, \Lambda_{R}^A)$. The 6D Gamma matrices can be decomposed as:
\begin{equation} \Gamma^a=\left(\begin{array}{cc}
0 & (\Sigma^a)_{AB} \\
(\bar{\Sigma}^a)^{AB} & 0
\end{array}\right),
\end{equation}
where the invariant symbols satisfy $\Sigma^a\bar{\Sigma}^b+\Sigma^b\bar{\Sigma}^a=2\delta^{ab}$. The matrices can be written in terms of the antisymmetric 't Hooft symbols
\be
\begin{aligned}
&\eta^a_{MN} = \bar{\eta}^a_{MN} = \epsilon_{a MN}, \quad M,N \in \{1,2,3\}, \\
&\eta_{4N}^a =  -\bar{\eta}_{4N}^a =  \delta_{a N}
\end{aligned}
\ee
as
\be
\begin{aligned}
&(\Sigma^a)_{AB} = \eta^a_{AB}, \quad (\bar{\Sigma}^a)^{AB} =-\eta^a_{AB}, \quad a = 1,2,3,  \\
&(\Sigma^a)_{AB} = i\bar{\eta}^{a-3}_{AB}, \quad (\bar{\Sigma}^{a})^{AB} = i\bar{\eta}^{a-3}_{AB}, \quad a = 4,5,6,  \\
\end{aligned}
\ee
from which numerous useful identities, for example
\be
(\Sigma^a)_{AB} (\Sigma_a)_{CD} = 2\epsilon_{ABCD}
\label{sigmasigmaeps}
\ee
can be derived by using identities for the 't Hooft symbols, such as
\be\begin{gathered}
\delta_{ab} \eta^a_{AB} \eta^b_{CD} = \delta_{AC} \delta_{BD} - \delta_{AD} \delta_{BC} + \epsilon_{ABCD}\\
\delta_{ab} \bar{\eta}^a_{AB} \bar{\eta}^b_{CD} = \delta_{AC} \delta_{BD} - \delta_{AD} \delta_{BC} - \epsilon_{ABCD}.
\end{gathered}\ee
The 6-dimensional charge conjugation matrix is given by:
\begin{equation} C_{(6)}=\left(\begin{array}{cc}
0 & C_A{ }^B \\
C_B^A & 0
\end{array}\right)=\left(\begin{array}{cc}
0 & -\mathrm{i} \delta_A^B \\
-\mathrm{i} \delta_B^A & 0
\end{array}\right).
\end{equation}

\subsubsection*{$\mathfrak{so}(10)$ spinors}

The 32-dimensional $SO(10)$ Dirac spinor can be decomposed as $\Psi=(\Psi_{\mathcal{A}}, \tilde{\Psi}_{\dot{\mathcal{A}}})$, and the Gamma matrices in this basis are given by:
\ie
\Gamma^M=
\left(\begin{matrix}
0 & (\Gamma^M)_{\mathcal{A}}{}^{\dot{\mathcal{B}}}\\
(\Gamma^M)_{\dot{\mathcal{A}}}{}^{{\mathcal{B}}} & 0
\end{matrix}\right).
\fe
Under the subgroup reduction $SO(10)\supset SO(6)\times SO(4)$, the 10D Gamma matrices, the 10D chirality matrix, and the 10D charge conjugation matrix are decomposed as:
\begin{equation} \Gamma^\mu=\gamma^\mu\otimes \mathbf{1},\quad \Gamma^a=\gamma_5\otimes \Gamma^a \end{equation}
\begin{equation} \Gamma_{11}=\gamma_5\otimes \Gamma_7,\quad C_{(10)}=C_{(4)}\otimes C_{(6)}.
\end{equation}

\subsubsection*{$\mathfrak{so}(8)$ and $\mathfrak{so}(7)$ spinors}

In Appendix \ref{appC}, we need to use the decomposition of a 10-dimensional spinor under the subgroup reduction $SO(10)\supset SO(8)\supset \text{Spin}(7)$, where the $SO(8)$ is obtained by considering 8 out of 10 directions of $SO(10)$, and the Spin$(7)$ is the subgroup that preserves the Cayley 4-form of $\mathbb{R}^8$. We first consider the subgroup reduction $SO(8)\times SO(2)\subset SO(10)$, under which the chiral and anti-chiral spinors $\mathbf{16}, \overline{\mathbf{16}}$ decompose into:
\ie
\mathbf{16}=(\mathbf{8_s})_+\oplus (\mathbf{8_c})_-,\quad \overline{\mathbf{16}}=(\mathbf{8_s})_-\oplus (\mathbf{8_c})_+
\fe
and we will denote $\Psi_{\mathcal{A}}=(\Psi_\mathbf{A},\Psi_{\dot{\mathbf{A}}})$, $\tilde{\Psi}_{\dot{\mathcal{A}}}=(\tilde{\Psi}_\mathbf{A},\tilde{\Psi}_{\dot{\mathbf{A}}})$. We will also rotate our basis such that the charge conjugate matrix is diagonal:
\ie
(C_{(10)})^{\dot{\mathcal{A}}\mathcal{B}}=\left(\begin{matrix}
\delta^{\mathbf{AB}} & 0\\
0 & \delta^{\dot{\mathbf{A}}\dot{\mathbf{B}}}
\end{matrix}\right),\quad (C_{(10)})^{\mathcal{A}\dot{\mathcal{B}}}=\left(\begin{matrix}
-\delta^{\mathbf{AB}} & 0\\
0 & -\delta^{\dot{\mathbf{A}}\dot{\mathbf{B}}}
\end{matrix}\right)
\fe
The Gamma matrices can be written as:
\ie
X_M(\Gamma^M)_{\dot{\mathcal{A}}}{}^{\mathcal{B}}&=
\left(\begin{matrix}
(-X_9+iX_{10})\delta_{\mathbf{A}}^{\mathbf{B}} & X_\mathbf{a}(\Gamma_8^\mathbf{a})_{\mathbf{A}}{}^{\dot{\mathbf{B}}}\\
X_\mathbf{a}(\Gamma_8^\mathbf{a})_{\dot{\mathbf{A}}}{}^{\mathbf{B}} & (X_9+iX_{10})\delta^{\dot{\mathbf{B}}}_{\dot{\mathbf{A}}}
\end{matrix}\right)
\label{G101}
\fe
\ie
X_M(\Gamma^M)_{\mathcal{A}}{}^{\dot{\mathcal{B}}}&=
\left(\begin{matrix}
(-X_9-iX_{10})\delta_{\mathbf{A}}^{\mathbf{B}} & X_\mathbf{a}(\Gamma_8^\mathbf{a})_{\mathbf{A}}{}^{\dot{\mathbf{B}}}\\
X_\mathbf{a}(\Gamma_8^\mathbf{a})_{\dot{\mathbf{A}}}{}^{\mathbf{B}} & (X_9-iX_{10})\delta^{\dot{\mathbf{B}}}_{\dot{\mathbf{A}}}
\end{matrix}
\right),
\label{G102}
\fe
where $(\Gamma_8^\mathbf{a})_{\mathbf{A}}{}^{\dot{\mathbf{B}}}$ and $(\Gamma_8^\mathbf{a})_{\dot{\mathbf{A}}}{}^{\mathbf{B}}$ are 8D gamma matrices.

Now we want to further reduce $SO(8)$ down to the Spin$(7)$ that preserves a unit chiral spinor $\zeta$. Under this $\mathrm{Spin}(7)\subset SO(8)$, the chiral spinor $\mathbf{8_s}$, anti-chiral spinor $\mathbf{8_c}$ and the vector $\mathbf{8_v}$ decompose as:
\ie
\mathbf{8_s}=\mathbf{1}\oplus \mathbf{7},\quad \mathbf{8_c}=\mathbf{8},\quad \mathbf{8_v}=\mathbf{8},
\fe
where the $\mathbf{1}$ is given by the unit spinor $\zeta$, and the $\mathbf{7}$ is the same representation as the vector of Spin(7). At the level of Gamma matrices, we have:
\ie
(\Gamma_8^8)_{\mathbf{A}}{}^{\dot{\mathbf{B}}}=(\Gamma_8^8)_{\dot{\mathbf{A}}}{}^{{\mathbf{B}}}=-\mathbf{1}_{8\times 8}
\label{G81}
\fe
\ie
(\Gamma_8^\mathbf{i})_{\mathbf{j}}{}^{\dot{\mathbf{k}}}=-(\Gamma_8^\mathbf{i})_{\dot{\mathbf{j}}}{}^{\mathbf{k}}=c_{\mathbf{ijk}},\quad (\Gamma_8^\mathbf{i})_\mathbf{j}{}^{\dot{8}}=-(\Gamma_8^\mathbf{i})_8{}^{\dot{\mathbf{j}}}=-(\Gamma_8^\mathbf{i})_{\dot{\mathbf{j}}}{}^{8}=(\Gamma_8^\mathbf{i})_{\dot{8}}{}^{\mathbf{j}}=-\delta_{\mathbf{ij}},
\label{G82}
\fe
where $\mathbf{i,j}=1,...,7$. $c_{\mathbf{ijk}}$ are the octonionic structure constants. They are completely asymmetric, and the non-vanishing ones are given by
\ie
c_{712}=c_{734}=c_{756}=c_{642}=c_{613}=c_{514}=c_{523}=1.
\fe
In this basis, the unit spinor $\zeta$ is given by $\zeta=(0,...,0,1)$, and the Spin(7) generators are then given by the 21 $\omega_{\mathbf{ab}}\Gamma^{\mathbf{ab}}_8$'s that satisfy $\omega_{\mathbf{ab}}\Gamma^{\mathbf{ab}}_8\zeta=0$.

\subsection{String perturbation theory}

Our conventions for string perturbation theory are summarized as the following:
\begin{enumerate}
\item[$\bullet$] The basic free field OPEs are given by:
\begin{equation}
\begin{gathered}
\partial X^\mu(z) \partial X^\nu(0)\sim -\frac{\alpha'}{2}\frac{\eta^{\mu\nu}}{z^2},\quad \psi^\mu(z)\psi^\nu(0)\sim \frac{\eta^{\mu\nu}}{z}\\
b(z)c(0)\sim \frac{1}{z},\quad \beta(z)\gamma(0)\sim -\frac{1}{z}.
\end{gathered}
\end{equation}
\item[$\bullet$] The $\beta\gamma$ system is re-bosonized into the $(\eta, \xi, \phi)$ system by the following dictionary:
\ie
\beta \sim e^{-\phi}\partial \xi,\quad \gamma \sim \eta e^{\phi},\quad \delta(\beta)\sim e^{\phi},\quad \delta(\gamma)\sim e^{-\phi}
\fe
with the OPEs
\be
\xi(z)\eta(0) \sim \frac{1}{z}, \quad \partial \phi(z) \partial \phi(0) \sim -\frac{1}{z^2}, \quad e^{q_1 \phi(z)} e^{q_2 \phi(0)} \sim z^{-q_1 q_2} e^{(q_1+q_2)\phi(0)}
\ee

\item[$\bullet$] The worldsheet BRST current is given by:
\ie
j_\text{B}(z)=cT^\text{m}-\frac{1}{2} \eta e^{\phi} G^\text{m}+bc\partial c+c\left(-\eta \partial \xi-\partial^2\phi-\frac{1}{2}(\partial \phi)^2\right)-\frac{1}{4}be^{2\phi}\eta \partial \eta+\frac{3}{2}\partial (c\partial \phi),
\fe
where the matter supercurrent is:
\ie
G^{\mathrm{m}} = i \sqrt{\frac{2}{\alpha'}} \psi_\mu \partial X^\mu
\fe
and the BRST charge $Q_\text{B}$ is defined by:
\ie
Q_\text{B}=\oint \frac{dz}{2\pi i}j_\text{B}(z).
\fe

\item[$\bullet$] The picture changing operator is defined as:
\ie
\mathcal{X}(z)=Q_\text{B}\cdot \xi(z)=-\frac{1}{2}e^{\phi}G^\text{m}+c\partial \xi-\frac{1}{4}e^{2\phi}\partial \eta b-\frac{1}{4}\partial(e^{2\phi}\eta b)
\fe
and the zero mode $\mathcal{X}_0$ of the PCO is defined to be:
\ie\label{zmpcodef}
\mathcal{X}_0=\oint \frac{dz}{2\pi i} \frac{1}{z} \mathcal{X}(z).
\fe

\item[$\bullet$] Sphere correlators of the matter and ghost CFTs are normalized as:
\begin{equation}
\begin{gathered}
\left\langle \prod_{i=1}^3 c(z_i)\tilde{c}(\bar{z}_i) \right\rangle_{bc,S^2}=|z_{12}z_{23}z_{13}|^2\\
\left\langle \prod_{i=1}^2 e^{-\phi-\tilde{\phi}}(z_i, \bar{z}_i) \right\rangle_{\beta\gamma, S^2}=\frac{1}{|z_{12}|^2}\\
\langle e^{ik\cdot X}\rangle_{\text{m}, S^2}=-i\frac{8\pi}{\alpha'} g_s^{-2} (2\pi)^{10}\delta^{10}(k).
\label{nclosed}
\end{gathered}
\end{equation}

\item[$\bullet$] Disk correlators with D$p$-brane boundary condition are normalized as:
\begin{equation}
\begin{gathered}
\langle c(z_1)c(z_2)c(z_3)\rangle_{bc,D^2}=z_{12}z_{13}z_{23}\\
\langle e^{-\phi}(z_1)e^{-\phi}(z_2)\rangle_{\beta \gamma, D^2}=\frac{1}{z_{12}}\\
\langle \dot{:}e^{ik\cdot X}\dot{:}\rangle_{\text{m}, D^2}=-(2\pi)^{p+1}\delta^{p+1}(k) C_{D^2}^{(p)},
\end{gathered}
\end{equation}
where the triple dot stands for boundary normal ordering. The factor $C_{D^2}^{(p)}$ is related to $g_s$ and $\alpha'$ by:
\ie
C_{D^2}^{(p)}=\pi^{3/2}g_s^{-1}(2\pi \sqrt{\alpha'})^{3-p}.
\fe
In the main text, $C_{D^2}$ denotes specifically the normalization factor on a D-instanton, namely the one with $p=(-1)$. Other normalization factors will be written with the superscript $(p)$.

\item[$\bullet$] The open string coupling is defined by $C_{D^2}^{(p)}=\alpha'^{-1}(g_o^{(p)})^{-2}$, and it is related to $g_s$ and $\alpha'$ by:
\ie
g_s=\frac{(g_o^{(p)})^2}{4\sqrt{\pi}}(2\pi \sqrt{\alpha'})^{5-p}.
\fe

\item[$\bullet$] The Yang-Mills coupling $g_\text{YM}^{(p)}$ on the D$p$-brane is related to $g_o^{(p)}$ by:
\ie
g_\text{YM}^{(p)}=2^{-3/2}\frac{g_o^{(p)}}{\sqrt{\alpha'}}.
\fe
Note that our Yang-Mills coupling $g_\text{YM}^{(p)}$ is defined such that the Yang-Mills action is normalized as $\frac{1}{4g_\text{YM}^2}\mathrm{Tr}(F_{\mu\nu} F^{\mu\nu})$, as opposed to the convention $\frac{1}{2g_\text{YM}^2}\mathrm{Tr}(F_{\mu\nu} F^{\mu\nu})$ that is often used in the literature.

\item[$\bullet$] The dimensionless string coupling $g_B$ of type IIB is defined to be the ratio between the F1-string and D1-brane tension. It can be expressed as:
\ie
\label{gB}
g_B=\frac{g_s}{16\pi^{5/2}\alpha'^2}=\frac{1}{\tau_2}.
\fe
\end{enumerate}

\subsection{Spin fields and twist fields}

In our notation, the Dirichlet-Neumann and Neumann-Dirichlet twist fields are normalized by the following OPE:
\ie
{\sigma}(z)\bar{\sigma}(0)\sim z^{-1/8}.
\fe
Importantly, we always contract the twist fields such that the Dirichlet boundary is on the outside, hence this OPE is only true for $z<0$.

The twist field $\Delta, \bar{\Delta}$ is defined by the product of 4 $\sigma$ or $\bar{\sigma}$'s, and they have the following OPE:
\ie
{\Delta}(z)\bar{\Delta}(0)\sim z^{-1/2}.
\fe

The excited twist fields are defined by the following OPE:
\ie
i\partial X(z)\sigma(0)\sim \frac{\sigma'(0)}{z^{1/2}},\quad i\partial X(z)\sigma'(0)=\frac{\alpha'}{4z^{3/2}}\sigma(0)+\frac{\alpha'}{z^{1/2}}\partial \sigma(0).
\fe
We will denote the product of 1 excited twist field in the $\mu$-direction and 3 ordinary twist fields by $\tau^\mu$. They are defined by the following OPEs:
\ie
i\partial X^\mu (z)\Delta(0)\sim \frac{\tau^\mu(0)}{z^{1/2}},\quad i\partial X^\mu (z)\tau^\nu (0)=\frac{\alpha'}{4z^{3/2}}\eta^{\mu\nu}\Delta(0)+\frac{\alpha'}{z^{1/2}}\eta^{\mu\nu}\partial \Delta(0)+\frac{\tau^{\mu\nu}(0)}{z^{1/2}}.
\fe

When calculating disk correlators, we'll always group the Neumann-Dirichlet or Dirichlet-Neumann twist field together with the SO(4) or SO(6) spin field to cancel the cocycle phases, if possible. In the Ramond sector, we will also group the $e^{-\phi/2}$ factor together. The operators $\Delta S_\alpha, \Delta S^{\dot{\alpha}}$ have conformal weight $\frac{1}{2}$, and $e^{-\phi/2}\Delta \Theta_A, e^{-\phi/2}\Delta \Theta^A$ have conformal weight $1$. Both $\Delta S_\alpha, \Delta S^{\dot{\alpha}}$ and $e^{-\phi/2}\Delta \Theta_A, e^{-\phi/2}\Delta \Theta^A$ should be regarded as fermionic objects.

The NS-sector spin fields have the following OPE:
\begin{equation} \begin{aligned}
\psi^\mu(z) \Delta S_\alpha(0)&\sim \frac{1}{\sqrt{2} z^{1/2}}(\sigma^\mu)_{\alpha \dot{\beta}}\Delta S^{\dot{\beta}}(0)\\
\psi^\mu(z) \Delta S^{\dot{\alpha}}(0)&\sim \frac{1}{\sqrt{2} z^{1/2}}(\bar{\sigma}^\mu)^{\dot{\alpha} {\beta}} \Delta S_{{\beta}}(0)\\
{\Delta} S_\alpha (z) \bar{\Delta} S_\beta(0) &\sim -\frac{\varepsilon_{\alpha\beta}}{z}\\
{\Delta} S^{\dot{\alpha}} (z)  \bar{\Delta} S^{\dot{\beta}}(0) &\sim -\frac{\varepsilon^{\dot{\alpha}\dot{\beta}}}{z} \\
{\Delta} S^{\dot{\alpha}} (z)  \bar{\Delta} S_{\beta}(0) &\sim -\frac{(\bar{\sigma}^\mu)^{\dot{\alpha}}{}_\beta}{\sqrt{2}z^{1/2}} \psi^\mu(0)\\
{\Delta} S_{\alpha} (z) \bar{\Delta} S^{\dot{\beta}}(0) &\sim -\frac{({\sigma}^\mu)_\alpha{}^{\dot{\beta}}}{\sqrt{2}z^{1/2}} \psi^\mu(0).
\end{aligned} \end{equation}

The R-sector spin fields have the following OPE:
\begin{equation} \begin{aligned}
\psi^a(z)e^{-\phi/2}\Delta \Theta_A(0)&\sim \frac{1}{\sqrt{2} z^{1/2}}({\Sigma}^a)_{AB} e^{-\phi/2}\Delta\Theta^B(0)\\
\psi^a(z)e^{-\phi/2}\Delta \Theta^A(0)&\sim \frac{1}{\sqrt{2} z^{1/2}}(\bar{\Sigma}^a)^{AB} e^{-\phi/2}\Delta\Theta_B(0)\\
e^{-\phi/2} {\Delta} \Theta_A(z) e^{-\phi/2} \bar{\Delta} \Theta_B(0) &\sim -i \frac{({\Sigma}^a)_{AB}}{\sqrt{2} z}e^{-\phi} \psi^a(0)\\
e^{-\phi/2} {\Delta} \Theta^A(z) e^{-\phi/2} \bar{\Delta} \Theta^B(0) &\sim -i \frac{(\bar{\Sigma}^a)^{AB}}{\sqrt{2} z}e^{-\phi} \psi^a(0)\\
e^{-\phi/2} {\Delta} \Theta^A(z) e^{-\phi/2} \bar{\Delta} \Theta_B(0) &\sim -i \frac{\delta^A_B}{z^{3/2}}e^{-\phi}(0)\\
e^{-\phi/2} {\Delta} \Theta_A(z) e^{-\phi/2} \bar{\Delta} \Theta^B(0) &\sim -i\frac{\delta^A_B}{z^{3/2}}e^{-\phi}(0).
\end{aligned} \end{equation}

The $(-\frac{3}{2})$-picture spin fields are defined by the following OPE:
\begin{equation} \begin{aligned}
e^{-\phi}(z) e^{-\phi/2}\Delta \Theta_A(0)&\sim \frac{1}{z^{1/2}}e^{-3\phi/2} \Delta \Theta_A(0)\\
e^{-\phi}(z) e^{-\phi/2}\Delta \Theta^A(0)&\sim \frac{1}{z^{1/2}}e^{-3\phi/2} \Delta \Theta^A(0),
\end{aligned} \end{equation}
and hence they have the following OPEs with the  $(-\frac{1}{2})$-picture spin fields.
\begin{equation} \begin{aligned}
e^{-3\phi/2} {\Delta} \Theta^A(z) e^{-\phi/2} \bar{\Delta} \Theta_B(0) &\sim -i \frac{\delta^A_B}{z^{2}}e^{-2\phi}(0)\\
e^{-3\phi/2} {\Delta} \Theta_A(z) e^{-\phi/2} \bar{\Delta} \Theta^B(0) &\sim -i\frac{\delta^A_B}{z^{2}}e^{-2\phi}(0)\\
e^{-\phi/2} {\Delta} \Theta^A(z) e^{-3\phi/2} \bar{\Delta} \Theta_B(0) &\sim i \frac{\delta^A_B}{z^{2}}e^{-2\phi}(0)\\
e^{-\phi/2} {\Delta} \Theta_A(z) e^{-3\phi/2} \bar{\Delta} \Theta^B(0) &\sim i\frac{\delta^A_B}{z^{2}}e^{-2\phi}(0).
\end{aligned} \end{equation}
Consistency of this set of OPEs can be checked by calculating the correlator of $ce^{-\phi}\psi^a(z_1)$, $ce^{-\phi/2}\Delta \Theta(z_2)$ and $ce^{-\phi/2}\bar{\Delta} \Theta(z_3)$ in different OPE channels.

The picture-raised version of the $(-\frac{3}{2})$-picture spin field is given by:
\begin{equation} \begin{aligned}
\mathcal{X}_0\cdot \left(e^{-3\phi/2}\Delta \Theta^A e^{ik\cdot X}(0)\right)&=-\frac{i}{2}\sqrt{\frac{2}{\alpha'}} \oint \frac{dz}{2\pi i}\ z^{-1} \left(e^{\phi}\psi^a \partial X_a(z)\right)\left(e^{-3\phi/2}\Delta \Theta^A e^{ik\cdot X}(0)\right)\\
&=\frac{\sqrt{\alpha'} k_a}{4}(\bar{\Sigma}^a)^{AB} e^{-\phi/2}\Delta \Theta_B e^{ik\cdot X}(0)\\
\mathcal{X}_0\cdot \left(e^{-3\phi/2}\Delta \Theta_A e^{ik\cdot X}(0)\right)&=\frac{\sqrt{\alpha'} k_a}{4}({\Sigma}^a)_{AB} e^{-\phi/2}\Delta \Theta^B e^{ik\cdot X}(0).
\end{aligned} \end{equation}

More correlators of spin fields and twist fields can be found in \cite{Vosmera:2019mzw, Hartl:2011tza, Mattiello:2018kue}, but note that their notation is different than ours.

\section{The $w^4$ interaction}

\label{w4}

The $w^4$ interaction term in the SFT action plays an important role when evaluating the zero mode integrals. To get this interaction from the worldsheet, we need to do an honest calculation in string field theory. To do so, we first need to specify the 3-point and 4-point vertex we are using.

\begin{figure}
\centering
\includegraphics[scale=0.5]{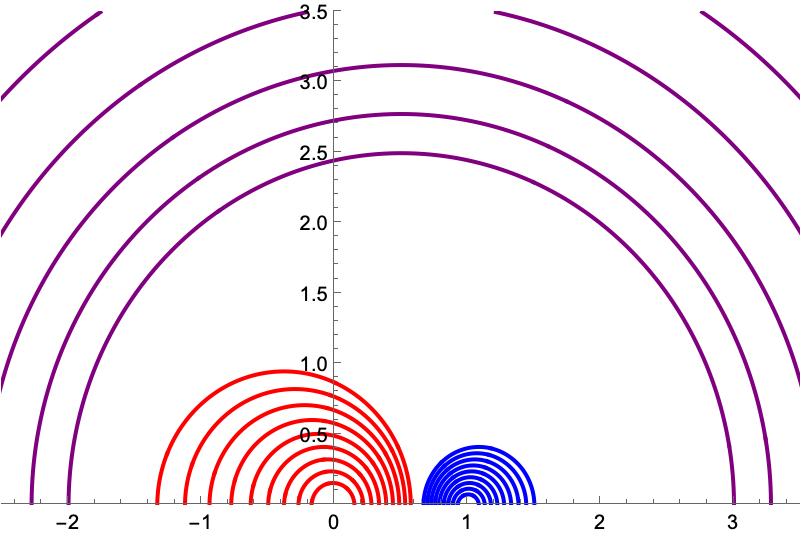}
\caption{Local coordinate maps $f_a(w_a)$ used to define the 3-vertex $\mathcal{V}_{0,3}$, with $\alpha=2.5$. Each half-circle has the same $|w_a|$, ranging from 0.2 to 1.}
\label{coordsys}
\end{figure}

Since the moduli space $\mathcal{M}_{0,3}$ of 3 boundary-punctured disk is a single point, the 3-point string vertex $\mathcal{V}_{0,3}$ is defined by specifying 3 local coordinate maps $f_{a}(w_a)$ for the off-shell punctures and 1 PCO location on the upper half plane. Following \cite{Agmon:2022vdj}, we take the local coordinate maps to be:
\begin{equation} f_1(w_1)=\frac{2w_1}{2\alpha+w_1},\quad f_2(w_2)=\frac{2\alpha+w_2}{2\alpha-w_2},\quad f_{3}(w_3)=\frac{w_3-2\alpha}{2w_3},
\end{equation}
which sends $w_a=0$ to $z_a=0, 1$ and $\infty$, as shown in figure \ref{coordsys}. The PCO is taken to be located at the place of the $\bar{w}$ vertex operator. The $\bar{w}w\xi^1$ 3-point amplitude is then given by:
\begin{equation} \begin{aligned}
\mathcal{A}_3=(-1)\frac{1}{C_{D^2}}\langle (\mathcal{X}_0\cdot ce^{-\phi}{\Delta}S^{\dot{\alpha}})(0)ce^{-\phi}\bar{\Delta}S^{\dot{\beta}}(1) c\partial ce^{-2\phi}\partial \xi(\infty)\rangle= -\frac{1}{4}\varepsilon^{\dot{\alpha}\dot{\beta}},
\end{aligned} \end{equation}
where the 0-picture vertex operator of $\bar{w}$ is given by:
\ie
\mathcal{X}_0\cdot ce^{-\phi}{\Delta}S^{\dot{\alpha}}=\frac{1}{2\sqrt{\alpha'}}c(\bar{\sigma}^\mu)^{\dot{\alpha}\beta}\tau_\mu S_{\beta}-\frac{1}{4}\eta e^{\phi}{\Delta}S^{\dot{\alpha}}.
\fe
Therefore, the 3-vertex contribution to the $w_{\dot{\alpha}}\bar{w}_{\dot{\beta}}w_{\dot{\gamma}}\bar{w}_{\dot{\delta}}$ amplitude is the following:
\begin{equation} \mathcal{A}_\text{3-vertex}=-\frac{1}{8}\frac{1}{C_{D^2}}(\varepsilon^{\dot{\alpha}\dot{\beta}}\varepsilon^{\dot{\gamma}\dot{\delta}}+\varepsilon^{\dot{\alpha}\dot{\delta}}\varepsilon^{\dot{\gamma}\dot{\beta}}), \end{equation}
where we have used the $\xi^1$ propagator $P_{\xi^1}=-\frac{2}{C_{D^2}}$.

The geometric BV equation requires the 4-vertex $\mathcal{V}_{0,4}$ to be the complement of the Feynman diagram regions, which are given by plumbing two 3-boundary-punctured disks. To do so, we choose two boundary punctures parametrized half-disk coordinates $w,w'$. We then make the identification $ww'=-q\ (0<q<1)$ around the half-circles $|w|=q^{1/2}, |w'|=q^{1/2}$. Let us consider the case where we plumb together $w_2,w_2'$. The plumbing map $w_2w_2'=-q$ then takes the following form for the $z,z'$ coordinates:
\begin{equation} 4\alpha^2 \frac{z-1}{z+1}\frac{z'-1}{z'+1}=-q. \end{equation}

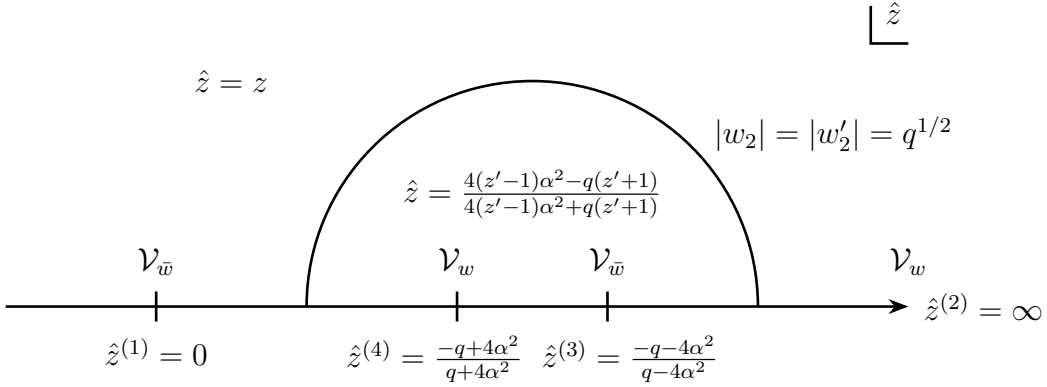
\begin{figure}
\center
\begin{tikzpicture}

\draw[color=black, line width = 1pt, arrows=-{Stealth}] (-7,0) -- (5,0);
\draw[color=black, line width = 1pt] (1,-0.2) -- (1,0.2);
\draw[color=black, line width = 1pt] (-1,-0.2) -- (-1,0.2);
\draw (-5,-0.6) node {$\hat{z}^{(1)}=0$};
\draw (-5,0.6) node {$\mathcal{V}_{\bar{w}}$};
\draw (5,0.6) node {$\mathcal{V}_{w}$};
\draw (1,0.6) node {$\mathcal{V}_{\bar{w}}$};
\draw (-1,0.6) node {$\mathcal{V}_{w}$};
\draw (6,0) node {$\hat{z}^{(2)}=\infty$};
\draw (1.3,-0.7) node {$\hat{z}^{(3)}=\frac{-q-4\alpha^2}{q-4\alpha^2}$};
\draw (-1.3,-0.7) node {$\hat{z}^{(4)}=\frac{-q+4\alpha^2}{q+4\alpha^2}$};
\draw[color=black, line width = 1pt] (-5,-0.2) -- (-5,0.2);
\draw[color=black, line width = 1pt] (5,3.5) -- (4.5,3.5) -- (4.5,4);
\draw (4.8,3.9) node {$\hat{z}$};
\draw (4,2.3) node {$|w_2|=|w_2'|=q^{1/2}$};
\draw[color = black, line width = 1pt] (3,0) arc[start angle=0, end angle = 180, radius=3];

\draw (-4,3) node {$\hat{z}=z$};
\draw (0,1.5) node {$\hat{z}=\frac{4(z'-1)\alpha^2-q(z'+1)}{4(z'-1)\alpha^2+q(z'+1)}$};

\end{tikzpicture}
\caption{The new coordinates $\hat{z}$ defined on the newly formed disk.}
\label{newcoord}
\end{figure}

\begin{figure}
\center
\begin{tikzpicture}

\draw[color=black, line width = 1pt] (0,0) -- (12,0);
\draw[color=black, line width = 1pt] (12,-0.3) -- (12,0.3);
\draw[color=black, line width = 1pt] (0,-0.3) -- (0,0.3);
\draw[color=black, line width = 1pt] (4,-0.3) -- (4,0.3);
\draw (13,0) node {$\mathcal{M}_{0,4}$};
\draw (12,-0.5) node {$x=1$};
\draw (0,-0.5) node {$x=0$};
\draw (4,-0.5) node {$x=\alpha^{-2}$};
\draw (2,0.5) node {Feynman region};
\draw (8,0.5) node {Vertex region};

\filldraw[color=black, fill=black!20, thick] (1,2.5) circle (0.6);
\filldraw[color=black, fill=black!20, thick] (3,2.5) circle (0.6);
\draw[densely dotted,line width = 1pt] (1.6,2.5) -- (2.4,2.5);
\filldraw[color=black, fill=black, thick] (1,3.1) circle (0.05);
\draw (1,3.5) node {$\mathcal{V}_{\bar{w}}$};
\draw (1,1.5) node {$\mathcal{V}_{{w}}$};
\draw (3,3.5) node {$\mathcal{V}_{{w}}$};
\draw (3,1.5) node {$\mathcal{V}_{\bar{w}}$};
\draw (2,2.1) node {$\xi^1$};
\filldraw[color=black, fill=black, thick] (1,1.9) circle (0.05);
\filldraw[color=black, fill=black, thick] (3,3.1) circle (0.05);
\filldraw[color=black, fill=black, thick] (3,1.9) circle (0.05);

\draw[dashed, line width = 1pt] (0,0) -- (0,4);
\draw[dashed, line width = 1pt] (4,0) -- (4,4);
\draw[dashed, line width = 1pt] (12,0) -- (12,4);

\filldraw[color=black, fill=black!20, thick] (8,2.5) circle (1.414*0.8);
\filldraw[color=black, fill=black, thick] (7.2, 3.3) circle (0.05);
\filldraw[color=black, fill=black, thick] (8.8, 3.3) circle (0.05);
\filldraw[color=black, fill=black, thick] (7.2, 1.7) circle (0.05);
\filldraw[color=black, fill=black, thick] (8.8, 1.7) circle (0.05);
\draw (6.8,3.5) node {$\mathcal{V}_{\bar{w}}$};
\draw (9.2,3.5) node {$\mathcal{V}_{{w}}$};
\draw (6.8,1.5) node {$\mathcal{V}_{{w}}$};
\draw (9.2,1.5) node {$\mathcal{V}_{\bar{w}}$};

\end{tikzpicture}
\caption{The moduli space $\mathcal{M}_{0,4}$ of 4-boundary-punctured disks, covered by the Feynman region and the vertex region.}
\label{moduli4}
\end{figure}
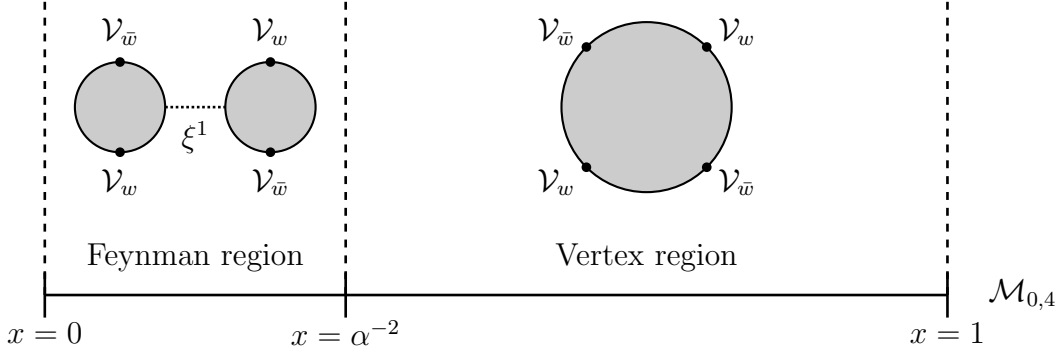

We can introduce a new coordinate system $\hat{z}$ on the newly formed 4-boundary-punctured disk by the following: Outside the image of $|w|=q^{1/2}$, we use the $z$-coordinate on the first disk, i.e. $\hat{z}=z$. Inside the image of $|w|=q^{1/2}$, we define the new $\hat{z}$ as the solution to the equation $4\alpha^2 \frac{\hat{z}-1}{\hat{z}+1}\frac{z'-1}{z'+1}=-q$. The plumbing map then ensures that our new coordinate is continuous on the intersection $|w|=q^{1/2}$. In the new coordinate system, the 4 punctures $z=0,z=\infty, z'=0,z'=\infty$ are located at:
\begin{equation} \begin{gathered}
\hat{z}^{(1)}=0,\quad \hat{z}^{(2)}=\infty,\quad \hat{z}^{(3)}=\frac{-q-4\alpha^2}{q-4\alpha^2},\quad \hat{z}^{(4)}=\frac{-q+4\alpha^2}{q+4\alpha^2}
\end{gathered} \end{equation}
as illustrated in figure \ref{newcoord}. It is important to keep track of which kind of boundary-changing operator we put on the 4 punctures. Since we can only plumb together punctures on the D(-1) boundary, the Neumann-Dirichlet twist field $\bar{\Delta}$ should be inserted at $\hat{z}^{(1)}, \hat{z}^{(3)}$, and the Dirichlet-Neumann twist field ${\Delta}$ are put at $\hat{z}^{(2)}, \hat{z}^{(4)}$. We then perform a $SL(2,\mathbb{R})$ transformation $\tilde{z}=1-\frac{\hat{z}^{(4)}}{\hat{z}}$ to set 3 of the 4 punctures to $0,1,\infty$:
\begin{equation} \begin{gathered}
\tilde{z}^{(1)}=\infty,\quad \tilde{z}^{(2)}=1,\quad \tilde{z}^{(3)}=\frac{16 q \alpha^2}{(q+4\alpha^2)^2}=\frac{q}{\alpha^2}+\mathcal{O}(\alpha^{-4}),\quad \tilde{z}^{(4)}=0.
\end{gathered} \end{equation}
Therefore, if one fixes the 2 DN twist fields at $\tilde{z}=0,1$ and a ND twist field at $\tilde{z}=\infty$, then the moduli space $\mathcal{M}_{0,4}$ will be parametrized by the position $x\in[0,1]$ of the remaining ND twist field. The discussion above tells us that the Feynman region covers the interval $[0, \alpha^{-2}]$, and hence the 4-vertex should cover the moduli region $[\alpha^{-2},1]$, as illustrated in figure \ref{moduli4}.

To define the 4-vertex, we need to specify 4 local coordinates $f_a$ and 2 PCO positions $p_1, p_2$. However, since all of our external states are on-shell, the vertex contribution is independent of these local coordinates. We will put the 2 PCOs on the 2 $\bar{w}$ vertex operators, as doing so will guarantee that the PCO locations match on the boundary of the vertex and Feynman regions, and hence there will be no additional terms from vertical integration. The 4-point vertex contribution to the $w_{\dot{\alpha}}\bar{w}_{\dot{\beta}}w_{\dot{\gamma}}\bar{w}_{\dot{\delta}}$ amplitude is then simply:
\be
\begin{aligned}
	\mathcal{A}_\text{4-vertex}=&-\frac{1}{(C_{D^2})^2}\int_{\alpha^{-2}}^1 dx\langle(\mathcal{X}_0\cdot ce^{-\phi}{\Delta}S^{\dot{\alpha}})(0)e^{-\phi}\bar{\Delta} S^{\dot{\beta}}(x)(\mathcal{X}_0\cdot ce^{-\phi}{\Delta}S^{\dot{\gamma}})(1)ce^{-\phi}\bar{\Delta} S^{\dot{\delta}}(\infty)\rangle \\
&+(\beta\leftrightarrow \delta).
\end{aligned}
\ee
The only term that contributes to the correlator is the $\tau_\mu$ term in the 0-picture vertex operator, and the integrand evaluates to \cite{Hartl:2011tza, Mattiello:2018kue}:
\ie
\begin{aligned}
-\frac{1}{4\alpha'}&(\bar{\sigma}^\mu)^{\dot{\alpha}\alpha}(\bar{\sigma}^\nu)^{\dot{\gamma}\gamma}\langle c {\tau}_\mu S_\alpha(0) e^{-\phi}\bar{\Delta} S^{\dot{\beta}}(x)c{\tau}_\nu S_{\gamma}(1) ce^{-\phi}\bar{\Delta} S^{\dot{\delta}}(w)\rangle\\
&=\frac{C_{D^2}}{16}(\bar{\sigma}^\mu)^{\dot{\alpha}\alpha}(\bar{\sigma}_\mu)^{\dot{\gamma}\gamma}\varepsilon_{\alpha \gamma}\varepsilon^{\dot{\beta}\dot{\delta}}\frac{1}{x}\left(\frac{\pi}{2K(x)}\right)^2\left(1-\frac{E(x)}{(1-x)K(x)}\right),
\end{aligned}
\fe
where $K(x)$ and $E(x)$ are the complete elliptic integrals of the first and second kinds, which comes from the 4-point correlator of the twist field. Upon summing over the 2 cyclic orders, the 4-point vertex contribution vanishes, and the total 4-pt amplitude is given by:
\begin{equation}
\mathcal{A}=-\frac{1}{8}\frac{1}{C_{D^2}}(\varepsilon^{\dot{\alpha}\dot{\beta}}\varepsilon^{\dot{\gamma}\dot{\delta}}+\varepsilon^{\dot{\alpha}\dot{\delta}}\varepsilon^{\dot{\gamma}\dot{\beta}}).
\end{equation}
Therefore, the 4-point coupling in the SFT effective action (with all the massive modes and $\xi^1$ integrated out) reads:
\begin{equation} S\supset \frac{1}{16}\frac{1}{C_{D^2}}(\bar{w}w)^2=\frac{\alpha'^2 (g_\text{YM}^{(-1)})^2}{2}({\bar{w}}{w})^2,
\end{equation}
where $g_\text{YM}^{(-1)}$ is the Yang-Mills coupling on the D-instanton.

\section{Localization computation of the matrix integral $\mathsf{Z}_k$}
\label{appC}

In this appendix, we review the localization computation of the matrix integral $\mathsf{Z}_k$, following the logic of \cite{Moore:1998et, Bruzzo:2002xf, Bruzzo:2003rw}. As mentioned in the main text, $\mathsf{Z}_k$ is defined as
\ie
\mathsf{Z}_k\equiv \int d^{4(k^2-1)}a d^{6k^2}\chi d^{8k^2}\lambda d^{8(k^2-1)}\theta d^{4k}w d^{8k}\mu \exp(-S_{k\text{-inst}}),
\label{Zk0}
\fe
where all the integrals over $U(k)$ adjoint matrices are defined via $d^{k^2}M=\prod_{\mathtt{a}=1}^{k^2} dM^\mathtt{a}$. The components $M^\mathtt{a}$ are defined via $M=M^\mathtt{a}t^\mathtt{a}$. $t^\mathtt{a}$'s are Hermitian generators of $U(k)$, normalized so that $\mathrm{Tr}(t^\mathtt{a} t^\mathtt{b})=\delta^{\mathtt{ab}}$. The $SU(k)$ adjoint integrals are defined in the same way, and the $U(k)$ fundamental integrals are defined by $d^kw=\prod_{i=1}^k dw^i$. The $k$-instanton action $S_{k\text{-inst}}$ is defined in \eqref{kaction}, with the auxiliary fields $D^c$ already integrated out.

\subsubsection*{$Q$-exact form of the matrix integral}

The key property of \eqref{Zk0} that allows us to use supersymmetric localization is that $S_{k\text{-inst}}$ can be written in a $Q$-exact form after introducing a set of auxiliary fields, where $Q$ is a conserved supercharge. For this purpose, we introduce $(\XX, \Psi, \mmu, \ww)$ by
\ie\begin{gathered}
\chi^a=2^{-1/2}\mathbf{X}^a,\quad a_\mu=\sqrt{2}(C_{D^2})^{1/2}\mathbf{X}_\mu\\
\lambda_{\dot{\alpha}A}=2^{1/4}\Psi_{\dot{\alpha}A},\quad
\theta_{\alpha}^A=2^{1/4}(C_{D^2})^{1/2}\Psi_{\alpha}^A\\
\mu^A=2^{-3/4}(C_{D^2})^{1/2}\boldsymbol{\mu}^A,\quad w=\sqrt{2}(C_{D^2})^{1/2}\boldsymbol{w}.
\end{gathered}\fe
Using the new fields, the $k$-instanton action \eqref{kaction} can then be written as
\ie\begin{aligned}
S_{k\text{-inst}}=C_{D^2}\mathrm{Tr}\bigg(&-\frac{1}{2}[\mathbf{X}^M, \mathbf{X}^N]^2-\Gamma^M_{\mathcal{AB}}\Psi^\mathcal{A}[\mathbf{X}_M ,\Psi^{\mathcal{B}}]+ \bar{\boldsymbol{w}}^{\dot{\alpha}}\boldsymbol{w}_{\dot{\alpha}}\mathbf{X}^a\mathbf{X}^a\\
&-\frac{1}{2}i(\Sigma^a)_{AB}\bar{\boldsymbol{\mu}}^A \boldsymbol{\mu}^B\mathbf{X}_a +i(\bar{\boldsymbol{\mu}}^A \boldsymbol{w}_{\dot{\alpha}}+\bar{\boldsymbol{w}}_{\dot{\alpha}}\boldsymbol{\mu}^A)\Psi^{\dot{\alpha}}_A+\frac{1}{4}\mathbf{W}^c \mathbf{W}^c+\frac{i}{2}\bar{\eta}^c_{\mu\nu}[\mathbf{X}_\mu, \mathbf{X}_\nu]\mathbf{W}^c\bigg)
\end{aligned}\fe
and the matrix integral $\mathsf{Z}_k$ is given by
\ie
\mathsf{Z}_k&=2^{-5k^2+8k}(C_{D^2})^{-2k^2-2k+2}\\
&\quad\times \int d^{4(k^2-1)}\XX^\mu d^{6k^2}\XX^a d^{8k^2}\Psi_{\dot{\alpha}A} d^{8(k^2-1)}\Psi_{\alpha}^A d^{4k}\ww d^{8k}\mmu \exp(-S_{k\text{-inst}}).
\fe

To introduce the supercharge $Q$, we pick $\XX_9, \XX_{10}$ out of the 10 directions among $\XX_M$. We also group $\XX_{2s-1}$ and $\XX_{2s}$ into complex coordinates $B_s$ as
\ie\begin{gathered}
\phi=\mathbf{X}_9-i\mathbf{X}_{10},\quad \bar{\phi}=\mathbf{X}_9+i\mathbf{X}_{10}\\
B_s=\mathbf{X}_{2s-1}-i\mathbf{X}_{2s},\quad B_s^\dagger = \mathbf{X}_{2s-1}+i\mathbf{X}_{2s},
\end{gathered}\fe
where $s=1,...,4$.

Picking out the $\XX_9, \XX_{10}$ directions also defines a $SO(8)$ subgroup of $SO(10)$, under which $\phi, \bar{\phi}\sim \XX_9, \XX_{10}$ are singlets. Under the subgroup reduction $SO(8)\subset SO(10)$, the chiral Weyl spinor $\mathbf{16}$ of $SO(10)$ can be decomposed into $\mathbf{16}=\mathbf{8_s}\oplus \mathbf{8_c}$. Furthermore, we consider a $\mathrm{Spin}(7)$ subgroup of $SO(8)$ that preserves a unit chiral spinor, under which the $SO(8)$ chiral spinor $\mathbf{8_s}$ further splits into $\mathbf{7} \oplus \mathbf{1}$. We can then split the adjoint fermion $\Psi^{\mathcal{A}}$ into $\Psi^{\mathcal{A}}=(\Psi_\mathbf{7}, \Psi_\mathbf{1}, \Psi_\mathbf{8})$, and the 10-dimensional Gamma matrix $(\Gamma^M)_{\mathcal{A}\mathcal{B}}$ are given by \eqref{G101}, \eqref{G102} in this basis.

Since the bases $(\Psi_\mathbf{7}, \Psi_\mathbf{1}, \Psi_\mathbf{8})$ and $(\Psi_{\alpha}^A, \Psi_{\dot{\alpha} A})$ are related by an orthogonal change of basis, the measure $d^{8k^2}\Psi_{\alpha}^A d^{8k^2}\Psi_{\dot{\alpha}A}$ equals $d^{7k^2}\Psi_\mathbf{7} d^{k^2}\Psi_\mathbf{1}d^{8k^2}\Psi_\mathbf{8}$, where both are defined by integrating over each component. We further define
\ie \begin{gathered}
\boldsymbol{\Theta}_{s}=-i\left[(\Psi_\mathbf{8})_{\dot{\mathbf{A}}=2s-1}-i(\Psi_\mathbf{8})_{\dot{\mathbf{A}}=2s}\right]\\
\boldsymbol{\Theta}_{s}^\dagger=-i\left[(\Psi_\mathbf{8})_{\dot{\mathbf{A}}=2s-1}+i(\Psi_\mathbf{8})_{\dot{\mathbf{A}}=2s}\right]
\\({\Psi}_\mathbf{7})_\mathbf{i}=i{\chi}_\mathbf{i},\quad \Psi_\mathbf{1}=\frac{i}{2} \eta,
\end{gathered}\fe
where we identified the $SO(8)$ anti-chiral spinor index $\dot{\mathbf{A}}$ with the $SO(8)$ vector index $s$. Among these new variables, the bosonic components $B_{1,2}, B_{1,2}^\dagger$ correspond to $\XX_{1,2,3,4}$, and the fermionic components $\boldsymbol{\Theta}_1, \boldsymbol{\Theta}_2, \boldsymbol{\Theta}_1^\dagger, \boldsymbol{\Theta}_2^\dagger, \chi_{1,2,3,4}$ correspond to $\Psi_{\alpha}^A$. The $U(1)$ component of these fields is not integrated over in $\mathsf{Z}_k$.

For the fields in the fundamental sector, we explicitly write their components as
\ie\begin{gathered}
\boldsymbol{w}_{\dot{\alpha}}=(I^\dagger, J),\quad \bar{\boldsymbol{w}}^{\dot{\alpha}}=(I, J^\dagger)\\
\boldsymbol{\mu}^A=\sqrt{2}(\mu_K, \mu_L^\dagger, \mu_I, \mu_J^\dagger)^\mathrm{T},\quad \boldsymbol{\bar{\mu}}^A=\sqrt{2}(\mu_I^\dagger, \mu_J, \mu_K^\dagger, \mu_L).
\end{gathered}\fe

In order to write $S_{k\text{-inst}}$ in the $Q$-exact form, we further introduce the auxiliary fields $(H_\mathbf{i}, K, K^\dagger, L, L^\dagger)$ and couple them with the following $k\times k$ matrices:
\ie\begin{gathered}
\mathcal{E}_{\mathbb{R}}=\sum_{i=1}^4 [B_i, B_i^\dagger]+II^\dagger-J^\dagger J,\quad \mathcal{E}_\mathbb{C}^{(1)}=[B_1, B_2]-[B_3^\dagger, B_4^\dagger]+IJ\\
\mathcal{E}_\mathbb{C}^{(2)}=-[B_1, B_3]-[B_2^\dagger, B_4^\dagger],\quad \mathcal{E}_\mathbb{C}^{(3)}=[B_1, B_4]-[B_2^\dagger, B_3^\dagger]\\
\mathcal{E}_K=B_3I-B_4^\dagger J^\dagger,\quad \mathcal{E}_{L^\dagger}=B_4I+B_3^\dagger J^\dagger.
\end{gathered}
\label{EEEEE}\fe
The matrices $\mathcal{E}_{\mathbb{R}},\mathcal{E}_\mathbb{C}^{(1,2,3)},\mathcal{E}_\mathbb{C}^{(1,2,3)\dagger}$ can also be reorganized into 7 matrices $\mathcal{E}_\mathbf{i}$ as
\ie\begin{gathered}
\mathcal{E}_1=\frac{i}{2}\left(\mathcal{E}_\mathbb{C}^{(3)}+\mathcal{E}_\mathbb{C}^{(3)\dagger}\right),\quad \mathcal{E}_2=\frac{1}{2}\left(\mathcal{E}_\mathbb{C}^{(3)}-\mathcal{E}_\mathbb{C}^{(3)\dagger}\right)\\
\mathcal{E}_3=\frac{i}{2}\left(\mathcal{E}_\mathbb{C}^{(2)}+\mathcal{E}_\mathbb{C}^{(2)\dagger}\right),\quad \mathcal{E}_4=\frac{1}{2}\left(\mathcal{E}_\mathbb{C}^{(2)}-\mathcal{E}_\mathbb{C}^{(2)\dagger}\right)\\
\mathcal{E}_5=\frac{i}{2}\left(\mathcal{E}_\mathbb{C}^{(1)}+\mathcal{E}_\mathbb{C}^{(1)\dagger}\right),\quad \mathcal{E}_6=\frac{1}{2}\left(\mathcal{E}_\mathbb{C}^{(1)}-\mathcal{E}_\mathbb{C}^{(1)\dagger}\right),\quad \mathcal{E}_7=\frac{1}{2i}\mathcal{E}_{\mathbb{R}}.
\end{gathered}\fe
We define $\{\chi_\mathbb{R},\chi_\mathbb{C}^{(r)}\}$ and $\{H_\mathbb{R}, H_\mathbb{C}^{(r)}\}$ in terms of $\chi_\mathbf{i}, H_\mathbf{i}$ analogously.

With the field redefinitions above, we can introduce the supercharge $Q$, which acts on fields by
\ie
\begin{gathered}
QB_s=\boldsymbol{\Theta}_{s},\quad Q \boldsymbol{\Theta}_{s}=[\phi, B_s],\quad QI=\mu_I, \quad Q\mu_I=\phi I,\quad QJ^\dagger = \mu_J^\dagger, \quad Q\mu_J^\dagger=\phi J^\dagger \\
Q\mu_K=K,\quad Q\mu_{L}^\dagger=L^\dagger, \quad QK=\phi\mu_K, \quad QL^\dagger = \phi \mu_L^\dagger\\
Q\chi_\mathbf{i}=H_\mathbf{i}, \quad QH_\mathbf{i}=[\phi, \chi_\mathbf{i}],\quad Q\phi=0,\quad Q\bar{\phi}=\eta,\quad Q\eta=[\phi, \bar{\phi}].
\end{gathered}
\fe
Note that the supercharge $Q$ does not square to zero. Instead, $Q^2$ is a gauge transformation with $\phi$ being the gauge parameter. After integrating out the auxiliary fields, the $k$-instanton action $S_{k\text{-inst}}$ can be written in $Q$-exact form:
\ie
S_{k\text{-inst}}=C_{D^2}Q\cdot\mathrm{Tr}\left[\frac{1}{4}\eta[\phi, \bar{\phi}]-2\vec{\chi}\cdot \vec{\mathcal{E}}+\vec{\chi}\cdot\vec{H}-\frac{1}{2}(\boldsymbol{\Theta}_\mathbf{s}^\dagger\bar{\phi}\cdot B_\mathbf{s}+\boldsymbol{\Theta}_\mathbf{s}\bar{\phi}\cdot B_\mathbf{s}^\dagger)\right].
\fe
Here, we group $(\mathcal{E}_K,\mathcal{E}_K^\dagger, \mathcal{E}_L,\mathcal{E}_{L^\dagger}, \mathcal{E}_\mathbf{i})$ into a vector $\vec{\mathcal{E}}$ with 11 real components, and we introduced $\vec{H}=(K,K^\dagger,L, L^\dagger, H_\mathbf{i})$ as auxiliary fields. We also group $B_\mathbf{s}=(I, J^\dagger, B_s)$, $\boldsymbol{\Theta}_\mathbf{s}=(\mu_I, \mu_J^\dagger, \boldsymbol{\Theta}_s)$, and $\vec{\chi}=(\mu_K,\mu_K^\dagger,\mu_L, \mu_{L}^\dagger, \chi_\mathbf{i})$. The Lie algebra action $\phi\cdot B$ gives $\phi I, \phi J^\dagger, -I^\dagger \phi, -J\phi$ for the fundamental components $I, J^\dagger, I^\dagger, J$ and $[\phi, B_s]$ for the adjoint components. The inner product $\vec{\mathcal{E}}\cdot \vec{H}$ between 11-dimensional vectors is defined as follows:
\ie\vec{\mathcal{E}}\cdot \vec{H}&=\sum_{\mathbf{i}=1}^7 \mathcal{E}_\mathbf{i} H_\mathbf{i}-\frac{1}{2}\left(\mathcal{E}_K^\dagger K+\mathcal{E}_L L^\dagger+\mathcal{E}_K K^\dagger+\mathcal{E}_L^\dagger L\right)\\
&=-\frac{1}{4}\mathcal{E}_\mathbb{R}H_\mathbb{R}-\frac{1}{2}\sum_{r=1}^3(\mathcal{E}_\mathbb{C}^{(r)\dagger}H_\mathbb{C}^{(r)}+\text{c.c.})-\frac{1}{2}\left(\mathcal{E}_K^\dagger K+\mathcal{E}_L L^\dagger+\mathcal{E}_K K^\dagger+\mathcal{E}_L^\dagger L\right).
\fe

We therefore rewrite the matrix integral $\mathsf{Z}_k$ as follows:
\ie\begin{aligned}
\mathsf{Z}_k=\ &2^{-5k^2+4k}(C_{D^2})^{-2k^2-2k+2} \mathcal{N}_\text{aux}\\
 \int &\bigg(\prod_{s=1}^2 \mathcal{D}'B_s \mathcal{D}'B_s^\dagger \mathcal{D}'\boldsymbol{\Theta}_s \mathcal{D}'\boldsymbol{\Theta}_s^\dagger\bigg) \bigg(\prod_{s=3}^4 \mathcal{D}B_s \mathcal{D}B_s^\dagger \mathcal{D}\boldsymbol{\Theta}_s \mathcal{D}\boldsymbol{\Theta}_s^\dagger\bigg)\\
&\quad \prod_{\mathbf{i}=1}^7\mathcal{D}H_\mathbf{i} \prod_{\mathbf{j}=1}^4\mathcal{D}'\chi_\mathbf{j} \prod_{\mathbf{k}=5}^7 \mathcal{D}\chi_{\mathbf{k}} \mathcal{D}\eta \mathcal{D}\phi \mathcal{D}\bar{\phi}\left(\prod_{\Phi=I,J,K,L}\mathfrak{D}\Phi \mathfrak{D}\Phi^\dagger \mathfrak{D}\mu_{\Phi}\mathfrak{D}\mu_{\Phi}^\dagger\right)\\
& \times\exp\left\{-C_{D^2}Q\cdot\mathrm{Tr}\left[\frac{1}{4}\eta[\phi, \bar{\phi}]-2\vec{\chi}\cdot \vec{\mathcal{E}}+\vec{\chi}\cdot\vec{H}-\frac{1}{2}(\boldsymbol{\Theta}_s^\dagger\bar{\phi}\cdot B_s+\boldsymbol{\Theta}_s\bar{\phi}\cdot B_s^\dagger)\right]\right\}.
\end{aligned}\fe
Here, $\mathcal{D}$ is a shorthand for the adjoint integral $d^{k^2}$ of $\mathfrak{u}(k)$, and $\mathcal{D}'$ stands for the adjoint integral $d^{(k^2-1)}$ of $\mathfrak{su}(k)$. $\mathfrak{D}$ stands for the fundamental integral $d^k$. The factor $\mathcal{N}_{\text{aux}}=(-2)^{-2k}(C_{D^2}/\pi)^{\frac{7}{2}k^2+2k}$ is included to cancel the Gaussian factor caused by integrating out the auxiliary fields $\vec{H}$.

\subsubsection*{Supersymmetric Localization}

The first step towards solving this matrix integral is to deform the supercharge $Q$. The matrix integral $\mathsf{Z}_k$ has a $U(1)^3$ subgroup within the full symmetry group, under which the fields are charged according to Table $\ref{u1charges}$.
\begin{table}[H]
\centering
\begin{tabular}{|c|c|c|c|c|c|c|c|c|}
\hline
Fields              & $B_1, \boldsymbol{\Theta}_1$ & $B_2, \boldsymbol{\Theta}_2$              & $B_3, \boldsymbol{\Theta}_3$ & $B_4, \boldsymbol{\Theta}_4$              & $\mathcal{\chi}_\mathbb{C}^{(1)}, {H}_\mathbb{C}^{(1)}$ & $ \mathcal{\chi}_\mathbb{C}^{(2)}, {H}_\mathbb{C}^{(2)}$ & $ \mathcal{\chi}_\mathbb{C}^{(3)}, {H}_\mathbb{C}^{(3)}$ & $\chi_{\mathbb{R}}, H_{\mathbb{R}}$ \\ \hline
$U(1)_m$            & 0             & 0                          & -1            & 1                          & 0                            & -1                           & 1                            & 0                                                             \\ \hline
$U(1)_{\epsilon_1}$ & 1             & 0                          & 0             & -1                         & 1                            & 1                            & 0                            & 0                                                             \\ \hline
$U(1)_{\epsilon_2}$ & 0             & 1                          & 0             & -1                         & 1                            & 0                            & -1                           & 0                                                             \\ \hline
Fields              & $I, \mu_I$    & $J^\dagger, \mu_J^\dagger$ & $K, \mu_K$    & $L^\dagger, \mu_L^\dagger$ & $\mathcal{E}_K$              & $\mathcal{E}_L^\dagger$      & $\bar{\phi}, \eta$           & $\phi$                                                        \\ \hline
$U(1)_m$            & 0             & 0                          & -1            & 1                          & -1                           & 1                            & 0                            & 0                                                             \\ \hline
$U(1)_{\epsilon_1}$ & 0             & -1                         & 0             & -1                         & 0                            & -1                           & 0                            & 0                                                             \\ \hline
$U(1)_{\epsilon_2}$ & 0             & -1                         & 0             & -1                         & 0                            & -1                           & 0                            & 0                                                             \\ \hline
\end{tabular}
\caption{$U(1)^3$ charges of open string fields}
\label{u1charges}
\end{table}

We choose an element $(\epsilon_1, \epsilon_2, m)$ in the Lie algebra of $U(1)^3$ and deform the supercharge $Q$ into the following $Q_\epsilon$:
\ie
\begin{gathered}
Q_\epsilon B_s=\boldsymbol{\Theta}_s,\quad Q_\epsilon \boldsymbol{\Theta}_s=[\phi, B_s]+\lambda_s B_s, \quad Q_\epsilon\chi_\mathbf{i}=H_\mathbf{i}, \quad Q_\epsilon H_\mathbf{i}=[\phi, \chi_\mathbf{i}]+\lambda_\mathbf{i} \chi_\mathbf{i}\\
Q_\epsilon I=\mu_I, \quad  Q_\epsilon \mu_I=\phi I,\quad Q_\epsilon J^\dagger = \mu_J^\dagger, \quad Q_\epsilon \mu_J^\dagger=\phi J^\dagger-\epsilon J^\dagger \\
Q_\epsilon \mu_K=K,\quad Q_\epsilon \mu_{L}^\dagger=L^\dagger, \quad Q_\epsilon K=\phi\mu_K-m\mu_K, \quad Q_\epsilon L^\dagger = \phi \mu_L^\dagger+(m-\epsilon)\mu_L^\dagger\\
Q_\epsilon \phi=0,\quad Q_\epsilon \bar{\phi}=\eta,\quad Q_\epsilon \eta=[\phi, \bar{\phi}],
\end{gathered}
\fe
where $\lambda_s=(\epsilon_1, \epsilon_2, -m, m-\epsilon)$, $\lambda_\mathbf{i}=(\lambda_\mathbb{R}; \lambda_{\mathbb{C}}^{(r)})=(0;\epsilon, \epsilon_1-m,m-\epsilon_2)$ and $\epsilon=\epsilon_1+\epsilon_2$. This deformation adds a mass term to $\mu_K, \mu_L$ and it changes the value of $\mathsf{Z}_k$ to $\mathsf{Z}_k(\epsilon_1, \epsilon_2, m)$. However, the $(\epsilon_1, \epsilon_2, m)$ dependence will vanish when we take the limit $m\rightarrow 0$ in the end.

The deformed matrix integral $\mathsf{Z}_k(\epsilon_1, \epsilon_2, m)$ is given by
\ie\begin{aligned}
\mathsf{Z}_k(\epsilon_1, \epsilon_2,m)=\ &2^{-5k^2+4k}(C_{D^2})^{-2k^2-2k+2} \mathcal{N}_\text{aux}\\
 \int &\bigg(\prod_{s=1}^2 \mathcal{D}'B_s \mathcal{D}'B_s^\dagger \mathcal{D}'\boldsymbol{\Theta}_s \mathcal{D}'\boldsymbol{\Theta}_s^\dagger\bigg) \bigg(\prod_{s=3}^4 \mathcal{D}B_s\mathcal{D}B_s^\dagger \mathcal{D}\boldsymbol{\Theta}_s \mathcal{D}\boldsymbol{\Theta}_s^\dagger\bigg)\\
&\quad \prod_{\mathbf{i}=1}^7\mathcal{D}H_\mathbf{i} \prod_{\mathbf{j}=1}^4\mathcal{D}'\chi_\mathbf{j} \prod_{\mathbf{k}=5}^7 \mathcal{D}\chi_{\mathbf{k}} \mathcal{D}\eta \mathcal{D}\phi \mathcal{D}\bar{\phi}\left(\prod_{\Phi=I,J,K,L}\mathfrak{D}\Phi \mathfrak{D}\Phi^\dagger \mathfrak{D}\mu_{\Phi}\mathfrak{D}\mu_{\Phi}^\dagger\right)\\
& \times\exp\left\{-C_{D^2}Q_\epsilon \cdot\mathrm{Tr}\left[\frac{1}{4}\eta[\phi, \bar{\phi}]-2\vec{\chi}\cdot \vec{\mathcal{E}}+\vec{\chi}\cdot\vec{H}-\frac{1}{2}(\boldsymbol{\Theta}_\mathbf{s}^\dagger\bar{\phi}\cdot B_\mathbf{s}+\boldsymbol{\Theta}_\mathbf{s}\bar{\phi}\cdot B_\mathbf{s}^\dagger)\right]\right\}.
\end{aligned}
\label{deformedZk}\fe
By the standard argument in supersymmetric localization, since the integrand is $Q_\epsilon$-closed, we can deform the action by any $Q_\epsilon$-exact term without changing the value of $\mathsf{Z}_k(\epsilon_1, \epsilon_2,m)$. Therefore,
\ie\begin{aligned}
\mathsf{Z}_k(\epsilon_1, \epsilon_2,m)=\ &2^{-5k^2+4k}(C_{D^2})^{-2k^2-2k+2} \mathcal{N}_\text{aux}\\
 \int &\bigg(\prod_{s=1}^2 \mathcal{D}'B_s \mathcal{D}'B_s^\dagger \mathcal{D}'\boldsymbol{\Theta}_s \mathcal{D}'\boldsymbol{\Theta}_s^\dagger\bigg) \bigg(\prod_{s=3}^4 \mathcal{D}B_s \mathcal{D}B_s^\dagger \mathcal{D}\boldsymbol{\Theta}_s \mathcal{D}\boldsymbol{\Theta}_s^\dagger\bigg)\\
&\quad \prod_{\mathbf{i}=1}^7\mathcal{D}H_\mathbf{i} \prod_{\mathbf{j}=1}^4\mathcal{D}'\chi_\mathbf{j} \prod_{\mathbf{k}=5}^7 \mathcal{D}\chi_{\mathbf{k}} \mathcal{D}\eta \mathcal{D}\phi \mathcal{D}\bar{\phi}\left(\prod_{\Phi=I,J,K,L}\mathfrak{D}\Phi \mathfrak{D}\Phi^\dagger \mathfrak{D}\mu_{\Phi}\mathfrak{D}\mu_{\Phi}^\dagger\right)\\
& \times\exp\left\{-C_{D^2}Q_\epsilon \cdot\mathrm{Tr}\left[\vec{\chi}\cdot\vec{H}+\frac{1}{2}(B_\mathbf{s}^\dagger \boldsymbol{\Theta}_\mathbf{s}- B_\mathbf{s} \boldsymbol{\Theta}_\mathbf{s}^\dagger)+\chi_\mathbb{R}\bar{\phi}\right]\right\}.
\end{aligned}\fe
The action is now Gaussian in every fields except $\phi$:
\ie
C_{D^2}Q_\epsilon &\cdot\mathrm{Tr}\left[\vec{\chi}\cdot\vec{H}+\frac{1}{2}(B_\mathbf{s}^\dagger \boldsymbol{\Theta}_\mathbf{s}- B_\mathbf{s} \boldsymbol{\Theta}_\mathbf{s}^\dagger)+\chi_\mathbb{R}\bar{\phi}\right]\\
&=C_{D^2}\mathrm{Tr}\left[\vec{H}\cdot \vec{H}-\vec{\chi}\cdot(\phi+\lambda_\mathbf{i})\vec{\chi}+B_\mathbf{s}^\dagger (\phi+\lambda_{\mathbf{s}})B_\mathbf{s}+\boldsymbol{\Theta}_\mathbf{s}^\dagger \boldsymbol{\Theta}_\mathbf{s}+H_\mathbb{R}\bar{\phi}-\chi_\mathbb{R}\eta\right]\\
&=C_{D^2}\mathrm{Tr}\Bigg[-\frac{1}{4}H_\mathbb{R}^2-\sum_{r=1}^3|H_\mathbb{C}^{(r)}|^2-K^\dagger K-LL^\dagger+\frac{1}{4}\chi_\mathbb{R}\mathrm{ad}(\phi) \chi_\mathbb{R}+\sum_{r=1}^3 \chi_{\mathbb{C}}^{(r)\dagger} (\mathrm{ad}(\phi)+\lambda_\mathbb{C}^{(r)}) \chi_{\mathbb{C}}^{(r)}
\\&\quad \quad \quad \quad +B_s^\dagger(\mathrm{ad}(\phi)+\lambda_s)B_s+\boldsymbol{\Theta}_s^\dagger \boldsymbol{\Theta}_s+ H_\mathbb{R}\bar{\phi}-\chi_\mathbb{R}\eta
\\&\quad \quad \quad \quad + I^\dagger \phi I+ J (\phi-\epsilon) J^\dagger+\mu_I^\dagger\mu_I+\mu_J\mu_J^\dagger+\mu_{K}^\dagger(\phi-m)\mu_{K}+\mu_L (\phi+m-\epsilon)\mu_L^\dagger\Bigg],
\fe
where $\lambda_\mathbf{s}, \lambda_\mathbf{I}$ on the second line are defined as $\lambda_\mathbf{s}=(\lambda_I, \lambda_J^\dagger, \lambda_s)$, $\lambda_\mathbf{I}=(\lambda_K, \lambda_L^\dagger, \lambda_\mathbf{i})$ while $\lambda_K, \lambda_L$ are given by $\lambda_K=-m, \lambda_L^\dagger = m-\epsilon$. $\mathrm{ad}(\phi)\equiv [\phi,\cdot]$ is the adjoint action of $\phi$ on the Lie algebra $\mathfrak{u}(k)$. The matrix integral $\mathsf{Z}_k(\epsilon_1, \epsilon_2,m)$ takes the following form after carrying out the Gaussian integrals explicitly:\footnote{Note that changing variable from $\chi_{5,6,7}$ to $\chi_{\mathbb{C}}^{(3)}, \chi_{\mathbb{C}}^{(3)\dagger}, \chi_\mathbb{R}$ in the $U(1)$ sector will introduce a minus sign from the Jacobian factor.}
\ie
\mathsf{Z}_k(\epsilon_1,\epsilon_2,m)&=(-i)^{k^2}2^{3k^2+6k-4}\pi^{\frac{9}{2}k^2+2k-2}(C_{D^2})^{\frac{3k^2}{2}}\\
&\times \frac{\epsilon}{m(m-\epsilon)} \int \mathcal{D}\phi\ \frac{\mathrm{det}_{\mathbf{3}\otimes \mathfrak{su}(k)}(\mathrm{ad}(\phi)+\lambda_\mathbb{C}^{(r)})}{\det_{\mathbf{4}\otimes \mathfrak{su}(k)}(\mathrm{ad}(\phi)+\lambda_s)} \frac{\mathrm{det}_{KL^\dagger\otimes \mathbf{k}}(\phi+\lambda_{K,L^\dagger})}{\mathrm{det}_{IJ^\dagger\otimes \mathbf{k}}(\phi+\lambda_{I,J^\dagger})}.
\label{Zk1}
\fe

We now proceed to compute the $\phi$-integral over the Lie algebra $\mathfrak{u}(k)$. It can be further reduced to an integral over the Cartan subalgebra $\mathfrak{t}=\mathfrak{u}(1)^k$ of $\mathfrak{u}(k)$ using the Weyl integral formula. To do so, we choose the basis of $\mathfrak{u}(k)$ to be $\{T^\mathtt{a}\}=\{T_\text{U(1)}, H^\mathtt{i}, E_{ij}\}$, where $\{T_{U(1)}, H^\mathtt{i}\}$ belongs to the Cartan subalgebra $\mathfrak{t}$ and $E^{ij}$ corresponds to the nonzero root $e^i-e^j$. We also split the $\mathfrak{u}(1)$ and $\mathfrak{su}(k)$ part of $\phi$ by $\phi=\phi^\mathtt{a} T^\mathtt{a}=\hat{\phi} T_{U(1)}+\phi'$. Note that every element $\phi'\in \mathfrak{su}(k)$ can be related to $\mathrm{Ad}(U)\phi'\in \mathfrak{t}$ by some adjoint action of $U\in SU(k)/U(1)^{k-1}$, where $U(1)^{k-1}$ is the maximal torus of $SU(k)$. Therefore, by an analogy to the Faddeev-Popov procedure, we have
\ie
\int &\mathcal{D}\phi\ f(\phi)=\int d\hat{\phi}\int \mathcal{D}'\phi'\ f(\hat{\phi}+\phi')\\
&=\int d\hat{\phi}\int \mathcal{D}'\phi'\left(\frac{1}{k!}\int_\mathfrak{t} d^{k-1}t \int_{\frac{SU(k)}{U(1)^{k-1}}}d\mu(U)\ \delta(\mathrm{Ad}(U)\phi'-t) \left|\det\left(\frac{\partial(\mathrm{Ad}(U)\phi'-t)}{\partial(X,t)}\right)\right| \right) f(\hat{\phi}+\phi')\\
&=\frac{1}{k!}\int d\hat{\phi}\int_{\mathfrak{t}} d^{k-1}t \int_{\frac{SU(k)}{U(1)^{k-1}}}d\mu(U)\left|\det\left(\frac{\partial(\mathrm{Ad}(U)\phi'-t)}{\partial(X,t)}\right)\right| f(\hat{\phi}+t),
\fe
where the integrand $f(\phi)$ is invariant under the adjoint action of $SU(k)$. $d\mu(U)$ is the measure on the coset space $SU(k)/U(1)^{k-1}$ induced by the Killing metric $(X,Y)=\mathrm{Tr}(XY)$ on $\mathfrak{su}(k)$. The integration measure $d^{k-1}t$ over $\mathfrak{t}$ is locally given by $d^{k-1}t=\prod_{i=1}^{k-1}dt_\mathtt{i}$ where $t=t_\mathtt{i} H^\mathtt{i}$. The factor of $(k!)^{-1}$ is introduced to compensate the overcounting due to the fact that the adjoint action $\mathrm{Ad}:SU(k)/U(1)^{k-1}\times \mathfrak{t}\rightarrow \mathfrak{su}(k)$ has mapping degree $k!$.

The Jacobian can be evaluated as
\ie
\left|\det\left(\frac{\partial(\mathrm{Ad}(U)\phi'-t)}{\partial(X,t)}\right)\right|=\left|(-1)^{k-1}\prod_{i\neq j}i(t_i-t_j)\right|=i^{k^2-k}\prod_{i\neq j}(t_i-t_j),
\fe
where $t_i$ is the diagonal element of $\hat{\phi}+t$. The volume of $SU(k)/U(1)^{k-1}$ is given by:
\ie
\int_{\frac{SU(k)}{U(1)^{k-1}}}d\mu(U)=\frac{\mathrm{Vol}(SU(k))}{\mathrm{Vol}(U(1)^{k-1})}=\frac{\mathrm{Vol}(SU(k))}{(2\pi)^{k-1}\sqrt{k}}.
\fe

Therefore, \eqref{Zk1} can be reduced to
\ie
\mathsf{Z}_k(\epsilon_1,\epsilon_2,m)&=i^{-k}2^{3k^2+6k-4}\pi^{\frac{9}{2}k^2+2k-2}(C_{D^2})^{\frac{3k^2}{2}}\frac{\mathrm{Vol}(SU(k))}{(2\pi)^{k-1}k!\sqrt{k}}\\
				     &\times \frac{\epsilon}{m(m-\epsilon)} \int d^{k}t\ \prod\limits_{i \neq j} t_{i j}\, \frac{\mathrm{det}_{\mathbf{3}\otimes \mathfrak{su}(k)}(\mathrm{ad}(t)+\lambda_\mathbb{C}^{(r)})}{\det_{\mathbf{4}\otimes \mathfrak{su}(k)}(\mathrm{ad}(t)+\lambda_s)} \frac{\mathrm{det}_{KL^\dagger\otimes \mathbf{k}}(t+\lambda_{K,L^\dagger})}{\mathrm{det}_{IJ^\dagger\otimes \mathbf{k}}(t+\lambda_{I,J^\dagger})},
\label{Zk2}
\fe
where we have combined the $\hat{\phi}$ and $\mathfrak{t}$-integrals into $\int d^k t$ over the Cartan subalgebra of $\mathfrak{u}(k)$, parameterized by the diagonal elements $t_i$. The measure $d^k t$ is normalized as $\prod_{i=1}^k dt_i$.

The determinants in the integrand are now diagonal and can be calculated explicitly as
\ie\begin{gathered}
\mathrm{det}_{\mathbf{3}\otimes \mathfrak{su}(k)}(\mathrm{ad}(t)+\lambda_\mathbb{C}^{(r)})=\left[\epsilon(\epsilon_1-m)(m-\epsilon_2)\right]^{k-1}\prod_{i\neq j} (\epsilon+t_{ij})(\epsilon_1-m+t_{ij})(m-\epsilon_2+t_{ij})\\
\mathrm{det}_{\mathbf{4}\otimes \mathfrak{su}(k)}(\mathrm{ad}(t)+\lambda_s)=\left[\epsilon_1\epsilon_2 m(\epsilon-m)\right]^{k-1}\prod_{i\neq j}(\epsilon_1+t_{ij})(\epsilon_2+t_{ij})(t_{ij}-m)(m-\epsilon+t_{ij})\\
\mathrm{det}_{KL^\dagger\otimes \mathbf{k}}(t+\lambda_{K,L^\dagger})=\prod_{i=1}^k(t_i-m)(m-\epsilon+t_i)\\
\mathrm{det}_{IJ^\dagger\otimes \mathbf{k}}(t+\lambda_{I,J^\dagger})=\prod_{i=1}^k t_i (t_i-\epsilon),
\end{gathered}\fe
where $t_{ij}=t_i-t_j$. Plugging the determinants into \eqref{Zk2}, we then get:
\ie
\mathsf{Z}_k(\epsilon_1, \epsilon_2, m)&=i^{-k}2^{3k^2+6k-4}\pi^{\frac{9}{2}k^2+2k-2}(C_{D^2})^{\frac{3k^2}{2}}\frac{\mathrm{Vol}(SU(k))}{(2\pi)^{k-1}k!\sqrt{k}}\\
&\times \frac{\epsilon}{m(m-\epsilon)} \left(\frac{P'(0)}{Q(0)}\right)^{k-1} \int \prod_{i=1}^k dt_i \prod_{i\neq j}\frac{P(t_{ij})}{Q(t_{ij})}\prod_{i=1}^k\frac{(t_i-m)(\epsilon-m-t_i)}{t_i(\epsilon-t_i)},
\label{Zk3}
\fe
where $P(x), Q(x)$ are polynomials
\ie\begin{gathered}
P(x)=x(\epsilon+x)(\epsilon_1-m+x)(m-\epsilon_2+x)\\
Q(x)=(\epsilon_1+x)(\epsilon_2+x)(x-m)(x+m-\epsilon).
\end{gathered}\fe

This integral should be regarded as a contour integral, and the residues at the multivariable poles are given by the Jeffrey–Kirwan residue prescription \cite{Jeffrey:1993cun} (We follow the notations of \cite{Benini:2013xpa})
\ie
\mathrm{JKRes}_{u=0} \frac{\mathrm{d} u_1 \wedge \cdots \wedge \mathrm{d} u_k}{Q_{j_1}(u) \cdots Q_{j_k}(u)}= \begin{cases}\frac{1}{\left|\operatorname{det}\left(\partial Q_{j}/\partial u_i\right)\right|} & \text { if } \vec{v} \in \operatorname{Cone}\left(Q_{j_1} \ldots Q_{j_k}\right) \\ 0 & \text { otherwise },\end{cases}
\fe
where $\vec{v}$ is an auxiliary vector in $(\mathbb{R}^k)^*$ of our choice, $u=0$ is the point where all the linear functionals $Q_j\in (\mathbb{R}^k)^*$ vanish, and $\operatorname{Cone}\left(Q_{j_1} \ldots Q_{j_k}\right)$ is the cone spanned by the $k$ vectors $Q_k$. Note that the integral depends on the chosen vector $\vec{v}$. More specifically, if we write $\vec{v}=(a_1,...,a_k)$, then the signs of $a_i$ will affect the set of multivariate poles we pick up. The physical choice of this vector $\vec{v}$ can be determined by requiring the poles picked up by the physical $\vec{v}$ to coincide with the fixed points of $Q^2_\epsilon$ on the ADHM moduli space.\footnote{The logic is that the integral should be viewed in the first place as the integral of a $U(1)^3$-equivariant form on the ADHM moduli space $\mathcal{M}_\text{ADHM}$, \textit{i.e.} the space of $(B_i, I, J)$ that satisfies the ADHM equations $\mathcal{E}_\mathbf{i}=\mathcal{E}_K=\mathcal{E}_L^\dagger = 0$, with the $U(k)$ gauge redundancy modded out. Such equivariant integrals localize onto fixed points of the $U(1)^3$ action generated by $Q^2_\epsilon$ on $\mathcal{M}_\text{ADHM}$. In physicists' terms, we start from \eqref{deformedZk} and take $C_{D^2}\rightarrow \infty$. After integrating out the multiplet $(\chi, H)$, the integral becomes a $Q_\epsilon$-closed integral on $\mathcal{M}_\text{ADHM}$. Then, in the limit where $C_{D^2}$ goes to infinity, only a discrete set of points where $Q_\epsilon(\text{fermions})=0$, \textit{i.e.} the fixed points of the vector field $Q^2_\epsilon$ on $\mathcal{M}_\text{ADHM}$, will contribute to the integral. The contour integration prescription used above is best understood as a bookkeeping device to keep track of normalization factors and the Gaussian determinants, and the poles picked up by the JK residue should be in one-to-one correspondence with these fixed points on $\mathcal{M}_\text{ADHM}$.} This fixes $\vec{v}$ to be $(1,\cdots,1)$.

The poles picked up by the JK prescription are in one-to-one correspondence with Young diagrams with $k$ boxes\footnote{The poles can be generated by the following: We pick a sequence $(s(1),...,s(k))$ such that $\{s(1),...,s(k)\}=\{1,...,k\}$. Start from the "anchor point" $s(1)$, we take $t_{s(1)}=0$ (which corresponds to the factor $t_i$ in the denominator), and then 'grow' other $t_i$ from the anchor point by taking $t_{s(j)}=t_{s(i)}-E_\alpha$ where $E_\alpha = \epsilon_1, \epsilon_2, -m, -\epsilon+m$ (which corresponds to the factor $t_{ij}+E_{\alpha}$ in $Q(x)$). The $j$-th point $t_{s(j)}$ can be grown from every $t_{s(i)}$ with $i<j$. Moreover, the contour integral has a $S_k$ permutation symmetry, hence if we have a pole at $(t_1, ..., t_k)$, then $(t_{\sigma(1)}, ..., t_{\sigma(k)})$ is also a pole of the integrand. Moreover, since our choice of $\vec{v}$ also enjoys the $S_k$ symmetry, both poles will get picked up by the JK prescription. Therefore, we can only consider the poles generated by the sequence $(s(1),...,s(k))=(1,...,k)$, and the full contour integral will be $k!$ times their contribution.

	Due to possible zero factors in the numerator, not all poles generated by this process contribute to the contour integral. If we grow in the $E_{3}=m, E_4=-m+\epsilon$ direction in the first step, namely taking $t_{2}=m$ or $t_2=-m+\epsilon$, then the numerator factor $(t_i-m)(m-\epsilon+t_i)$ will be 0. Moreover, if we want to grow in the $E_\beta$ direction from $t_{i_2}=t_{i_1}-E_\alpha$ by taking $t_{i_3}=t_{i_2}-E_\beta$, then the numerator factor $P(t_{i_1i_3})$ will be 0. In this case, the pole will contribute to the multivariate residue only if both the points $t_{i_1}-E_\alpha$ and $t_{i_1}-E_\beta$ are already in the set of poles when we grow the point $t_{i_3}=t_{i_1}-E_\alpha-E_\beta$, since then growing the point $t_{i_3}$ introduces a double zero in $Q(x)$, which balances the single zero in $P(x)$. Importantly, this means the $E_3, E_4$ directions will never be grown on, since the first step of growing is prohibited.

After eliminating all the poles with zero contributions, we find that each pole corresponds to a Young diagram $Y$ with $k$ boxes. If we label the boxes in $Y$ by $i\in\{1,...,k\}$ in lexicographic order, each number $i\in\{1,...,k\}$ will be associated with the coordinates $(a_1^{(i)}, a_2^{(i)})$ of the corresponding box. The multivariate pole corresponding to $Y$ is then located at $\{t_{j}=-(a_1^{(j)}-1)\epsilon_1-(a_2^{(j)}-1)\epsilon_2\}$.}\footnote{In equation \eqref{Zk3}, we wrote the integrand in a way that the fundamental determinant contributes $t_i(\epsilon-t_i)$ in the denominator, and hence the poles anchored at $t_i=\epsilon$ will not be picked up by the JK prescription with our choice of $\vec{v}$. This is consistent with the fixed point analysis on the ADHM moduli space, where the fixed points are labeled by Young diagrams and there are no fixed points with the anchor $t_i=\epsilon$.}. In the limit where $m\rightarrow 0$, each Young diagram will contribute a constant factor $k!$ to the JK residue, and hence the $t$-integral evaluates to
\ie
\lim_{m\rightarrow 0}\frac{\epsilon}{m(m-\epsilon)}\left(\frac{P'(0)}{Q(0)}\right)^{k-1} \int \prod_{i=1}^{k} dt_i \prod_{i\neq j} \frac{P(t_{ij})}{Q(t_{ij})} \prod_{i=1}^k \frac{(t_i-m)(m-\epsilon+t_i)}{t_i(t_i-\epsilon)}=(2\pi i)^k k! p(k),
\fe
where the factor $k!$ comes from the $S_k$ symmetry of poles, and the partition number $p(k)$ of $k$ comes from summing over all Young tableaux with $k$ boxes. Note that this result does not depend on the equivariant deformation parameters $\epsilon_1, \epsilon_2$ as expected.

Plugging this into \eqref{Zk3}, we can arrive at the final result of $\mathsf{Z}_k$:
\ie
\mathsf{Z}_k=2^{3k^2+6k-3}\pi^{\frac{9}{2}k^2+2k-1}(C_{D^2})^{\frac{3k^2}{2}}\sqrt{k}p(k)\mathrm{Vol}(SU(k)/\mathbb{Z}_k).
\fe

\clearpage

\printbibliography

@article{Jeffrey:1993cun,
	archiveprefix = {arXiv},
	author = {Jeffrey, L. C. and Kirwan, F. C.},
	eprint = {alg-geom/9307001},
	month = {7},
	title = {{Localization for nonabelian group actions}},
	year = {1993}}

@article{Benini:2013xpa,
	archiveprefix = {arXiv},
	author = {Benini, Francesco and Eager, Richard and Hori, Kentaro and Tachikawa, Yuji},
	doi = {10.1007/s00220-014-2210-y},
	eprint = {1308.4896},
	journal = {Commun. Math. Phys.},
	number = {3},
	pages = {1241--1286},
	primaryclass = {hep-th},
	reportnumber = {IPMU-13-0146, UT-13-29},
	title = {{Elliptic Genera of 2d ${\mathcal{N}}$ = 2 Gauge Theories}},
	volume = {333},
	year = {2015}}

@article{Bergshoeff:1996tu,
	archiveprefix = {arXiv},
	author = {Bergshoeff, E. and Townsend, P. K.},
	doi = {10.1016/S0550-3213(97)00072-2},
	eprint = {hep-th/9611173},
	journal = {Nucl. Phys. B},
	pages = {145--162},
	reportnumber = {DAMTP-R-96-53, UG-8-96},
	title = {{Super D-branes}},
	volume = {490},
	year = {1997}}

@article{Cederwall:1996ri,
	archiveprefix = {arXiv},
	author = {Cederwall, Martin and von Gussich, Alexander and Nilsson, Bengt E. W. and Sundell, Per and Westerberg, Anders},
	doi = {10.1016/S0550-3213(97)00075-8},
	eprint = {hep-th/9611159},
	journal = {Nucl. Phys. B},
	pages = {179--201},
	reportnumber = {GOTEBORG-ITP-96-14, CTP-TAMU-59-96},
	title = {{The Dirichlet super p-branes in ten-dimensional type IIA and IIB supergravity}},
	volume = {490},
	year = {1997}}

@article{Cederwall:1996pv,
	archiveprefix = {arXiv},
	author = {Cederwall, Martin and von Gussich, Alexander and Nilsson, Bengt E. W. and Westerberg, Anders},
	doi = {10.1016/S0550-3213(97)00071-0},
	eprint = {hep-th/9610148},
	journal = {Nucl. Phys. B},
	pages = {163--178},
	reportnumber = {GOTEBORG-ITP-96-13},
	title = {{The Dirichlet super three-brane in ten-dimensional type IIB supergravity}},
	volume = {490},
	year = {1997}}

@article{Aganagic:1996nn,
	archiveprefix = {arXiv},
	author = {Aganagic, Mina and Popescu, Costin and Schwarz, John H.},
	doi = {10.1016/S0550-3213(97)00180-6},
	eprint = {hep-th/9612080},
	journal = {Nucl. Phys. B},
	pages = {99--126},
	reportnumber = {CALT-68-2088},
	title = {{Gauge invariant and gauge fixed D-brane actions}},
	volume = {495},
	year = {1997}}

@article{Aganagic:1996pe,
	archiveprefix = {arXiv},
	author = {Aganagic, Mina and Popescu, Costin and Schwarz, John H.},
	doi = {10.1016/S0370-2693(96)01643-7},
	eprint = {hep-th/9610249},
	journal = {Phys. Lett. B},
	pages = {311--315},
	reportnumber = {CALT-68-2081A},
	title = {{D-brane actions with local kappa symmetry}},
	volume = {393},
	year = {1997}}

@article{Sen:2024zqr,
	archiveprefix = {arXiv},
	author = {Sen, Ashoke},
	doi = {10.1007/JHEP06(2025)225},
	eprint = {2407.06278},
	journal = {JHEP},
	pages = {225},
	primaryclass = {hep-th},
	title = {{D-instanton induced effective action and its gauge invariance}},
	volume = {06},
	year = {2025}}

@article{Bruzzo:2003rw,
	archiveprefix = {arXiv},
	author = {Bruzzo, Ugo and Fucito, Francesco},
	doi = {10.1016/j.nuclphysb.2003.11.033},
	eprint = {math-ph/0310036},
	journal = {Nucl. Phys. B},
	pages = {638--655},
	title = {{Superlocalization formulas and supersymmetric Yang-Mills theories}},
	volume = {678},
	year = {2004}}

@article{Moore:1998et,
	archiveprefix = {arXiv},
	author = {Moore, Gregory W. and Nekrasov, Nikita and Shatashvili, Samson},
	doi = {10.1007/s002200050016},
	eprint = {hep-th/9803265},
	journal = {Commun. Math. Phys.},
	pages = {77--95},
	reportnumber = {CERN-TH-98-83, HUTP-98-A008, ITEP-TH-8-98, NSF-ITP-98-031, YCTP-P6-98},
	title = {{D particle bound states and generalized instantons}},
	volume = {209},
	year = {2000}}

@article{Hollowood:2002zv,
	archiveprefix = {arXiv},
	author = {Hollowood, Timothy J.},
	doi = {10.1016/S0550-3213(02)00558-8},
	eprint = {hep-th/0202197},
	journal = {Nucl. Phys. B},
	pages = {66--94},
	title = {{Testing Seiberg-Witten theory to all orders in the instanton expansion}},
	volume = {639},
	year = {2002}}

@article{Hollowood:2002ds,
	archiveprefix = {arXiv},
	author = {Hollowood, Timothy J.},
	doi = {10.1088/1126-6708/2002/03/038},
	eprint = {hep-th/0201075},
	journal = {JHEP},
	pages = {038},
	reportnumber = {SWAT-328},
	title = {{Calculating the prepotential by localization on the moduli space of instantons}},
	volume = {03},
	year = {2002}}

@article{Dorey:2000zq,
	archiveprefix = {arXiv},
	author = {Dorey, Nick and Hollowood, Timothy J. and Khoze, Valentin V.},
	doi = {10.1088/1126-6708/2001/03/040},
	eprint = {hep-th/0011247},
	journal = {JHEP},
	pages = {040},
	title = {{The D instanton partition function}},
	volume = {03},
	year = {2001}}

@article{Bruzzo:2002xf,
	archiveprefix = {arXiv},
	author = {Bruzzo, Ugo and Fucito, Francesco and Morales, Jose F. and Tanzini, Alessandro},
	doi = {10.1088/1126-6708/2003/05/054},
	eprint = {hep-th/0211108},
	journal = {JHEP},
	pages = {054},
	reportnumber = {ROM2F-2002-27, SISSA-75-2002-FM, PAR-LPTHE-02-54, STR-02-042},
	title = {{Multiinstanton calculus and equivariant cohomology}},
	volume = {05},
	year = {2003}}

@article{Nekrasov:2002qd,
	archiveprefix = {arXiv},
	author = {Nekrasov, Nikita A.},
	doi = {10.4310/ATMP.2003.v7.n5.a4},
	eprint = {hep-th/0206161},
	journal = {Adv. Theor. Math. Phys.},
	number = {5},
	pages = {831--864},
	reportnumber = {ITEP-TH-22-02, IHES-P-04-22},
	title = {{Seiberg-Witten prepotential from instanton counting}},
	volume = {7},
	year = {2003}}

@article{Dorey:2001ym,
	archiveprefix = {arXiv},
	author = {Dorey, Nick and Hollowood, Timothy J. and Khoze, Valentin V.},
	eprint = {hep-th/0105090},
	month = {5},
	reportnumber = {SWAT-IPPP-01-20},
	title = {{Notes on soliton bound state problems in gauge theory and string theory}},
	year = {2001}}

@article{Ishibashi:1996xs,
	archiveprefix = {arXiv},
	author = {Ishibashi, N. and Kawai, H. and Kitazawa, Y. and Tsuchiya, A.},
	doi = {10.1016/S0550-3213(97)00290-3},
	eprint = {hep-th/9612115},
	journal = {Nucl. Phys. B},
	pages = {467--491},
	reportnumber = {KEK-TH-503},
	title = {{A Large N reduced model as superstring}},
	volume = {498},
	year = {1997}}

@article{Dorey:2002ik,
	archiveprefix = {arXiv},
	author = {Dorey, Nick and Hollowood, Timothy J. and Khoze, Valentin V. and Mattis, Michael P.},
	doi = {10.1016/S0370-1573(02)00301-0},
	eprint = {hep-th/0206063},
	journal = {Phys. Rept.},
	pages = {231--459},
	title = {{The Calculus of many instantons}},
	volume = {371},
	year = {2002}}

@article{Sethi:1999qv,
	archiveprefix = {arXiv},
	author = {Sethi, Savdeep and Stern, Mark},
	doi = {10.1088/1126-6708/1999/06/004},
	eprint = {hep-th/9903049},
	journal = {JHEP},
	pages = {004},
	reportnumber = {DUK-CGTP-99-01, IASSNS-HEP-99-17},
	title = {{Supersymmetry and the Yang-Mills effective action at finite N}},
	volume = {06},
	year = {1999}}

@article{Billo:2005fg,
	archiveprefix = {arXiv},
	author = {Billo, Marco and Frau, Marialuisa and Sciuto, Stefano and Vallone, Giuseppe and Lerda, Alberto},
	doi = {10.1088/1126-6708/2006/05/069},
	eprint = {hep-th/0511036},
	journal = {JHEP},
	pages = {069},
	reportnumber = {DFTT-35-2005},
	title = {{Non-commutative (D)-instantons}},
	volume = {05},
	year = {2006}}

@article{Witten:1995gx,
	archiveprefix = {arXiv},
	author = {Witten, Edward},
	doi = {10.1016/0550-3213(95)00625-7},
	eprint = {hep-th/9511030},
	journal = {Nucl. Phys. B},
	pages = {541--559},
	reportnumber = {IASSNS-HEP-95-87},
	title = {{Small instantons in string theory}},
	volume = {460},
	year = {1996}}

@article{Douglas:1995bn,
	archiveprefix = {arXiv},
	author = {Douglas, Michael R.},
	editor = {Baulieu, L. and Kazakov, V. and Picco, M. and Windey, Paul and Di Francesco, P. and Douglas, Michael R.},
	eprint = {hep-th/9512077},
	journal = {NATO Sci. Ser. C},
	pages = {267--275},
	reportnumber = {RU-95-92},
	title = {{Branes within branes}},
	volume = {520},
	year = {1999}}

@article{Bianchi:2015cta,
	archiveprefix = {arXiv},
	author = {Bianchi, Massimo and Morales, Jose Francisco and Wen, Congkao},
	doi = {10.1007/JHEP11(2015)006},
	eprint = {1508.00554},
	journal = {JHEP},
	pages = {006},
	primaryclass = {hep-th},
	reportnumber = {ROM2F-2015-11},
	title = {{Instanton corrections to the effective action of $ \mathcal{N}=4 $ SYM}},
	volume = {11},
	year = {2015}}

@article{Balthazar:2019rnh,
	archiveprefix = {arXiv},
	author = {Balthazar, Bruno and Rodriguez, Victor A. and Yin, Xi},
	eprint = {1907.07688},
	month = {7},
	primaryclass = {hep-th},
	title = {{ZZ Instantons and the Non-Perturbative Dual of c = 1 String Theory}},
	year = {2019}}

@article{Billo:2004zq,
	archiveprefix = {arXiv},
	author = {Billo, Marco and Frau, Marialuisa and Pesando, Igor and Lerda, Alberto},
	doi = {10.1088/1126-6708/2004/05/023},
	eprint = {hep-th/0402160},
	journal = {JHEP},
	pages = {023},
	reportnumber = {DFTT-7-2004},
	title = {{N = 1/2 gauge theory and its instanton moduli space from open strings in RR background}},
	volume = {05},
	year = {2004}}

@article{Green:1997tv,
	archiveprefix = {arXiv},
	author = {Green, Michael B. and Gutperle, Michael},
	doi = {10.1016/S0550-3213(97)00269-1},
	eprint = {hep-th/9701093},
	journal = {Nucl. Phys.},
	pages = {195-227},
	primaryclass = {hep-th},
	reportnumber = {DAMTP-96-104},
	slaccitation = {%%CITATION = HEP-TH/9701093;%%},
	title = {{Effects of D instantons}},
	volume = {B498},
	year = {1997}}

@article{Green:1998by,
	archiveprefix = {arXiv},
	author = {Green, Michael B. and Sethi, Savdeep},
	doi = {10.1103/PhysRevD.59.046006},
	eprint = {hep-th/9808061},
	journal = {Phys. Rev. D},
	pages = {046006},
	reportnumber = {DAMTP-98-96, IASSNS-HEP-98-70},
	title = {{Supersymmetry constraints on type IIB supergravity}},
	volume = {59},
	year = {1999}}

@article{Maccaferri:2018vwo,
	archiveprefix = {arXiv},
	author = {Maccaferri, Carlo and Merlano, Alberto},
	doi = {10.1007/JHEP03(2018)112},
	eprint = {1801.07607},
	journal = {JHEP},
	pages = {112},
	primaryclass = {hep-th},
	title = {{Localization of effective actions in open superstring field theory}},
	volume = {03},
	year = {2018}}

@article{Polchinski:1994fq,
	archiveprefix = {arXiv},
	author = {Polchinski, Joseph},
	doi = {10.1103/PhysRevD.50.R6041},
	eprint = {hep-th/9407031},
	journal = {Phys. Rev.},
	pages = {R6041-R6045},
	primaryclass = {hep-th},
	reportnumber = {NSF-ITP-94-73},
	slaccitation = {%%CITATION = HEP-TH/9407031;%%},
	title = {{Combinatorics of boundaries in string theory}},
	volume = {D50},
	year = {1994}}

@article{Sen:1990hh,
	author = {Sen, Ashoke},
	doi = {10.1016/0550-3213(90)90400-8},
	journal = {Nucl. Phys. B},
	pages = {551--583},
	reportnumber = {TIFR/TH/90-7},
	title = {{On the Background Independence of String Field Theory}},
	volume = {345},
	year = {1990}}

@article{Sen:1993kb,
	archiveprefix = {arXiv},
	author = {Sen, Ashoke and Zwiebach, Barton},
	doi = {10.1016/0550-3213(94)90145-7},
	eprint = {hep-th/9311009},
	journal = {Nucl. Phys. B},
	pages = {580--630},
	reportnumber = {MIT-CTP-2244, TIFR-TH-93-37, IP-BBSR-93-56},
	title = {{Quantum background independence of closed string field theory}},
	volume = {423},
	year = {1994}}

@article{Sen:2017szq,
	archiveprefix = {arXiv},
	author = {Sen, Ashoke},
	doi = {10.1007/JHEP02(2018)155},
	eprint = {1711.08468},
	journal = {JHEP},
	pages = {155},
	primaryclass = {hep-th},
	title = {{Background Independence of Closed Superstring Field Theory}},
	volume = {02},
	year = {2018}}

@article{Sen:2019qqg,
	archiveprefix = {arXiv},
	author = {Sen, Ashoke},
	doi = {10.1007/JHEP03(2020)005},
	eprint = {1908.02782},
	journal = {JHEP},
	pages = {005},
	primaryclass = {hep-th},
	title = {{Fixing an Ambiguity in Two Dimensional String Theory Using String Field Theory}},
	volume = {03},
	year = {2020}}

@article{Sen:2020eck,
	archiveprefix = {arXiv},
	author = {Sen, Ashoke},
	eprint = {2012.11624},
	month = {12},
	primaryclass = {hep-th},
	title = {{D-instantons, String Field Theory and Two Dimensional String Theory}},
	year = {2020}}

@article{Sen:2021jbr,
	archiveprefix = {arXiv},
	author = {Sen, Ashoke},
	eprint = {2104.15110},
	month = {4},
	primaryclass = {hep-th},
	title = {{Muti-instanton Amplitudes in Type IIB String Theory}},
	year = {2021}}

@article{Wang:2015jna,
	archiveprefix = {arXiv},
	author = {Wang, Yifan and Yin, Xi},
	doi = {10.1103/PhysRevD.92.041701},
	eprint = {1502.03810},
	journal = {Phys. Rev. D},
	number = {4},
	pages = {041701},
	primaryclass = {hep-th},
	reportnumber = {MIT-CTP-4644},
	title = {{Constraining Higher Derivative Supergravity with Scattering Amplitudes}},
	volume = {92},
	year = {2015}}

@article{Sen:2021qdk,
	archiveprefix = {arXiv},
	author = {Sen, Ashoke},
	doi = {10.1007/JHEP11(2021)077},
	eprint = {2101.08566},
	journal = {JHEP},
	pages = {077},
	primaryclass = {hep-th},
	title = {{Normalization of D-instanton amplitudes}},
	volume = {11},
	year = {2021}}

@article{Sen:2021tpp,
	archiveprefix = {arXiv},
	author = {Sen, Ashoke},
	doi = {10.1007/JHEP12(2021)146},
	eprint = {2104.11109},
	journal = {JHEP},
	pages = {146},
	primaryclass = {hep-th},
	title = {{Normalization of type IIB D-instanton amplitudes}},
	volume = {12},
	year = {2021}}

@article{Agmon:2022vdj,
	archiveprefix = {arXiv},
	author = {Agmon, Nathan B. and Balthazar, Bruno and Cho, Minjae and Rodriguez, Victor A. and Yin, Xi},
	eprint = {2205.00609},
	month = {5},
	primaryclass = {hep-th},
	title = {{D-instanton Effects in Type IIB String Theory}},
	year = {2022}}

@article{Sen:2020cef,
	archiveprefix = {arXiv},
	author = {Sen, Ashoke},
	doi = {10.1007/JHEP08(2020)075},
	eprint = {2002.04043},
	journal = {JHEP},
	pages = {075},
	primaryclass = {hep-th},
	title = {{D-instanton Perturbation Theory}},
	volume = {08},
	year = {2020}}

@article{Green:2000ke,
	archiveprefix = {arXiv},
	author = {Green, Michael B. and Gutperle, Michael},
	doi = {10.1088/1126-6708/2000/02/014},
	eprint = {hep-th/0002011},
	journal = {JHEP},
	pages = {014},
	reportnumber = {DAMTP-2000-6, HUTP-00-A001},
	title = {{D instanton induced interactions on a D3-brane}},
	volume = {02},
	year = {2000}}

@article{Lin:2015ixa,
	archiveprefix = {arXiv},
	author = {Lin, Ying-Hsuan and Shao, Shu-Heng and Wang, Yifan and Yin, Xi},
	doi = {10.1103/PhysRevD.92.125017},
	eprint = {1503.02077},
	journal = {Phys. Rev. D},
	number = {12},
	pages = {125017},
	primaryclass = {hep-th},
	reportnumber = {MIT-CTP-4648},
	title = {{Higher derivative couplings in theories with sixteen supersymmetries}},
	volume = {92},
	year = {2015}}

@article{Vosmera:2019mzw,
	archiveprefix = {arXiv},
	author = {Vo\v{s}mera, Jakub},
	doi = {10.1007/JHEP12(2019)118},
	eprint = {1910.00538},
	journal = {JHEP},
	pages = {118},
	primaryclass = {hep-th},
	title = {{Generalized ADHM equations from marginal deformations in open superstring field theory}},
	volume = {12},
	year = {2019}}

@phdthesis{Hartl:2011tza,
	author = {H\"artl, Daniel},
	school = {Munich U.},
	title = {{Correlators of Ramond-Neveu-Schwarz Fields in String Theory}},
	year = {2011}}

@article{Billo:2002hm,
	archiveprefix = {arXiv},
	author = {Billo, Marco and Frau, Marialuisa and Pesando, Igor and Fucito, Francesco and Lerda, Alberto and Liccardo, Antonella},
	doi = {10.1088/1126-6708/2003/02/045},
	eprint = {hep-th/0211250},
	journal = {JHEP},
	pages = {045},
	reportnumber = {DFTT-38-2002, ROM2F-2002-28, DSF-23-2002},
	title = {{Classical gauge instantons from open strings}},
	volume = {02},
	year = {2003}}

@article{Mattiello:2018kue,
	archiveprefix = {arXiv},
	author = {Mattiello, Luca and Sachs, Ivo},
	doi = {10.1007/JHEP07(2018)099},
	eprint = {1803.07500},
	journal = {JHEP},
	pages = {099},
	primaryclass = {hep-th},
	reportnumber = {LMU-ASC 14/18, LMU-ASC-14-18},
	title = {{$\mathbb{Z}_2$ boundary twist fields and the moduli space of D-branes}},
	volume = {07},
	year = {2018}}

@article{Douglas:1996uz,
	archiveprefix = {arXiv},
	author = {Douglas, Michael R.},
	doi = {10.1016/S0393-0440(97)00024-7},
	eprint = {hep-th/9604198},
	journal = {J. Geom. Phys.},
	pages = {255--262},
	reportnumber = {RU-96-24},
	title = {{Gauge fields and D-branes}},
	volume = {28},
	year = {1998}}

@article{Mattiello:2019gxc,
	archiveprefix = {arXiv},
	author = {Mattiello, Luca and Sachs, Ivo},
	doi = {10.1007/JHEP11(2019)118},
	eprint = {1902.10955},
	journal = {JHEP},
	pages = {118},
	primaryclass = {hep-th},
	reportnumber = {LMU-ASC 09/19},
	title = {{On Finite-Size D-Branes in Superstring Theory}},
	volume = {11},
	year = {2019}}

@article{Berkovits:2021eny,
	archiveprefix = {arXiv},
	author = {Berkovits, Nathan and Juliatto, Vilson Fabricio and Portugal, Ulisses M.},
	doi = {10.1007/JHEP09(2022)005},
	eprint = {2110.07645},
	journal = {JHEP},
	pages = {005},
	primaryclass = {hep-th},
	title = {{Instanton solutions in open superstring field theory}},
	volume = {09},
	year = {2022}}

@article{Maccaferri:2019ogq,
    author = "Maccaferri, Carlo and Merlano, Alberto",
    title = "{Localization of effective actions in open superstring field theory: small Hilbert space}",
    eprint = "1905.04958",
    archivePrefix = "arXiv",
    primaryClass = "hep-th",
    doi = "10.1007/JHEP06(2019)101",
    journal = "JHEP",
    volume = "06",
    pages = "101",
    year = "2019"
}
\end{document}